\documentclass[12pt]{article}

\usepackage{hyperref}
\hypersetup{
    colorlinks=true,
    linktocpage=true,
    linkcolor=blue,
    urlcolor=blue,
    citecolor=blue,
    filecolor=blue,
    bookmarksopen=true,
    }
\usepackage{amsmath,amssymb,amsthm,mathtools,mathrsfs,amsfonts,latexsym}
\usepackage[italicdiff]{physics}
\usepackage{bm}
\usepackage[all]{xy}
\usepackage{cite}
\usepackage{tikz,tikz-cd}
\usetikzlibrary{intersections,calc}
\usepackage{here}
\usepackage{color}
\usepackage{ulem}
\usepackage[hang,small,bf]{caption}
\usepackage[subrefformat=parens]{subcaption}
\makeatletter
	
	\@addtoreset{equation}{section}

	\@addtoreset{figure}{section}

	\@addtoreset{table}{section}
\makeatother

\newcommand{\sectiono}[1]{\section{#1}\setcounter{equation}{0}}

\newcommand{\bd}[1]{\boldsymbol{#1}}
\newcommand{\id}{\mathbb{I}}

\begin{document}

\baselineskip=17pt

\begin{titlepage}
\rightline{\tt YITP-24-161}
\rightline{\tt KUNS-3026}


\begin{center}
\vskip 2.5cm
{\Large \bf {A consistent light-cone-gauge superstring field theory
}}
\vskip 1.0cm
{\large {Yuji Ando,$^{1\star\dag}$
\renewcommand\thefootnote{\fnsymbol{footnote}}
\footnotetext[2]{Present Address: Mathematical Science Center for Co-creative Society, Tohoku University, Sendai, Miyagi 980-0845, Japan}
\renewcommand\thefootnote{\arabic{footnote}}
Ryota Fujii,$^{2\star}$ Hiroshi Kunitomo,$^{3\star}$\\ 
and Jojiro Totsuka-Yoshinaka$^{4\star}$}}
\vskip 1.0cm

{\it{}$^1$ Degree Programs in Pure and Applied Sciences, Graduate School of Science and Technology,
University of Tsukuba, Tsukuba, Ibaraki 305-8571, Japan}\\
{\it{}$^2$ Department of Physics, Ibaraki University, Mito 310-8512, Japan}\\
{\it {{}$^3$Center for Gravitational Physics and Quantum Information}},\\ 
{\it {Yukawa Institute for Theoretical Physics}}
{\it {Kyoto University, Kyoto 606-8502, Japan}},\\
{\it{}$^4$ Department of Physics, Kyoto University, Kyoto 606-8502, Japan}\\
\vspace{2mm}

{}$^{\star}$\textrm{Email}: ando@het.ph.tsukuba.ac.jp (Y.A.);
22nd102n@vc.ibaraki.ac.jp (R.F.); kunitomo@yukawa.kyoto-u.ac.jp (H.K.);\\
george.yoshinaka@gauge.scphys.kyoto-u.ac.jp (J.T-Y.)

\vskip 1.0cm

{\bf Abstract}
\end{center}
\noindent
Extending a recent development in the bosonic string field theory, we construct a map from 
the Witten-type gauge-invariant superstring field theory based on an $A_{\infty}$ structure to a light-cone-gauge 
superstring field theory via two intermediate theories, which we call the Kaku-type and 
Kugo-Zwiebach-type superstring field theories. We find that a naive extension only gives us 
an inconsistent light-cone-gauge theory that suffers from the well-known problem caused 
by divergence due to collisions of local operators. 
However, we also find that this difficulty may be resolved by considering the stubbed theory 
and propose it as a consistent light-cone-gauge superstring field theory. 
The result possibly gives a proof of the unitarity of the Witten-type superstring field theory.

\renewcommand{\thefootnote}{\fnsymbol{footnote}}

\end{titlepage}

\tableofcontents

\newpage
\renewcommand{\thefootnote}{\arabic{footnote}}

\sectiono{Introduction}

There are two formulations in the string field theory:\footnote{
In this paper, we restrict our discussion to open string field theories.} 
the light-cone gauge formulation \cite{Kaku:1974zz,Kaku:1974fc,Green:1984fu} 
and gauge-invariant formulation \cite{Witten:1985cc,Witten:1986qs,Erler:2016ybs}. 
The former is based only on the physical degrees of freedom, which makes the unitarity of 
the theory manifest. The latter, on the other hand, includes unphysical degrees 
of freedom, although they decouple from the physical sector thanks to the gauge invariance.
Although it seems reasonable to consider the former is obtained from the latter by imposing 
an appropriate gauge condition and eliminating the unphysical degrees of freedom, 
it is not so easy because the interactions in the two formulations have completely different 
forms. Clarifying the relation between these two formulations
has been a significant and long-standing problem in string field theory.

Recently, however, the equivalence of two formulations was explicitly shown 
for the bosonic string field theory \cite{Erler:2020beb} in two steps.
The first step uses the string field theory 
they call the Kugo-Zwiebach theory \cite{Kugo:1992md} 
(HIKKO theory \cite{Hata:1985zu,Hata:1986jd} along with setting $\alpha=p^+$) as 
an intermediate theory. Although this theory is slightly noncovariant, 
it is gauge-invariant theory with light-cone-type interactions.
Since the forms of interactions of the light-cone-gauge and 
Kugo-Zwiebach theories are the same, we can derive the former by
imposing the light-cone gauge condition followed by integrating out 
the nondynamical degrees of freedom. 
This procedure is the same as that in the gauge theory:
imposing the gauge condition $A^-=0$ and then integrating $A^+$.
Another step to show the equivalence is to relate two gauge-invariant theories: 
the gauge-invariant theory with cubic interactions (Witten's theory) 
and the Kugo-Zwiebach theory. 
A field redefinition using a one-parameter family of vertices \cite{Kaku:1987jx}, 
which connects the Witten-type and light-cone-type vertices, 
allows the two theories to be related.

The purpose of this paper is to extend these considerations to the case of 
superstring theory and construct a consistent light-cone-gauge superstring field theory.
Although the same methods and ingredients as the bosonic theory can be used 
in several parts, this extension is not so straightforward 
due to some characteristic differences between the superstring and bosonic 
string cases.
On the side of gauge-invariant formulation, the picture-changing operator (PCO)
is inserted in interaction vertices (string products) so that the superstring field theory has 
an $A_\infty$ algebra structure \cite{Erler:2013xta}. 
In addition, the Ramond string field satisfies an extra constraint restricting 
the ghost-zero-mode dependence. Correspondingly, the symplectic form for 
the Ramond sector is modified to one suitable for the restricted string-field space.
These differences complicate the problem.
On the light-cone-gauge formulation, on the other hand, 
the naive interaction vertices involve the insertion of 
a local operator (supercurrent) \cite{Mandelstam:1974hk,Kaku:1974fc} 
at the interaction points. 
The collision of these local operators causes problematic divergences \cite{Greensite:1986gv,Greensite:1987hm,Green:1987qu},
which makes the theory inconsistent (super Poincar\'{e} noncovariant).
For this reason, a consistent light-cone gauge superstring field theory has not existed for a long time; making it possible 
has also been a long-standing problem.
In this paper, extending the development in the bosonic string field theory,
we attempt to construct a light-cone-gauge superstring field theory
from the gauge-invariant theory by fixing the gauge invariance and eliminating unphysical
degrees of freedom. 
However, we find that a naive extension of the results of the bosonic theory only gives us an inconsistent theory that suffers from the well-known problem 
caused by divergence due to collisions of local operators. So we consider the stubbed theory, in which the operator collisions do not appear since 
the moduli point at which the interaction points coincide is contained in the quartic and higher interactions, in which the divergence is canceled out and does not appear in total.
We claim that the stubbed theory almost certainly avoids the divergence problem and propose it as a consistent 
light-cone-gauge superstring field theory\footnote{A different approach to construct consistent light-cone-gauge
superstring field theory has been given in Ref.~\cite{Baba:2009kr,Baba:2009fi,Baba:2009zm}.}.

This paper is organized as follows. 
In Section \ref{two vector}, we consider two vector spaces, the covariant vector space (in Section \ref{cov H})
and the light-cone-gauge vector space (in Section \ref{LCG H}), in which the gauge-invariant 
and light-cone-gauge superstring fields take values, respectively. The light-cone-gauge vector space is enlarged
by including trivial quartets of unphysical fields to be isomorphic to the covariant space.
Each of vector spaces is associated with a BRST charge, forming the isomorphic BRST complex.
We make an isomorphism in Section \ref{map S} by using a similarity transformation $S$ following the technique 
developed in Ref.~\cite{Aisaka:2004ga}. In Section \ref{tf S}, we give a conjecture that this map gives a specific 
superconformal transformation. Concrete transformations of oscillators are also given. This isomorphism is uplifted
in Section \ref{free theory} to the free superstring field theories. After summarizing free gauge-invariant (in Section \ref{free gauge}) 
and light-cone gauge theories (in Section \ref{free lc}), we show in Section \ref{free g to l} 
how gauge-fixing followed by integration of nondynamical fields yields free light-cone gauge theories from free gauge invariant theories.
The extra constraint in the Ramond sector makes the correspondence a little complicated, 
which we clarify in detail. This consideration is extended to the interacting theories in Section \ref{LCG effective}. 
In Section \ref{gen int}, we briefly summarize the general structure of interacting theories, a cyclic $A_{\infty}$ 
algebra structure. Then, in Section \ref{i-o long}, we show the way to completely integrate out nondynamical fields 
in a generic covariant theory using homological perturbation theory. We obtain 
a theory containing only physical degrees of freedom in a closed form that we call the light-cone effective theory.
In particular, if we start from the Kugo-Zwiebach-type gauge-invariant theory we obtain a light-cone gauge effective theory 
with light-cone-type interactions, which is a possible candidate for a light-cone-gauge superstring field theory.
This theory is investigated in detail in Section \ref{cons lgt}. We confirm that the cubic interactions between
massless states (in Section \ref{ev cub massless}) and general cubic interactions (in Section \ref{gen cubic}) coincide with
those of known light-cone-gauge superstring field theory, and thus, it suffers from a divergence problem. 
Since this divergence problem is caused by a collision of local operators at interaction points, 
we point out that there is a possibility to circumvent it by considering the stubbed theory.
In the stubbed theory, points in the moduli space at which the interaction points coincide are in the region 
of the moduli integral contained in the quartic or higher order interactions. 
In Section \ref{quartic stubbed}, we show that, as an example, the four-Ramond vertex of the massless states is finite. 
We further argue that this is not unique to the four-Ramond vertex of the massless states but is a common property for general 
higher vertices and claim that the stubbed theory has no divergence.
Then, we propose the stubbed light-cone-gauge effective theory constructed in this way as a consistent light-cone-gauge superstring field theory.
This may be sufficient for constructing a consistent light-cone gauge theory, but one may be concerned that the Kugo-Zwiebach-type theory is not completely covariant.
So, in Section \ref{L to W}, we finally relate the Kugo-Zwiebach-type theory to the Witten-type theory using the Kaku-type theory, a one-parameter theory connecting the two.
After briefly summarizing what is the Kaku-type theory in Section \ref{kaku type},
we relate the Kugo-Zwiebach-type theory to the Witten-type theory by field redefinition in Section \ref{field redef}. 
It provides a possible proof of the unitarity of the Witten-type superstring field theory.
Section \ref{concl} is devoted to conclusion and discussion. Three appendices are appended. In Appendix \ref{conventions}, we summarize conventions used in the paper: metric and light-cone coordinate 
(\ref{coord}), oscillators (\ref{oscillators}), spin operators (\ref{spin op}), Fock spaces (\ref{Fock sp}), superconformal field theory (\ref{SCFT}), and GSO projection (\ref{GSO}). 
In Appendix \ref{proof}, we give evidence that the conjecture (\ref{conjecture}) given in Section \ref{tf S} is true.
The method of constructing general superstring products with own picture numbers is given in Appendix \ref{const M}. 

 \section{Covariant versus light-cone gauge quantization}\label{two vector}

Let us consider two vector spaces that result from covariant and light-cone-gauge first quantization. 
These vector spaces define the space of the superstring field, which is the most fundamental building block of 
superstring field theory. We will consider a space-filling $D$-9 brane in a flat space for simplicity.

\subsection{Covariant vector space \texorpdfstring{$\mathcal{H}_{\textrm{cov}}$}{Hcov}}\label{cov H}

In the covariant quantization, the worldsheet boundary conformal field theory consists of 10 free worldsheet bosons,  
the conformal ghost ($b,c$) with central charge $-26$,  10 free chiral worldsheet fermions, and the superconfromal 
ghost ($\beta,\gamma$) with central charge $11$:
\begin{alignat}{2}
X^\mu(z,\bar{z})=&\left(X^\pm(z,\bar{z}),X^i(z,\bar{z})\right),\quad& \psi^\mu(z)=&\left(\psi^\pm(z),\psi^i(z)\right),\quad i\ =\ 1,\cdots,8;\nonumber\\
& \left(b(z), c(z)\right),\quad& &\left(\beta(z), \gamma(z)\right).\nonumber
\end{alignat}
The transverse coordinates, $8$ bosons $X^i(z,\bar{z})$ and $8$ fermions $\psi^i(z)$,
describe physical degrees of freedom. The nonzero modes\footnote{
Zero-modes $x_0^+$ and 
$\alpha_0^+
=\sqrt{2}p^{+}
$ are identified as the light-cone time 
and the string length, respectively.}
of longitudinal coordinates,
$X^\pm(z,\bar{z})$ and $\psi^\pm(z)$ become unphysical together with the ghost systems $(b(z),c(z))$ and $(\beta(z),\gamma(z))$
by the BRST quartet mechanism. Each of the matter and ghost sectors is described by superconformal field theory (SCFT) with central charge $c_{\textrm{m}}=15$ or $c_{\textrm{gh}}=-15$, respectively, as is given in Appendix~\ref{SCFT}. We have two types of such SCFTs, depending on the periodic and anti-periodic boundary conditions for the world-sheet fermions, $\psi^\mu,\beta$ and $ \gamma$.
The vector space $\mathcal{H}^{\textrm{NS}}_{\textrm{cov}}$ consisting of anti-periodic fermions is called the Neveu-Schwartz (NS) sector and the vector space $\mathcal{H}_{\textrm{cov}}^{\textrm{R}}$ consisting of periodic fermions is called the Ramond (R) sector.
By imposing the GSO projection given in Appendix~\ref{GSO}, the spectrum of the two sectors combined becomes supersymmetric.
The NS-sector represents the space-time bosons, and
the R-sector represents the space-time fermions.
Both vector spaces, $\mathcal{H}_{\textrm{cov}}^{\textrm{NS}}$ and $\mathcal{H}_{\textrm{cov}}^{\textrm{R}}$, have an infinite number 
of equivalent representations labeled by an integer and a half-integer picture number, 
respectively. We choose the representation with picture number $-1$ for $\mathcal{H}_{\textrm{cov}}^{\textrm{NS}}$ and the one with picture number $-1/2$ for $\mathcal{H}_{\textrm{cov}}^{\textrm{R}}$. 

The BRST operator
\begin{align}
    Q\ =&\ \oint\frac{dz}{2\pi i}\left( c\left(T_{\textrm{m}}+\frac{1}{2}T_{\textrm{gh}}\right)+\gamma\left(G_{\textrm{m}}+\frac{1}{2}G_{\textrm{gh}}\right)\right),
    \label{BRST operator}
\end{align}
defines BRST chain complex $(\mathcal{H}_{\textrm{cov}}, Q)$, and the physical subspace is obtained as (relative) BRST cohomology class \cite{Lian:1989cy}. It is known that a representative of physical subspace is spanned by the states obtained by applying the DDF operators\footnote{
The fractional power of $\sqrt{2}i\partial X^{+}$ is defined by expansion around the zero mode $\alpha^{+}_{0}$ by assuming $\alpha_0^{+}\neq0$.} 
\cite{Brower:1973iz,Goddard:1972iy,Schwarz:1972asw},
\begin{align}
A^{i}_{n}&=\oint\frac{dz}{2\pi i}\left(\sqrt{2}i\partial X^{i}+\frac{n}{\alpha^{+}_{0}}\psi^{+}\psi^{i}\right)e^{i\frac{n}{p^{+}}X^{{+}}},\\
B^{i}_{r}&=\frac{1}{\sqrt{\alpha^{+}_{0}}}\oint\frac{dz}{2\pi i}\left[
\psi^{i}(\sqrt{2}\,i\partial X^{+})^{1/2}-\frac{\psi^{+}i\partial X^{i}}{(\sqrt{2}\,i\partial X^{+})^{1/2}}+\frac{\psi^{+}\partial\psi^{+}\psi^{i}}{2(\sqrt{2}\,i\partial X^{+})^{3/2}}
\right]e^{i\frac{r}{p^{+}}X^{{+}}},
\end{align}
on the on-shell tachyon state in the NS sector and the on-shell massless spinor state in the R sector. 
For convenience, we split $\mathcal{H}_{\textrm{cov}}$ into the space $\mathcal{H}_{\textrm{DDF}}$ spanned by off-shell DDF states\footnote{
We also halve the degrees of freedom in $\mathcal{H}_{\textrm{DDF}}$ of the R sector so that $G_0=0$ is satisfied when $L_0=0$.
} and its complementary subspace $\mathcal{H}_{\textrm{long}}$:
\begin{equation}
    \mathcal{H}_{\textrm{cov}}=\mathcal{H}_{\textrm{DDF}}\oplus\mathcal{H}_{\textrm{long}}.
\end{equation}

\subsection{Light-cone-gauge vector space \texorpdfstring{$\mathcal{H}_{\textrm{lc}}^{\perp}$}{Hlc}}\label{LCG H}

In light-cone gauge quantization, we use only the physical degrees of freedom: $(X^i(z),\psi^i(z))$ and zero modes $(x^\pm, p^\pm)$.
The system is described by a SCFT with $c=12$, and its energy-momentum (EM) tensor and super-current are given, respectively, by
\begin{align}
    T^{\perp} =\  -\partial X^i\partial X^i - \frac{1}{2}\psi^i\partial\psi^i,\qquad
    G^{\perp} =\  \sqrt{2}i\psi^i\partial X^i.
    \label{lc EM perp}
\end{align}
The vector space $\mathcal{H}_{\textrm{lc}}^{\perp}$ is isomorphic to the subspace $\mathcal{H}_{\textrm{DDF}}$ in the covariant quantization
as shown in the next subsection.
Since the discrepancy in the size of vector spaces, $\mathcal{H}_{\textrm{cov}}$ and $\mathcal{H}_{\textrm{lc}}^{\perp}$, is not convenient for our purpose, we extend $\mathcal{H}_{\textrm{lc}}^{\perp}$ to
\begin{equation}
    \mathcal{H}_{\textrm{lc}}\ =\ \mathcal{H}_{\textrm{lc}}^{\perp}\oplus\mathcal{H}_{\textrm{lc}}^\parallel,
\end{equation}
by including longitudinal modes $(X^\pm(z), \psi^\pm(z))$ and (twisted) ghosts $(\tilde{b}(z),\tilde{c}(z))$, and $(\tilde{\beta}(z),\tilde{\gamma}(z))$. The space $\mathcal{H}_{\textrm{lc}}^\perp$, in which the light-cone gauge superstring field takes value, is obtained as (relative) cohomology class of BRST chain complex $(\mathcal{H}_{\textrm{lc}}, Q^{\textrm{lc}})$ with \textit{the light-cone-gauge BRST operator}, 
\begin{align}
    Q^{\textrm{lc}}\ =\ \alpha_0^+\delta^- + c_0L_0\ \equiv q^{\textrm{lc}}+c_0L_0,\qquad
        \delta^-\ =\ \oint\frac{dz}{2\pi i}\left(\tilde{c}\sqrt{2}i\partial\tilde{X}^{-}+\tilde{\gamma}\psi^-\right),
        \label{lc BRST}
\end{align}
where $\tilde{X}^\pm(z)$ are the non-zero-mode part of longitudinal coordinate $X^\pm(z)$:
\begin{equation}
    X^{\pm}(z)=\frac{1}{2}x^{\pm}-ip^{\pm}\ln(z)+\tilde{X}^{\pm}(z).
\end{equation}
The longitudinal subspace $\mathcal{H}_{\textrm{lc}}^\parallel$ is described by a SCFT with $c=0$. The EM tensor and super-current
are given by
\begin{align}
    T^{\parallel} =&\ -2\partial \tilde{X}^+\partial \tilde{X}^- 
    - \frac{1}{2}\psi^+\partial\psi^- - \frac{1}{2}\psi^-\partial\psi^+
 -\frac{1}{2}\tilde{\beta}\partial\tilde{\gamma}+\frac{1}{2}\partial\tilde{\beta}\tilde{\gamma}-\tilde{b}\partial\tilde{c},\label{lc EM para}\\
    G^{\parallel} =&\ \sqrt{2}\psi^+i\partial \tilde{X}^- + \sqrt{2}\psi^-i\partial \tilde{X}^+
 +\tilde{\beta}\partial\tilde{c}-\tilde{b}\tilde{\gamma},\label{lc SC para}
\end{align}
respectively, so as to be commutative with $Q^{\textrm{lc}}$. The ghosts $(\tilde{b},\tilde{c})$ and $(\tilde{\beta},\tilde{\gamma})$ have dimension $(1,0)$ and $(1/2,1/2)$, respectively.
The twisted and ordinary ghosts are related in conformal frame $z$ of the radial quantization as
\begin{equation}
\tilde{b}(z)= z b(z),\quad \tilde{c}(z)=\frac{1}{z}c(z),\qquad \tilde{\beta}(z)= z\beta(z),\quad \tilde{\gamma}(z)=\frac{1}{z}\gamma(z).
\end{equation}
The total system with the EM tensor and the super-current
\begin{equation}
    T^{\textrm{lc}}\ =\ T^\perp + T^\parallel,\qquad G^{\textrm{lc}}\ =\ G^\perp + G^\parallel,
\end{equation}
is the SCFT with $c=12$. We can characterize the genuine light-cone-gauge vector space $\mathcal{H}_{\textrm{lc}}^{\perp}$ by using
the \textit{mass operator}  $L_0^\parallel$ of the longitudinal states, 
\begin{align}
 L_0^\parallel\ =&\ \sum_{m>0}^\infty\left(\alpha^+_{-m}\alpha^-_m + \alpha^-_{-m}\alpha^+_m\right)
+\sum_{m>0}^\infty m\left(b_{-m}c_m+c_{-m}b_m\right)
\nonumber\\
&\ +\sum_{r>0}^\infty r\left(\psi^+_{-r}\psi^-_r+\psi^-_{-r}\psi^+_r\right)
+\sum_{r>0}^\infty r\left(\beta_{-r}\gamma_r - \gamma_{-r}\beta_r\right),
\end{align}
where $r\in\mathbb{Z}+\kappa$ with $\kappa=1/2$ for the NS sector and with $\kappa=0$ for the R sector. 
It is BRST exact \cite{Erler:2020beb} and can be written as 
\begin{equation}
    L^{\parallel}_{0}=\{Q^{\textrm{lc}}, b^{\parallel}_{0}\}=\{q^{\textrm{lc}}, b^{\parallel}_{0}\},
    \label{BRS exact}
\end{equation}
by introducing
\begin{align}
b^{\parallel}_{0}&=
\frac{1}{\alpha^{+}_{0}}
\left(
\sum_{n\neq0}\alpha^{+}_{-n}b_{n}
-\sum_{r\in\mathbb{Z}+\kappa}r\psi^{+}_{-r}\beta_{r}
\right).\label{b-para}
\end{align} 
We can show the relations
\begin{equation}
(b_0^{\parallel})^2 = 0,\qquad [L_0^{\parallel}, b_0^{\parallel}] = 0.
    \label{b0 relations}
\end{equation}
The subspace $(\mathcal{H}_{\textrm{lc}}^{\textrm{NS}})^{\perp}$ in the NS sector is defined by a constraint
\begin{equation}\label{perp NS}
    L_0^\parallel A\ =\ 0,\quad \leftrightarrow\quad A \in (\mathcal{H}_{\textrm{lc}}^{\textrm{NS}})^{\perp}\subset\mathcal{H}_{\textrm{lc}}^{\textrm{NS}}.
\end{equation}
Denoting the projection operator onto states with $L_0^{\parallel}=0$ as $\delta(L_0^{\parallel})$,
the decomposition of $\mathcal{H}_{\textrm{lc}}^{\textrm{NS}}$ is given by
\begin{equation}
    (\mathcal{H}_{\textrm{lc}}^{\textrm{NS}})^{\perp}\ =\ \delta(L_0^{\parallel})\mathcal{H}_{\textrm{lc}}^{\textrm{NS}},\qquad
        (\mathcal{H}_{\textrm{lc}}^{\textrm{NS}})^{\parallel}\ =\ \left(1-\delta(L_0^{\parallel})\right)\mathcal{H}_{\textrm{lc}}^{\textrm{NS}}.
\end{equation}
On the other hand, the subspace $(\mathcal{H}_{\textrm{lc}}^{\textrm{R}})^{\perp}$ in the R sector is defined by three constraints
\begin{equation}\label{perp R}
    L_0^\parallel A\ =\ \psi^-_0A\ =\ \beta_0A\ =\ 0,\quad \leftrightarrow\quad A \in (\mathcal{H}_{\textrm{lc}}^{\textrm{R}})^{\perp}\subset\mathcal{H}_{\textrm{lc}}^{\textrm{R}},
\end{equation}
and the decomposition is given by
\begin{align}
        (\mathcal{H}_{\textrm{lc}}^{\textrm{R}})^{\perp}\ =&\ 
        \psi_0^{-}\psi_0^{+}\delta(\beta_0)\delta(\gamma_0)\delta(L_0^{\parallel})\mathcal{H}_{\textrm{lc}}^{\textrm{R}},\\
        (\mathcal{H}_{\textrm{lc}}^{\textrm{R}})^{\parallel}\ =&\ \left(1 - \psi_0^{-}\psi_0^{+}\delta(\beta_0)\delta(\gamma_0)\right)\delta(L_0^{\parallel})\,\mathcal{H}_{\textrm{lc}}^{\textrm{R}} 
        \oplus \left(1-\delta(L_0^{\parallel})\right)\mathcal{H}_{\textrm{lc}}^{\textrm{R}}\ 
    =\ (\mathcal{H}_{\textrm{lc}}^{\textrm{R}})^{\parallel}_0 \oplus(\widetilde{\mathcal{H}}_{\textrm{lc}}^{\textrm{R}})^{\parallel}.
\end{align}
We define the projection operators onto $\mathcal{H}_{\textrm{lc}}^{\perp}$ and $\mathcal{H}_{\textrm{lc}}^{\parallel}$ as
\begin{align}
    \mathcal{P}_{\textrm{lc}}^{\perp}\ =&\ \left(\pi^0 + \psi_0^-\psi_0^+\delta(\beta_0)\delta(\gamma_0)\pi^1\right)\delta(L_0^{\parallel}),\label{Plc perp}\\
    \mathcal{P}_{\textrm{lc}}^{\parallel}\ =&\ \left(1-\delta(L_0^{\parallel})\right) + \Big(1-\psi_0^-\psi_0^+\delta(\beta_0)\delta(\gamma_0)\Big)\pi^1\delta(L_0^{\parallel})
    \label{Plc para}.
\end{align}
Note that $(\mathcal{H}_{\textrm{lc}}^{\textrm{R}})^{\parallel}$ has a subspace $(\mathcal{H}_{\textrm{lc}}^{\textrm{R}})^{\parallel}_0$,
where $L_0^{\parallel}=0$. 
We can define a homotopy operator\footnote{
Note that $\gamma_0\beta_0-\psi_0^{+}\psi_0^{-}\ne0$ when acting on the subspace projected by $(1-\psi_0^{-}\psi_0^{+}\delta(\beta_0)\delta(\gamma_0))$.
}
\begin{align}
    Q_{\textrm{lc}}^{+}\ =&\ 
    \frac{b_0^{\parallel}}{L_0^{\parallel}}\left(1-\delta(L_0^{\parallel})\right) 
  +  \frac{\psi_0^+\beta_0}{\alpha_0^+(\gamma_0\beta_0-\psi_0^+\psi_0^-)}\Big(1-\psi_0^-\psi_0^+\delta(\beta_0)\delta(\gamma_0)\Big)\delta(L_0^{\parallel})\pi^1,
\end{align} 
so that it satisfies $\{Q^{\textrm{lc}}, Q_{\textrm{lc}}^{+}\}=\mathcal{P}^{\parallel}_{\textrm{lc}}$.
It guarantees the triviality of the BRST cohomology in $\mathcal{H}_{\textrm{lc}}^{\parallel}$:
\begin{equation}
    Q^{\textrm{lc}}A\ =\ 0,\quad  A\in\mathcal{H}_{\textrm{lc}}^{\parallel}\quad \longrightarrow\quad A\ =\ Q^{\textrm{lc}}(Q_{\textrm{lc}}^{+}A).
\end{equation}

\subsection{Mapping between \texorpdfstring{$\mathcal{H}_{\textrm{cov}}$}{Hcov} and \texorpdfstring{$\mathcal{H}_{\textrm{lc}}$}{Hlc}}\label{map S}

Let us find the isomorphism between two BRST chain complexes $(\mathcal{H}_{\textrm{cov}}, Q)$ and $(\mathcal{H}_{\textrm{lc}}, Q^{\textrm{lc}})$ by straightforward extension of the method used in \cite{Erler:2020beb} for the bosonic case, a similarity transformation $S$ introduced by Aisaka-Kazama\cite{Aisaka:2004ga}:
\begin{equation}\label{similarity tr}
Q=SQ^{^\textrm{lc}}S^{-1}.
\end{equation}
In order to find the similarity transformation $S$, we introduce the grading on the light-cone vector space by an $N$-number counted by the $Q^{\textrm{lc}}$-exact operator,
\begin{align}
N=\{Q^{\textrm{lc}},K\}=\sum_{n\neq0}:(\frac{1}{n}\alpha^{+}_{-n}\alpha^{-}_{n}+c_{-n}b_{n}):
+\sum_{r\in\mathbb{Z}+\kappa}:(\psi^{+}_{-r}\psi^{-}_{r}-\gamma_{-r}\beta_{r}):,
\end{align}
where
\begin{align}
K\equiv\frac{1}{\alpha_0^+}
\left(
\sum_{n\neq0}\frac{1}{n}\alpha^{+}_{-n}b_{n}
-\sum_{r\in\mathbb{Z}+\kappa}\psi^{+}_{-r}\beta_{r}
\right).
\end{align}
Nonzero-mode oscillators $(\alpha^{+}_{n}, \psi^{+}_{n}, c_{n}, \gamma_{r})$ and zero-modes $(\psi_0^+, \gamma_0)$ in the R sector are assigned $N$-number $+1$, while non-zero modes  $(\alpha^{-}_{n}, \psi^{-}_{r}, b_{n}, \beta_{r})$ and zero modes $(\psi_0^-, \beta_0)$ in the R-sector are assigned the $N$-number $-1$.
The BRST operator in Eq.\,(\ref{BRST operator}) can be decomposed to the $N$-number $0, 1$, and $2$ part as:\footnote{
After this work was completed, we found that the same decomposition as (\ref{q decomp}) is used in \cite{Dedushenko:2012ui} to prove the No-Ghost Theorem
for an arbitrary picture number.}
\begin{align}
Q = Q_{0}+Q_{1}+Q_{2},\label{q decomp}
\end{align}
where
\begin{align}
Q_{0}&=\alpha_0^+\delta^-+c_0L_0=Q^{\textrm{lc}},\\
Q_{1}&=\tilde{Q},\\
Q_{2}&=b_{0}M+\alpha_0^-\delta^{+},
\end{align}
where
\begin{align}
L_{0}&=L^{(m)}_{0}+\sum_{n=1}^{\infty}n\left(b_{-n}c_{n}+c_{-n}b_{n}\right)
+\sum_{\substack{r\in\mathbb{Z}+\kappa\\r>0}}r\left(\beta_{-r}\gamma_{r}-\gamma_{-r}\beta_{r}\right)-\kappa,\\
\delta^{\pm}&=\left(
\sum_{n\neq0}c_{-n}\alpha^{\pm}_{n}
+\sum_{r\in\mathbb{Z}+\kappa}\gamma_{-r}\psi^{\pm}_{r}
\right),\\
M&=-\sum_{n}nc_{-n}c_{n}-\sum_{r\in\mathbb{Z}+\kappa}\gamma_{-r}\gamma_{r},\\
\tilde{Q}&=
\left.
\left(
\sum_{n\neq0}c_{-n}L^{(m)}_{n}+\sum_{r\in\mathbb{Z}+\kappa}\gamma_{-r}G^{(m)}_{r}
\right)
\right|_{p_{\pm}=0}
+\sum_{\substack{n,m\neq0\\n+m\neq0}}\frac{1}{2}(n-m)b_{-n-m}c_{n}c_{m}\nonumber\\
&\hspace{5mm}+\sum_{n\neq0}\sum_{r\in\mathbb{Z}+\kappa}
\left(
(r+\frac{n}{2})c_{-n}\beta_{n-r}\gamma_{r}-b_{-n}\gamma_{n-r}\gamma_{r}
\right),
\end{align}
with
\begin{align}
L^{(m)}_{n}&=\frac{1}{2}\sum_{m}:\alpha^{\mu}_{n-m}\alpha_{\mu,m}:
+\sum_{r\in\mathbb{Z}+\kappa}\frac{1}{2}(r-\frac{n}{2}):\psi^{\mu}_{n-r}\psi_{\mu,r}:\\
G^{(m)}_{r}&=\sum_{n}\psi^{\mu}_{r-n}\alpha_{\mu,n}.
\end{align}
We can see the $N$-number zero part $Q_0$ is nothing but the light-cone-gauge BRST operator $Q^{\textrm{lc}}$ in Eq.\,(\ref{lc BRST}).
From the nilpotency of the BRST operator, we have the relations
\begin{subequations}\label{Q in N}
\begin{align}
Q^{2}_{0}&=0,\\
\{Q_{0},Q_{1}\}&=0,\\
Q_{1}^{2}+\{Q_{0},Q_{2}\}&=0,\\
\{Q_{1},Q_{2}\}&=0,\\
Q_{2}^{2}&=0.
\end{align}\end{subequations}
Introducing an $N$-number counting parameter $t$, we define a one-parameter family of BRST operators
\begin{align}
Q(t)=t^{N}Qt^{-N}=Q_{0}+tQ_{1}+t^2Q_{2},
\end{align}
connecting the light-cone BRST operator at $t=0$ to the ordinary BRST operator at $t=1$. 
If there is a similarity transformation in Eq.\,(\ref{similarity tr}) we must have
\begin{align}\label{QSQ}
Q(t)=S(t)Q_{0}S(t)^{-1}, 
\end{align}
with $S(t)=t^{N}St^{-N}$.
This implies the differential equation,
\begin{align}\label{differential eq.}
\frac{d}{dt}Q(t)=[Q(t),r(t)],
\end{align}
where
\begin{align}\label{r=s}
r(t)=S(t)\frac{d}{dt}S(t)^{-1}
\end{align}
holds. We solve Eq.\,(\ref{differential eq.}) to determine $r(t)$, and then, obtain $S(t)$ as a path-ordered exponential. 

Let us solve the differential equation Eq.\,(\ref{differential eq.}). 
From Eq.\,(\ref{QSQ}), we know that $S(0)=\id$, so we can expand $S(t)$ as
\begin{align}
S(t)=\mathbb{I}+tS_{1}+t^{2}S_{2}+\text{higher orders}.
\end{align}
This implies that $r(t)$ has an $N$-number expansion
\begin{align}
r(t)=r_{1}+tr_{2}+t^{2}r_{3}+t^{3}r_{4}+\text{higher orders}.
\end{align}
Equating the same powers of $t$ in the differential Eq.\,(\ref{differential eq.}), we have
\begin{align}\label{eq Q_n}
\begin{split}
    Q_{1}=&\ [Q_{0},r_{1}],\\
2Q_{2}=&\ [Q_{0},r_{2}]+[Q_{1},r_{1}],\\
0=&\ [Q_{0},r_{3}]+[Q_{1},r_{2}]+[Q_{2},r_{1}],\\
0=&\ [Q_{0},r_{4}]+[Q_{1},r_{3}]+[Q_{2},r_{2}],\\
0=&\ [Q_{0},r_{5}]+[Q_{1},r_{4}]+[Q_{2},r_{3}],\\
\vdots&.
\end{split}
\end{align}
Substituting $N=\{Q_0,K\}$ into $Q_1=[N,Q_1]$ and $2Q_2=[N,Q_2]$ and using the relations in Eqs.\,(\ref{Q in N}), we readily see that
\begin{align}
r_{1}&=\{K,Q_{1}\},\\
r_{2}&=\{K,Q_{2}\},
\end{align}
from the first two equations in Eqs.\,(\ref{eq Q_n}). It is not difficult to see that the remaining equations 
in Eqs.\,(\ref{eq Q_n}) are satisfied by setting $r_{n}=0$ for $n\ge3$.
Concretely, we find that
\begin{align}
r_{1}
=&\frac{1}{\alpha_0^+}\left\{
\sum_{n\neq0}\frac{1}{n}\alpha^{+}_{-n}\left(
L^{(m)}_{n}|_{p_{\pm}=0}+\sum_{m\neq0,n}mb_{n-m}c_{m}+\sum_{r\in\mathbb{Z}+\kappa}(r-\frac{n}{2})\beta_{n-r}\gamma_{r}
\right)
\right.\nonumber\\
&\hspace{20mm}
\left.
+\sum_{r\in\mathbb{Z}+\kappa}\psi^{+}_{-r}\left(
G_{r}^{(m)}|_{p_{\pm}=0}-\sum_{n\neq0}b_{-n}\gamma_{n+r}+\sum_{n\neq0}nc_{-n}\beta_{n+r}
\right)
\right\},\\
r_{2}=&\frac{2}{\alpha_0^+}
\left\{
\sum_{n\neq0}\alpha^{+}_{-n}b_{0}c_{n}-\sum_{r\in\mathbb{Z}+\kappa}\psi^{+}_{-r}b_{0}\gamma_{r}
\right\}.
\end{align}
Although the similarity transformation $S(t)$ is obtained as the path-ordered exponential
\begin{equation}
    S(t) = P\exp{-\int_0^t ds\left(r_1+sr_2\right)},
\end{equation}
ordered from right to left as t gets larger, the path-ordering is not necessary since one can verify that $r_{1}$ and $r_{2}$ commute: $[r_{1},r_{2}]=0$.
Then we find that
\begin{align}
S=e^{-R},
\end{align}
with
\begin{align}
R=&\ r_{1}+\frac{1}{2}r_{2}\nonumber\\
=&\frac{1}{\alpha_0^+}\left\{
\sum_{n\neq0}\frac{1}{n}\alpha^{+}_{-n}\left(
L^{(m)}_{n}|_{p_{\pm}=0}+\sum_{m\neq0}mb_{n-m}c_{m}+\sum_{r\in\mathbb{Z}+\kappa}(r-\frac{n}{2})\beta_{n-r}\gamma_{r}
\right)
\right.\nonumber\\
&\hspace{20mm}
\left.
+\sum_{r\in\mathbb{Z}+\kappa}\psi^{+}_{-r}\left(
G_{r}^{(m)}|_{p_{\pm}=0}-\sum_{n}b_{-n}\gamma_{n+r}+\sum_{n\neq0}nc_{-n}\beta_{n+r}
\right)
\right\}
\nonumber\\
=&\frac{1}{\alpha_0^+}
\left\{
\sum_{n\neq0}\frac{1}{n}\alpha^{+}_{-n}L^{\text{lc}}_{n}+\sum_{r\in\mathbb{Z}+\kappa}\psi^{+}_{-r}G^{\text{lc}}_{r}
\right\}
\nonumber\\
=& \frac{1}{\alpha_0^+}\oint\frac{dz}{2\pi i}z\left(\tilde{X}^+T^{\textrm{lc}}+\psi^+G^{\textrm{lc}}\right)(z). 
\end{align}
The isomorphism between the vector spaces $\mathcal{H}_{\textrm{cov}}$ and $\mathcal{H}_{\textrm{lc}}$
is given by the mapping $S$: $\mathcal{H}_{\textrm{cov}}=S\mathcal{H}_{\textrm{lc}}$
with
\begin{equation}
    \mathcal{H}_{\textrm{DDF}}\ =\ S\mathcal{H}_{\textrm{lc}}^{\perp},\qquad
    \mathcal{H}_{\textrm{long}}\ =\ S\mathcal{H}^{\parallel}_{\textrm{lc}}.
\end{equation}

\subsection{Transformation by isomorphism \texorpdfstring{$S$}{S}}\label{tf S}

We use the worldsheet super-space formalism in this section.
To consider the isomorphism $S$ in more detail, let us consider what transformations it induces on oscillators in the light-cone gauge. 
Extending the claim given in \cite{Erler:2020beb}, we make the following conjecture:\footnote{
Although we have not yet proved this conjecture, it seems to be quite natural and 
the Appendix~\ref{proof} provides evidence that supports that it is true.
}

\begin{description}
 \item[\textbf{Conjecture.}]
 Let $\mathcal{O}$ be an operator without contractions with $X^{+}$ or $\psi^{+}$.
Then
 \begin{align}
 S\mathcal{O}S^{-1}=\mathcal{F}^{-1}\circ_{\text{lc}}\mathcal{O},
 \label{conjecture}
 \end{align}
 where $\mathcal{F}^{-1}\circ_{\text{lc}}$ is the inverse of the super-conformal transformation
 $Z'\equiv(z', \theta')=(F_z(Z), F_{\theta}(Z))$ with
 \begin{align}\label{sctr.}
F_z(Z)\ =&\ e^{-i\frac{x^+}{\sqrt{2}\alpha_0^+}}\exp\left[
 \frac{i}{\alpha^{+}_0}\sqrt{2}\boldsymbol{X}^+(Z) \right]\,,\qquad
  F_{\theta}(Z)\ =\ \frac{D_{\theta}F_z(Z)}{(\partial_z F_z(Z))^{\frac{1}{2}}},
 \end{align}
generated by $(T^{\textrm{lc}}, G^{\textrm{lc}})$. We denote a super-field as a boldface symbol
and defined $\boldsymbol{X}^{\mu}(Z)$ by $\sqrt{2}\boldsymbol{X}^{\mu}(Z)=\sqrt{2}X^{\mu}(z)-i\theta\psi^{\mu}(z)$.
\end{description}
From the form of $T^{\textrm{lc}}$ and $G^{\textrm{lc}}$, 
we define three light-cone-gauge super-fields as
\begin{align}
\sqrt{2}iD_\theta\tilde{\boldsymbol{X}}^{\pm}(z) =&\ \psi^{\pm}(z) + \theta \sqrt{2}i\partial \tilde{X}^{\pm}(z),\\
\boldsymbol{\tilde{B}}(Z)\ =&\ -\tilde{\beta}(z)+\theta\tilde{b}(z),\qquad
\boldsymbol{\tilde{C}}(Z)\ =\ \tilde{c}(z)-\theta\tilde{\gamma}(z).
\end{align}
They are the primary super-fields of dimension $1/2,\ 1/2$, and $0$, respectively.
Although $F_z(Z)$ and $F_\theta(Z)$ depend on the operator $\boldsymbol{X}^{+}$,
they behave as ordinary functions when $\mathcal{O}$ has no contractions with $\boldsymbol{X}^{+}$.
The oscillator-mode of each component is obtained by integration over the super-space as
\begin{alignat}{2}
\alpha^{i}_{n}\ =&\ \sqrt{2}\oint dZz^{n}iD_{\theta}\boldsymbol{X}^{i}(Z),&\qquad 
\psi^{i}_{r}\ =&\ \sqrt{2}\oint dZ\theta z^{r} iD_{\theta}\boldsymbol{X}^{i}(Z),\\
\alpha^{\pm}_{n}\ =&\ \sqrt{2}\oint dZz^{n}iD_{\theta}\tilde{\boldsymbol{X}}^{\pm}(Z),\ (n\ne0)&\qquad 
\psi^{\pm}_{r}\ =&\ \sqrt{2}\oint dZ\theta z^{r} iD_{\theta}\tilde{\boldsymbol{X}}^{\pm}(Z),\\
\tilde{b}_{n}\ =&\ \oint dZz^{n}\tilde{\boldsymbol{B}}(Z),&\qquad 
\tilde{\beta}_{r}\ =&\ - \oint dZ\theta z^{r-1/2} \tilde{\boldsymbol{B}}(Z),\\
\tilde{\gamma}_{r}\ =&\ - \oint dZz^{r-1/2}\tilde{\boldsymbol{C}}(Z),&\qquad 
\tilde{c}_{n}\ =&\ \oint dZ\theta z^{n-1}\tilde{\boldsymbol{C}}(Z),
\end{alignat}
where $dZ$ denotes $\frac{dz}{2\pi i}d\theta$.
If we follow the Conjecture, the transformations of the transverse oscillators coincide with the DDF operators: 
\begin{align}
S\alpha^{i}_{n}S^{-1}
&= \mathcal{F}^{-1}\circ_{lc} \left(
\sqrt{2}\oint dZ'(z')^{n}\,iD_{\theta'}\boldsymbol{X}^{i}(Z')\right)\nonumber\\
&= \sqrt{2}\oint dZ'(z')^{n}(D_{\theta'}\theta)\, iD_{\theta}\boldsymbol{X}^{i}(Z)\nonumber\\
&= \sqrt{2}\oint dZ(F_z(Z))^{n} iD_{\theta}\boldsymbol{X}^{i}(Z)\nonumber\\
&=e^{-i\frac{n}{2p^{+}}x^+}\oint\frac{dz}{2\pi i}\left(\sqrt{2}i\partial X^{i}+\frac{n}{\alpha^{+}_{0}}\psi^{+}\psi^{i}\right)e^{i\frac{n}{p^{+}}X^{{+}}}\nonumber\\
&=\ e^{-i\frac{n}{2p^{+}}x^+}A^{i}_{n},\\
S\psi^{i}_{r}S^{-1}&=\mathcal{F}^{-1}\circ_{lc}\left(\sqrt{2}\oint dZ'\theta'(z')^{r-1/2}(iD_{\theta'}\boldsymbol{X}^{i}(Z'))\right)\nonumber\\
&=\sqrt{2}\oint dZ F_{\theta}(Z)(F_z(Z))^{r-1/2}\,iD_{\theta}\boldsymbol{X}^{i}(Z)\nonumber\\
&=e^{-i\frac{r}{2p^{+}}x^{+}}\frac{1}{\sqrt{\alpha^{+}_{0}}}\oint\frac{dz}{2\pi i}
\left[
\psi^{i}(\sqrt{2}i\partial X^{+})^{1/2}-\frac{\sqrt{2}\psi^{+}i\partial X^{i}}{(\sqrt{2}i\partial X^{+})^{1/2}}+\frac{\psi^{+}\partial\psi^{+}\psi^{i}}{2(\sqrt{2}i\partial X^{+})^{3/2}}.
\right]
e^{i\frac{r}{p^{+}}X^{{+}}}\nonumber\\
&=e^{-i\frac{r}{2p^{+}}x^+}B^{i}_{r}.
\end{align}
Furthermore, we can find that the ground state of $\mathcal{H}_{\textrm{lc}}$ is transformed as
\begin{alignat}{3}
S\ket{\ket{k}}\ =&\ \ket{\ket{k}},&\qquad&\textrm{for}&\quad &\mathcal{H}_{\textrm{lc}}^{\textrm{NS}}\,,\\
S\ket{\ket{(-, a);\,k}}\ =&\ \ket{\ket{(-, a);\,k}} - \frac{k^i}{\sqrt{2}k^+}\gamma^i\ket{\ket{(+, a);\,k}},&\qquad&\textrm{for}&\quad &\mathcal{H}_{\textrm{lc}}^R.
\end{alignat}
It is easy to show that $G_0(S\ket{\ket{(-, a);\,k}})=0$ when $k^2=0$\,.
The vector space $\mathcal{H}_{\textrm{DDF}}^{\textrm{NS}}$ with the ghost number one is spanned by the states
\begin{equation}
     B^{i_1}_{-r_1}\cdots B^{i_l}_{-r_l}A^{j_1}_{-n_1}\cdots A^{j_m}_{-n_m}\ket{\ket{k}},\qquad (l=\textrm{odd}).
\end{equation}
The vector space $\mathcal{H}_{\textrm{DDF}}^{\textrm{R}}$ with ghost number one is spanned by the states
\begin{equation}
     B^{i_1}_{-r_1}\cdots B^{i_l}_{-r_l}A^{j_1}_{-n_1}\cdots A^{j_m}_{-n_m}
     \Big(\ket{\ket{(-, a);\,k}} - \frac{k^i}{\sqrt{2}k^+}(\gamma)_{ab}^i\ket{\ket{(+, b);\,k}}\Big),
\end{equation}
where the chirality of the ground state is $\Gamma_{11}=1$ for $l=$ even or 
$\Gamma_{11}=-1$ for $l=$ odd.

The transformation of the longitudinal operators, except for $\alpha^{-}_{n},\psi^{-}_{r}$, can also be calculated according to the Conjecture (\ref{conjecture}) as
\begin{align}
S\alpha^{+}_{n}S^{-1}&=e^{-i\frac{n}{2p^{+}}x^+}A^{+}_{n},\ (n\ne0)
&S\psi^{+}_{r}S^{-1}&=e^{-i\frac{r}{2p^{+}}x^+}B^{+}_{r},\\
Sb_{n}S^{-1}&=e^{-i\frac{n}{2p^{+}}^+}B_{n},
&S\beta_{r}S^{-1}&=e^{-i\frac{r}{2p^{+}}x^+}\mathfrak{B}_{r},\\
S\gamma_{r}S^{-1}&=e^{-i\frac{r}{2p^{+}}x^+}\Gamma_{r}, 
&Sc_{n}S^{-1}&=e^{-i\frac{n}{2p^{+}}x^+}C_{n}, 
\end{align}
where 
\begin{align}
A^{+}_{n}&=-\alpha^{+}_{0}\oint\frac{dz}{2\pi i}\frac{1}{z}e^{i\frac{n}{p^{+}}X^{{+}}},\\
B^{+}_{r}&=\sqrt{\alpha^{+}_{0}}\oint\frac{dz}{2\pi i}
\frac{1}{z}\frac{\psi^{+}}{(\sqrt{2}i\partial X^{+})^{1/2}}
e^{i\frac{r}{p^{+}}X^{{+}}},\\
B_{n}&=\oint\frac{dz}{2\pi i}\left(\tilde{b}-\frac{n}{\alpha^{+}_{0}}\psi^{+}\tilde{\beta}\right)e^{i\frac{n}{p^{+}}X^{+}},\\
\mathfrak{B}_{r}&=\frac{1}{\sqrt{\alpha^{+}_{0}}}\oint\frac{dz}{2\pi i}
\left[
(\sqrt{2}i\partial X^{+})^{1/2}\tilde{\beta}+\frac{\psi^{+}\tilde{b}}{(\sqrt{2}i\partial X^{+})^{1/2}}+\frac{1}{2}\frac{\psi^{+}\partial\psi^{+}\tilde{\beta}}{(\sqrt{2}i\partial X^{+})^{3/2}}
\right]
e^{i\frac{r}{p^{+}}X^{{+}}},\\
\Gamma_{r}&=\frac{1}{\sqrt{\alpha^{+}_{0}}}\oint\frac{dz}{2\pi i}
\left[
(\sqrt{2}i\partial X^{+})^{1/2}\tilde{\gamma}+\frac{\psi^{+}\partial\tilde{c}}{(\sqrt{2}i\partial X^{+})^{1/2}}+\frac{1}{2}\frac{\psi^{+}\partial\psi^{+}\tilde{\gamma}}{(\sqrt{2}i\partial X^{+})^{3/2}}
\right]
e^{i\frac{r}{p^{+}}X^{+}},\\
C_{n}&=\frac{1}{\alpha^{+}_{0}}\oint\frac{dz}{2\pi i}\left(\sqrt{2}i\partial X^{+}\tilde{c}+\psi^{+}\tilde{\gamma}\right)e^{i\frac{n}{p^{+}}X^{+}}.
\end{align}
We also find $Sp^{\pm}S^{-1}=p^{\pm}$ by direct calculation.
We cannot use the Conjecture (\ref{conjecture}) for the transformations of the minus operators $\alpha^{-}_{n}$, $\psi^{-}_{r}$,
which have contraction with $X^+$ or $\psi^+$. Instead, we can calculate them from the relations
\begin{align}
\{Q^{\textrm{lc}},b_{n}\}=\alpha^{+}_0\alpha^{-}_{n}-nc_{0}b_{n},\qquad
[Q^{\textrm{lc}},\beta_{r}]=\alpha^{+}_0\psi^{-}_{r}-rc_{0}\beta_{r}.
\end{align}
Using the fact that
\begin{equation}
    C_0\ =\ Sc_0S^{-1}=\frac{i}{\sqrt{2}\alpha_0^+}[Q,x^+],
\end{equation}
we find that
\begin{align}
S\alpha^{-}_{n}S^{-1}&=e^{-i\frac{n}{2p^{+}}x^+}A^{-}_{n},\qquad
A^-_n\ =\ \{Q, B_n\},\quad(n\ne0),\\
S\psi^{-}_{r}S^{-1}&=
e^{-i\frac{r}{2p^{+}}x^+}B^{-}_{r},\qquad B^-_n\ =\ \{Q, \mathfrak{B}_n\},
\end{align}
and obtain
\begin{align}
A^{-}_{n}
=&\frac{1}{\alpha^{+}_{0}}\oint\frac{dz}{2\pi i}z\Bigg\{
:\!Te^{i\frac{n}{p^{+}}X^{+}}\!:-(bc+\beta\gamma)\partial e^{i\frac{n}{p^{+}}X^{+}}
\nonumber\\
&\hspace{28mm}
-\frac{n}{\alpha^{+}_{0}}:\!\!\left(G+\frac{1}{2}\beta\partial c+b\gamma\right)\psi^+e^{i\frac{n}{p^{+}}X^{+}}\!\!:
-\frac{n}{\alpha_0^+}\beta c\partial\left(\psi^+e^{i\frac{n}{p^{+}}X^{+}}\right)\Bigg\},\\
B^{-}_{r}=&
\frac{1}{(\alpha^{+}_{0})^{\frac{3}{2}}}\oint\frac{dz}{2\pi i}z
\left\{
:\!\!\left((\sqrt{2}i\partial X^{+})^{\frac{1}{2}}+\frac{1}{2}\frac{\psi^{+}\partial\psi^{+}}{(\sqrt{2}i\partial X^{+})^{\frac{3}{2}}}\right)\left(G+\frac{1}{2}\beta\partial c+b\gamma\right)e^{i\frac{r}{p^{+}}X^{+}}\!\!:
\right.\nonumber\\
&\hspace{28mm}
+\beta c\partial\left(\left[(\sqrt{2}i\partial X^{+})^{\frac{1}{2}}+\frac{1}{2}\frac{\psi^{+}\partial\psi^{+}}{(\sqrt{2}i\partial X^{+})^{\frac{3}{2}}}\right]e^{i\frac{r}{p^{+}}X^{+}}\right)\nonumber\\
&\hspace{28mm}
\left.
-:\!T\frac{\psi^{+}}{(\sqrt{2}i\partial X^{+})^{\frac{1}{2}}}e^{i\frac{r}{p^{+}}X^{+}}\!:
+(bc+\beta\gamma)\partial\left(\frac{\psi^{+}}{(\sqrt{2}i\partial X^{+})^{\frac{1}{2}}}e^{i\frac{r}{p^{+}}X^{+}}\right)
\right\}.
\end{align}
It may be worth pointing out that they can also be written as
\begin{align}
    A^-_n\ =&\ \frac{1}{\alpha_0^+}\oint\frac{dz}{2\pi i}\Big\{:\tilde{T}^{\textrm{lc}}e^{i\frac{n}{p^+}X^+}\!:
    -\frac{n}{\alpha_0^+}:\tilde{G}^{\textrm{lc}}\psi^+ e^{i\frac{n}{p^+}X^+}:\Big\}\,,\\
    B^-_r\ =&\ \frac{1}{(\alpha_0^+)^{\frac{3}{2}}}\oint\frac{dz}{2\pi i}\Bigg\{:\left((\sqrt{2}i\partial X^+)^{\frac{1}{2}}+\frac{1}{2}\frac{\psi^+\partial\psi^+}{(\sqrt{2}i\partial X^+)^{\frac{3}{2}}}\right)\tilde{G}^{\textrm{lc}}e^{i\frac{r}{p^+}X^+}:
    \nonumber\\
&\hspace{25mm}    
- :\tilde{T}^{\textrm{lc}}\frac{\psi^+}{(\sqrt{2}i\partial X^+)^{\frac{1}{2}}}e^{i\frac{r}{p^+}X^+}:\Bigg\},
\end{align}
using
\begin{alignat}{2}
    \tilde{T}^{\textrm{lc}}\ =&\ T+\partial(bc+\beta\gamma),&\qquad
    \tilde{G}^\textrm{lc}\ =&\ G-\frac{1}{2}\partial c\beta-c\partial\beta+b\gamma,
    \nonumber
\end{alignat}
which are nothing but the twisted EM tensor and supercurrent without setting $p^{\pm}=0$, respectively.

In order to identify the complementary subspace $\mathcal{H}_{\textrm{long}}$, it is also necessary to know the transformation of $L_0^{\parallel}$. We can calculate it using the relation $L_0^{\parallel}=L_0^{\textrm{lc}}-L_0^{\perp}$ as
\begin{align}
    L_{\textrm{long}}\ =&\ SL_0^{\parallel}S^{-1}\nonumber\\ 
    =& \ L_0^{\textrm{lc}} - \alpha_0^+\oint dZ \frac{1}{(D\Psi^+)^2}\Big(
   \boldsymbol{T}^{\perp}-2\boldsymbol{\mathcal{S}}(Z, \boldsymbol{X}^+)\Big)(Z) - \frac{1}{2},
\end{align}
where $\boldsymbol{T}^{\perp}(Z)$ is the EM tensor superfield defined by
\begin{equation}
    \boldsymbol{T}^{\perp}(Z)\ =\ \frac{1}{2}G^{\perp}(z)+\theta T^{\perp}(z),
\end{equation}
and $\boldsymbol{\mathcal{S}}(Z, \boldsymbol{X}^+)$ denotes the super-Schwarzian derivative,
\begin{equation}
       \boldsymbol{\mathcal{S}}(Z, \boldsymbol{X}^+)\ =\ 
    \frac{D^4\Psi^+(Z)}{D\Psi^+(Z)}-2\frac{D^3\Psi^+(Z)D^2\Psi^+(Z)}{(D\Psi^+(Z))^2},
\end{equation}
with
\begin{equation}
        \Psi^+(Z) =\ \frac{\sqrt{2}iD\boldsymbol{X}^+(Z)}{(\sqrt{2}i\partial\boldsymbol{X}^+(Z))^{\frac{1}{2}}}.
\end{equation}
We can also obtain the image of $b_0^{\parallel}$ in $\mathcal{H}_{\textrm{cov}}$ as
\begin{align}
b_{\text{long}}&=Sb^{\parallel}_{0}S^{-1}\nonumber\\
&=\oint dZ\frac{1}{(D\Psi^+)^2}\left(
    \frac{1}{\sqrt{2}}i\partial\tilde{\boldsymbol{X}}^+\tilde{\boldsymbol{B}} + \frac{1}{\sqrt{2}}iD\tilde{\boldsymbol{X}}^+D\tilde{\boldsymbol{B}}\right)(Z) \nonumber\\
&=b_{0}-\alpha^{+}_{0}\oint\frac{dz}{2\pi i}\left(
\frac{b}{\sqrt{2}i\partial X^{+}}-\frac{1}{4}\frac{\psi^{+}\partial\beta-3\partial\psi^{+}\beta}{(\sqrt{2}i\partial X^{+})^{2}}-\frac{\psi^{+}\partial\psi^{+}b}{(\sqrt{2}i\partial X^{+})^{3}}
\right)(z).
\end{align}
using the fact that $b_0^{\parallel}$ in Eq.\,(\ref{b-para}) is written in the form of the super-space integral of a primary super-field as
\begin{align}
        b^{\parallel}_0\ 
   &= \frac{1}{\alpha_0^+}\oint dZz\left(
    \frac{1}{\sqrt{2}}i\partial\tilde{\boldsymbol{X}}^+\tilde{\boldsymbol{B}} + \frac{1}{\sqrt{2}}iD\tilde{\boldsymbol{X}}^+D\tilde{\boldsymbol{B}}\right)(Z).
\end{align}
The relations
\begin{align}
L_{\text{long}}=\{Q_{B},b_{\text{long}}\},\qquad
(b_{\text{long}})^{2}=0,\qquad [L_{\text{long}},b_{\text{long}}]=0,
\end{align}
in $\mathcal{H}_{\textrm{cov}}$ are mapped from Eqs.\,(\ref{BRS exact}) and (\ref{b0 relations}).

The triviality of the cohomology in $\mathcal{H}_{\textrm{long}}$ can be shown by using the homotopy operator
\begin{align}
    Q^{+}\ =&\ SQ_{\textrm{lc}}^{+}S^{-1}
    \nonumber\\
    =&\ \frac{b_{\textrm{long}}}{L_{\textrm{long}}}\Big(1-\delta(L_{\textrm{long}})\Big)
+ \frac{B_0^+\mathfrak{B}_0}{\alpha_0^+(\Gamma_0\mathfrak{B}_0-B_0^+B_0^-)}\Big(1-B_0^-B_0^+\delta(\mathfrak{B}_0)\delta(\Gamma_0)\Big)\delta(L_{\textrm{long}})\pi^1,
\label{Q+}
\end{align}
satisfying $\{Q,Q^{+}\}=\mathcal{P}_{\textrm{long}}$ with
 \begin{align}
     \mathcal{P}_{\textrm{long}}\ =&\ S\mathcal{P}_{\textrm{lc}}^{\parallel}S^{-1}
     \nonumber\\
     =&\ \Big(1-\delta(L_{\textrm{long}})\Big) + \Big(1-B_0^-B_0^+\delta(\mathfrak{B}_0)\delta(\Gamma_0)\Big)\delta(L_{\textrm{long}})\pi^1.
 \end{align}

\section{Free superstring field theory}\label{free theory}

So far, we have considered the vector spaces of the RNS superstring theory both in the covariant gauge and in the light-cone gauge, and relate them by isomorphism. In this section, we uplift it to the free superstring field theory and clarify how the light-cone-gauge theory is derived from the gauge-invariant theory.

\subsection{Gauge-invariant theory}\label{free gauge}

Let us first consider a gauge-invariant open superstring field theory.
The superstring field $\Psi$ has two components:
\begin{equation}
 \Psi_{\textrm{cov}}\ =\ \Psi_{\textrm{cov}}^{\textrm{NS}} + \Psi_{\textrm{cov}}^{\textrm{R}}\ \in\ 
\mathcal{H}_{\textrm{cov}}^{\textrm{(res)}}\ =\ \mathcal{H}_{\textrm{cov}}^{\textrm{NS}} \oplus \mathcal{H}_{\textrm{cov}}^{\textrm{R(res)}},
\end{equation}
with ghost number one.
The vector space $\mathcal{H}_{\textrm{cov}}^{\textrm{R}(\textrm{res})}$ is the subspace of $\mathcal{H}_{\textrm{cov}}^{\textrm{R}}$ restricted by an extra constraint,
\begin{equation}
 XYA\ =\ A,\qquad A\in\mathcal{H}_{\textrm{cov}}^{\textrm{R}},
\label{constraint}
\end{equation}
to realize the relative BRST cohomology as on-shell physical states\cite{Kunitomo:2015usa,Erler:2016ybs}.
Here, $X$ and $Y$ are the PCO and its inverse, respectively,
defined by
\begin{equation}
 X\ =\ G_0\delta(\beta_0)+b_0\delta'(\beta_0),\qquad Y\ =\ -c_0\delta'(\gamma_0).
\end{equation} 
They satisfy the relations
\begin{equation}\label{XY const}
 XYX\ = X,\quad YXY\ =\ Y,\qquad [Q, X]\ = 0,
\end{equation}
which guarantee that the restriction (\ref{constraint}) preserves the BRST cohomology, $XYQXY=QXY$.
If we extend the (small) vector space $\mathcal{H}_{\textrm{cov}}$ to the large vector space $\mathcal{H}_{\textrm{l}}$,
the PCO $X$ is written as $X=[Q,\Xi]$ with\footnote{
For more precise definition of $\Xi$ in the large vector space, see Ref.\cite{Erler:2016ybs}.} 
$\Xi=\Theta(\beta_0)$. It is crucial for constructing interaction terms with an $A_\infty$-algebra structure.
We can show  that
\begin{align}
[b_{\text{long}}, X]\ =\ 0,
\label{b0 X}
\end{align}
from the concrete calculation, which derives
\begin{equation}
    0\ =\ \{Q,[b_{\textrm{long}}, X]\}\ =\ [L_{\textrm{long}}, X].
    \label{L0 X}
\end{equation}
These relations will be necessary later for the consistency between the constraint (\ref{constraint}) and the light-cone-gauge condition.


To construct an action of superstring field theory, we introduce three symplectic forms $\omega$, $\Omega$, and $\omega_{\textrm{l}}$. The first one, $\omega$ is a natural symplectic form of $\mathcal{H}_{\textrm{cov}}$ defined using the BPZ inner product as
\begin{align}
 \omega(A,B)\ =\ (-1)^{|A|+1}\langle A|B\rangle,\qquad
A,\ B\ \in \mathcal{H}_{\textrm{cov}}.
\end{align}
Introducing the projector $\pi^0$ $(\pi^1)$ onto $\mathcal{H}_{\textrm{cov}}^{\textrm{NS}}$ $(\mathcal{H}_{\textrm{cov}}^{\textrm{R}})$ and defining
$\mathcal{G}$ and $\mathcal{G}^{-1}$ as
\begin{equation}
 \mathcal{G} = \pi^0 + X\pi^1,\qquad
 \mathcal{G}^{-1} = \pi^0 + Y\pi^1,
\end{equation}
we define a natural symplectic form $\Omega$ of $\mathcal{H}_{\textrm{cov}}^{\textrm{(res)}}$ by
\begin{equation}
 \Omega(A,B)\ =\ \omega(A,\mathcal{G}^{-1}B),\qquad 
A,\ B\ \in \mathcal{H}_{\textrm{cov}}^{\textrm{(res)}}.
\end{equation}
The third symplectic form $\omega_{\textrm{l}}$ is the one of the large vector space $\mathcal{H}_{\textrm{l}}$,
\begin{align}
 \omega_{\textrm{l}}(A,B)\ =\ {(-1)^{|A|}}_{\textrm{l}}\langle A|B\rangle_{\textrm{l}},\qquad
A,\ B\ \in \mathcal{H}_{\textrm{l}}, 
\end{align}
defined by the BPZ inner product of $\mathcal{H}_{\textrm{l}}$. It is related to $\omega$ as
\begin{align}
    \omega(A, B)\ =\ \omega_{\textrm{l}}(\xi_0A, B),\qquad A, B\in \mathcal{H}_{\textrm{cov}}\subset \mathcal{H}_{\textrm{l}}.
\end{align}
We also define $\langle\Omega|$, $\bra{\omega}$, and $\langle\omega_{\textrm{l}}|$ by
\begin{align}
 \langle\Omega|(|A\rangle\otimes|B\rangle)\ =&\ \Omega(A,B),\qquad A,B \in \mathcal{H}_{\textrm{cov}}^{(\textrm{res})},\\
  \langle\omega|(|A\rangle\otimes|B\rangle)\ =&\ \omega(A,B),\qquad A,B \in \mathcal{H}_{\textrm{cov}},\\
 \langle\omega_{\textrm{l}}|(|A\rangle\otimes|B\rangle)\ =&\ \omega_{\textrm{l}}(A,B),\qquad A,B \in \mathcal{H}_{\textrm{l}}.
\end{align}

Using the symplectic form $\Omega$, an action for gauge-invariant free theory is given by
\begin{equation}
    S\ =\ \frac{1}{2}\Omega(\Psi_{\textrm{cov}}, Q\Psi_{\textrm{cov}}),
    \label{GI action}
\end{equation}
which is invariant under the gauge-transformation
\begin{equation}
    \delta\Psi_{\textrm{cov}}\ =\ Q\Lambda_{\textrm{cov}},
    \label{G-tf}
\end{equation}
with a gauge parameter $\Lambda_{\textrm{cov}}\in\mathcal{H}_{\textrm{cov}}^{\textrm{(res)}}$. The parameter $\Lambda_{\textrm{cov}}$ has ghost number zero.

\subsection{Light-cone-gauge theory}\label{free lc}

Using an isomorphism found in the previous section, the light-cone-gauge (free) theory can also be written 
in gauge-invariant form. The extended superstring field $\Psi_{\textrm{lc}}$
has ghost number one and takes value in the vector space $\mathcal{H}_{\textrm{lc}}^{\textrm{(res)}}$ restricted by an extra constraint 
\begin{equation}\label{const in lc}
    \mathcal{G}_{\textrm{lc}}\mathcal{G}_{\textrm{lc}}^{-1}\Psi_{\textrm{lc}}\ =\  \Psi_{\textrm{lc}},\qquad
    \Psi_{\textrm{lc}} = \Psi_{\textrm{lc}}^{\perp}+\Psi_{\textrm{lc}}^{\parallel}\in \mathcal{H}_{\textrm{lc}} = \mathcal{H}_{\textrm{lc}}^{\perp} \oplus \mathcal{H}_{\textrm{lc}}^{\parallel},
\end{equation}
where 
\begin{alignat}{2}
        \mathcal{G}_{\textrm{lc}}\ =&\ S^{-1}\mathcal{G}S\ =\ \pi^0+X^{\textrm{lc}}\pi^1,&\qquad \mathcal{G}_{\textrm{lc}}^{-1}\ =&\ S^{-1}\mathcal{G}^{-1}S\ =\ 
    \pi^0+Y^{\textrm{lc}}\pi^1,\\
X^{\textrm{lc}}\ =&\ S^{-1}XS,&\qquad
Y^{\textrm{lc}}\ =&\ S^{-1}YS,\label{XY in lc}
\end{alignat}
so that the map $S$ is also an isomorphism between the restricted BRST chain complexes $(\mathcal{H}_{\textrm{cov}}^{\textrm{(res)}}, Q)$ and $(\mathcal{H}_{\textrm{lc}}^{\textrm{(res)}}, Q^{\textrm{lc}})$.
Defining a symplectic form of $\mathcal{H}_{\textrm{lc}}^{\textrm{(res)}}$ as
\begin{equation}
    \Omega^{\textrm{lc}}(A, B)\ =\ \omega(A,\mathcal{G}_{\textrm{lc}}^{-1}B),\qquad
    A, B\in\mathcal{H}_{\textrm{lc}}^{\textrm{(res)}},
\end{equation}
a gauge-invariant free action for the light-cone-gauge superstring field theory is given by
\begin{align}\label{gauged lc}
    S^{\textrm{lc}}\ =&\ \frac{1}{2}\Omega^{\textrm{lc}}(\Psi_{\textrm{lc}}^{\perp}, c_0L_0\Psi_{\textrm{lc}}^{\perp})
    + \frac{1}{2}\Omega^{\textrm{lc}}(\Psi_{\textrm{lc}}^{\parallel}, Q^{\textrm{lc}}\Psi_{\textrm{lc}}^{\parallel}),
\end{align}
which is invariant under the gauge transformation
\begin{equation}
    \delta\Psi_{\textrm{lc}}^{\perp}\ =\ 0,\qquad \delta\Psi_{\textrm{lc}}^{\parallel}\ =\ Q^{\textrm{lc}}\Lambda_{\textrm{lc}}^{\parallel}.
    \label{gauge tf in lc}
\end{equation}

We can derive genuine light-cone gauge superstring field theory from this gauge-invariant theory
by fixing gauge invariance to the light-cone gauge,
\begin{equation}
    b_0^{\parallel}\left(1-\delta(L_0^{\parallel})\right)
    \Psi_{\textrm{lc}}^{\parallel}\ =\ 0\,,\label{NS lc}
\end{equation}
and then integrating out the remaining nondynamical longitudinal fields.
Note that it is not necessary to impose any condition on a piece 
\begin{equation}
    \big(1-\psi_0^-\psi_0^+\delta(\beta_0)\delta(\gamma_0)\big)\delta(L_0^{\parallel})\pi^1\Psi_{\textrm{lc}}^{\parallel},
\end{equation}
of $\Psi_{\textrm{lc}}^{\parallel}$ because it is invariant under the gauge transformation due to the absence of a ghost number zero state
in $(\mathcal{H}_{\textrm{lc}}^{\textrm{R}})_0$. 
If we take a projection operator onto this gauge slice as
\begin{equation}
    \mathcal{P}_{\textrm{lc}}\ =\ \frac{b_0^{\parallel}}{L_0^{\parallel}}Q^{\textrm{lc}}\left(1-\delta(L_0^{\parallel})\right),
\label{slice NS}
\end{equation}
the gauge-fixed action is given by
\begin{align}
    S_{\textrm{fix}}\ =&\ 
\frac{1}{2}\omega\Big(\Psi_{\textrm{lc}}^{\perp}, \mathcal{K}\Psi_{\textrm{lc}}^{\perp}\Big) 
+ \frac{1}{2}\Omega^{\textrm{lc}}\left(\overline{\Psi}_{\textrm{lc}}^{\parallel}, Q^{\textrm{lc}}\overline{\Psi}_{\textrm{lc}}^{\parallel}\right),
\end{align}
where $\overline{\Psi}_{\textrm{lc}}^{\parallel}$ is the gauge-fixed longitudinal field satisfying Eq.\,(\ref{NS lc}), and the kinetic operator $\mathcal{K}$ becomes
\begin{equation}
    \mathcal{K}\ =\ \left(c_0L_0\right)\pi^0 + \left(c_0\delta(\gamma_0)\frac{\psi_0^+}{\alpha_0^+}L_0\right)\pi^1,
\end{equation}
by using 
\begin{equation}
    Y^{\textrm{lc}}\delta(L_0^{\parallel})\ =\ - \left(
    c_0\delta'(\gamma_0) + \frac{\psi_0^+}{\alpha_0^+}\delta(\gamma_0)
    \right) \delta(L_0^{\parallel}).
    \label{Y in lc}
\end{equation}
Although not clear from the action, the gauge-fixed longitudinal field $\overline{\Psi}_{\textrm{lc}}^{\parallel}$ is nondynamical, 
and we can integrate it out with a (nondynamical) propagator 
\begin{equation}
    \Pi_{\textrm{lc}}^{\parallel}\ =\ Q_{\textrm{lc}}^{+}\mathcal{G}_{\textrm{lc}},
\end{equation}
which satisfies
\begin{equation}
    \{\Pi_{\textrm{lc}}^{\parallel}, Q^{\textrm{lc}}\}\ =\ \mathcal{P}_{\textrm{lc}}^{\parallel}\mathcal{G}_{\textrm{lc}}.
\end{equation}

At the end of this subsection, we point out that, like $Y^{\textrm{lc}}$ in (\ref{Y in lc}), $X^{\textrm{lc}}$ and thus the projector $X^{\textrm{lc}}Y^{\textrm{lc}}$ have simple forms in $(\mathcal{H}_{\textrm{lc}}^{\textrm{R}})_0$\,:
\begin{align}
    X^{\textrm{lc}}\delta(L_0^{\parallel})\ =&\ \Big(
        \big(\alpha_0^+\psi_0^-+\frac{\psi_0^+}{\alpha_0^+}L_0\big)\delta(\beta_0)
        -b_0\left(\psi_0^-\psi_0^+\gamma_0\delta(\beta_0)+\psi_0^+\psi_0^-
        \delta(\beta_0)\gamma_0\right)
        \Big)\delta(L_0^{\parallel}),\\
    X^{\textrm{lc}}Y^{\textrm{lc}}\delta(L_0^{\parallel})\ =&\ \Bigg((\psi_0^-\psi_0^+ + b_0c_0\psi_0^+\psi_0^-)\delta(\beta_0)\delta(\gamma_0)
    \nonumber\\
&\
    +\left(c_0\left(\alpha_0^+\psi_0^-+\frac{\psi_0^+}{\alpha_0^+}L_0\right)-\gamma_0b_0c_0\psi_0^-\psi_0^+\right)\delta(\beta_0)\delta(\gamma_0)\beta_0
    \Bigg)\delta(L_0^{\parallel}).
\end{align}
Taking into account the fact that ghost number one states in $(\mathcal{H}_{\textrm{lc}}^{\textrm{R}})_0$ are only in the space further restricted by the projector $b_0c_0\delta(\beta_0)\delta(\gamma_0)$,\footnote{
General states projected by $\delta(L_0^{\parallel})$ have the form $\sum_n(\gamma_0)^n(\ket{A_n}+c_0\ket{B_n})$, where $\ket{A_n}$ and $\ket{B_n}$ denote the states with only transverse excitation. Since $\ket{A_n}$ and $\ket{B_n}$ themselves have ghost number one, the states with ghost number one have the form $\ket{A_0}$.
} the constraint (\ref{XY in lc}) does not impose any restriction on $(\mathcal{H}_{\textrm{lc}}^{\textrm{R}})_0$:
\begin{equation}
    X^{\textrm{lc}}Y^{\textrm{lc}}\delta(L_0^{\parallel})b_0c_0\delta(\beta_0)\delta(\gamma_0)\ =\ \delta(L_0^{\parallel})b_0c_0\delta(\beta_0)\delta(\gamma_0).  
\end{equation}
Therefore, as long as we consider classical actions, there is no need to consider the constraint (\ref{XY in lc}) in the subspace $(\mathcal{H}_{\textrm{lc}}^{\textrm{R}})_0$.

\subsection{From gauge-invariant theory to light-cone-gauge theory}\label{free g to l}

Now, we are ready to consider the gauge fixing from the gauge-invariant theory to the light-cone-gauge theory.
Using the isomorphism $S$ we identify $\Psi_{\textrm{lc}}^{\perp}$
in $\Psi_{\textrm{cov}}$ as
\begin{equation}
    \Psi_{\textrm{cov}}\ =\ S\Psi_{\textrm{lc}}^{\perp} + \Psi_{\textrm{long}}.
\end{equation}
The gauge conditions mapped from Eq.\,(\ref{NS lc}) can be written as
\begin{align} 
    b_{\textrm{long}}\Big(1-\delta(L_{\textrm{long}})\Big)\Psi_{\textrm{long}}\ =&\ 0\,.\label{gauge fixing condition}
\end{align}
The projector onto the gauge slice is similarly mapped from Eq.\,(\ref{slice NS}) to
\begin{align}
    \mathcal{P}_{\textrm{cov}}\ =&\ \frac{b_{\textrm{long}}}{L_{\textrm{long}}}Q\left(1-\delta(L_{\textrm{long}})\right).
\end{align}
The latter is consistent with the constraint (\ref{constraint}): $[\mathcal{P}_{\textrm{cov}}, \mathcal{G}]=0$.
The gauge-fixed action in this gauge becomes
\begin{align}
    S_{\textrm{fix}}\ =&\ \frac{1}{2}\omega\left(\Psi_{\textrm{lc}}^{\perp}, \mathcal{K}\Psi_{\textrm{lc}}^{\perp}\right)
    + \frac{1}{2}\Omega(\overline{\Psi}_{\textrm{long}}, Q\overline{\Psi}_{\textrm{long}}).
    \label{GF action}
\end{align}
The longitudinal fields are integrating out using the propagator $\Pi_{\textrm{long}}=Q^+\mathcal{G}$ 
that satisfies $\{\Pi_{\textrm{long}}, Q\}=\mathcal{P}_{\textrm{long}}\mathcal{G}$.

To examine what this general argument actually does, we consider the gauge-fixed action for the massless fields.
The massless fields with ghost number one are the gauge field $A_{\mu}$ and an auxiliary field $B$ given by
\begin{align}
\Psi_{\textrm{lc}}^{\perp\,\textrm{NS}}\ =&\ \int\frac{d^{10}k}{(2\pi)^{10}}\ A_{i}(k)\psi^{i}_{-1/2}\ket{\ket{k}},\\
\Psi_{\text{long}}^{\textrm{NS}}\ =&\ \int\frac{d^{10}k}{(2\pi)^{10}}
\left\{
A_{+}(k)\psi^{+}_{-1/2}+A_{-}(k)\psi^{-}_{-1/2}-\frac{i}{\sqrt{2}}B(k)c_0\beta_{-1/2}
\right\}\ket{\ket{k}},
\end{align} 
in the NS-sector and the gaugino $\lambda_{\pm}^a$ given by
\begin{align}\label{massless R}
\Psi_{\text{lc}}^{\perp\,\textrm{R}}\
=\ \sqrt{\alpha_0^+}\int\frac{d^{10}k}{(2\pi)^{10}}\lambda_{-}^a(k)\ket{\ket{(-,a);k}},\qquad
\Psi_{\text{long}}^{\textrm{R}}\ =\ \int\frac{d^{10}k}{(2\pi)^{10}}\lambda_{+}^a(k)\ket{\ket{(+,a);k}},
\end{align}
in the R-sector. The gauge condition (\ref{gauge fixing condition}) imposes the light-cone-gauge condition in the NS sector: 
\begin{align}
A_{-}=0.
\end{align}
There is no gauge-variant field in the R-sector, and the gauge condition (\ref{gauge fixing condition}) requires nothing to the gaugino
in the R sector.

The gauge-fixed action (\ref{GF action}) for the massless fields becomes
\begin{align}
S_{\textrm{fix}}\ =\ \int d^{10}x\left\{
\frac{1}{2}A_{i}\Box A^{i}+B\partial_-A_+ -\frac{1}{2}B^{2}
+ \frac{1}{2}\lambda_{-}^a\Box\lambda_{-}^a 
- \frac{1}{2}\lambda_{+}^a\,i\partial_-\lambda_{+}^a
\right\}.
\end{align}
The remaining longitudinal fields $A_+$, $B$, and $\lambda_{+}^a$ are non-dynamical and can be integrated out
when we assume $k^{+}\neq0$; in other words $\partial_-$ is invertible.

\section{Light-cone-gauge effective field theory}\label{LCG effective}

After briefly introducing general interacting superstring field theory,
let us consider how to derive an interacting light-cone-gauge theory by eliminating the longitudinal degrees of freedom.
According to \cite{Erler:2020beb},
we call the resulting theory containing only physical degrees of freedom the light-cone-gauge effective field theory.
Much of this section is a straightforward extension of similar arguments given in \cite{Erler:2020beb} for bosonic string field theory, but it also includes some discussions specific to the superstring case.

\subsection{Interacting theory}\label{gen int}

The superstring field theories discussed in this paper, whether gauge-invariant or light-cone-gauge theories, interactions are governed by a cyclic $A_\infty$ algebra structure.
Their actions are given in the same form as
\begin{equation}
 S = \sum_{n=0}^\infty\frac{1}{n+2}\Omega(\Psi,M_{n+1}(\underbrace{\Psi,\cdots,\Psi}_{n+1})),
\end{equation}
with $M_1=Q$, or using coalgebra representation
\begin{equation}
 S = \int_0^1 dt\, \Omega\left(\Psi, \pi_1\bd{M}\frac{1}{1-t\Psi}\right),
\end{equation}
with the definition $\frac{1}{1-A}=\sum_{n=0}^\infty A^{\otimes n}$.
The superstring product $M_n$ is a multilinear map
\begin{equation}
 M_{n}:(\mathcal{H}^{(\textrm{res})})^{\otimes n}\rightarrow\mathcal{H}^{(\textrm{res})}.
\end{equation}
It is important that the product $M_n$ itself has own ghost and picture numbers for consistency.
A degree-odd coderivation $\bd{M}$ is defined from the string products $\{M_n\}$ as an operator acting on a tensor algebra $\mathcal{TH}^{(\textrm{res})}=\oplus_{n=0}^\infty(\mathcal{H}^{(\textrm{res})})^{\otimes n}$ as
\begin{equation}
\bd{M}\ =\ \sum_{n,i,j=0}^\infty \id^{\otimes i}\otimes M_n\otimes\id^{\otimes j}.
\end{equation}
A superstring field theory is characterized by its cyclic $A_\infty$ algebra structure
defined by the relations
\begin{equation}
 [\bd{M}, \bd{M}] = 0,\qquad \langle\Omega|\pi_2\bd{M} = 0\,,
\end{equation}
where $\pi_n$ denotes a projector onto the $n$-th component 
$(\mathcal{H}^{(\textrm{res})})^{\otimes n}\subset\mathcal{TH}^{\textrm{(res)}}$.
A genuine light-cone-gauge superstring field theory after eliminating unphysical degrees of freedom has
also the same form 
\begin{equation}
 S_{\textrm{lc}} = \sum_{n=0}^\infty\frac{1}{n+2}\Omega^{\textrm{lc}}
(\Psi_{\textrm{lc}},M^{\textrm{lc}}_{n+1}(\underbrace{\Psi_{\textrm{lc}},\cdots,\Psi_{\textrm{lc}}}_{n+1})),
\label{lc action}
\end{equation}
by using light-cone-gauge superstring products
\begin{equation}
  M_{n}^{\textrm{lc}}:(\mathcal{H}_{\textrm{lc}}^{\perp})^{\otimes n}\rightarrow\mathcal{H}_{\textrm{lc}}^{\perp},
\end{equation}
with $M_1^{\textrm{lc}}=c_0L_0$.
Let us discuss it in more detail next.

\subsection{Integrating out longitudinal fields}\label{i-o long}

\begin{figure}
\centering
    \begin{tikzpicture}
\draw [ultra thick] (8,2.5) -- (8.5,3) -- (8,3.5);
\draw [blue, ultra thick] (8.5,3) -- (9.5,3);
\draw [ultra thick](10,2.5) -- (9.5,3) -- (10,3.5);
\node (+) at (10.5,3) {+};
\draw [ultra thick] (11,2) -- (11.5,2.5) -- (12,2);
\draw [blue, ultra thick] (11.5,2.5) -- (11.5,3.5);
\draw [ultra thick](11,4) -- (11.5,3.5) -- (12,4);
\draw [ultra thick] (5.5,2.5) -- (6,3) -- (6.5,2.5);
\draw [ultra thick] (5.5,3.5) -- (6,3) -- (6.5,3.5);
\node (+) at (7,3) {+};
\coordinate (O') at (6,3);
\fill (O') circle (3pt);
\draw [blue, ultra thick, dashed] (7.75,2) [rounded corners] --(7.75,1.75) -- (12.25,1.75) -- (12.25,4.25) -- (7.75,4.25) -- (7.75,2);
\end{tikzpicture}
\caption {The light-cone-gauge effective four-string vertex has two contributions: original vertex and new contributions (in the blue dotted box) appeared by integrating out the longitudinal fields.}
\label{Fig 1}
\end{figure}
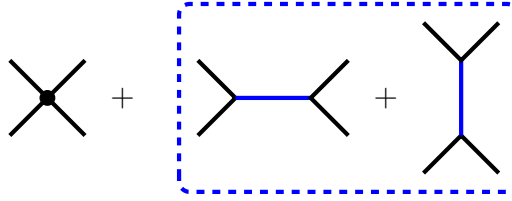

We consider a generic covariant open superstring field theory defined by using the superstring products $\{M_n\}$ which satisfies the relations of a cyclic $A_{\infty}$ algebra $(\mathcal{H}_{\text{cov}},\Omega, \{M_{n}\})$. To eliminate the longitudinal fields, we first fix the gauge invariance by imposing the light-cone-gauge condition (\ref{NS lc}) and then integrating out the remaining longitudinal fields.  It is equivalent to solving the equation of motion for the gauge-fixed longitudinal field,
\begin{equation}
 Q\overline{\Psi}_{\textrm{long}}\ =\ - \mathcal{P}_{\textrm{long}}\Big(M_{2}(\Psi_{\textrm{cov}},\Psi_{\textrm{cov}}) + M_3(\Psi_{\textrm{cov}}, \Psi_{\textrm{cov}}, \Psi_{\textrm{cov}}) + \cdots \Big),    
 \label{long eq}
\end{equation}
and then substituting it in the gauge-fixed action.
Using the homotopy operator $Q^{+}$ introduced in Eq.\,(\ref{Q+}), we have
\begin{align}
    \overline{\Psi}_{\textrm{long}}\ =&\ - Q^{+}\Big(M_2(S\Psi_{\textrm{lc}}^{\perp}+\overline{\Psi}_{\textrm{long}},S\Psi_{\textrm{lc}}^{\perp}+\overline{\Psi}_{\textrm{long}}) 
    \nonumber\\
    &\ + M_3(S\Psi_{\textrm{lc}}^{\perp}+\overline{\Psi}_{\textrm{long}}, S\Psi_{\textrm{lc}}^{\perp}+\overline{\Psi}_{\textrm{long}}, S\Psi_{\textrm{lc}}^{\perp}+\overline{\Psi}_{\textrm{long}}) + \cdots\Big),
\end{align}
where we used $Q^+\overline{\Psi}_{\textrm{long}}=0$.
It can be solved for $\overline{\Psi}_{\textrm{long}}$ in terms of $\Psi_{\textrm{lc}}^{\perp}$ recursively:
\begin{align}
    \overline{\Psi}_{\textrm{long}}\ =&\ -Q^{+}\Big(M_2(S\Psi_{\textrm{lc}}^{\perp}, S\Psi_{\textrm{lc}}^{\perp}) +M_3(S\Psi_{\textrm{lc}}^{\perp},S\Psi_{\textrm{lc}}^{\perp},S\Psi_{\textrm{lc}}^{\perp}) 
\nonumber\\
&\
   - M_2(S\Psi_{\textrm{lc}}^{\perp},Q^{+}M_2(S\Psi_{\textrm{lc}}^{\perp},S\Psi_{\textrm{lc}}^{\perp})) 
    - M_2(Q^{+}M_2(S\Psi_{\textrm{lc}}^{\perp},S\Psi_{\textrm{lc}}^{\perp}),S\Psi_{\textrm{lc}}^{\perp}) + \cdots
    \Big).
\end{align}
Plugging this solutions into the covariant action,
\begin{align}
S_{\text{cov}}\ =\
\sum_{n=1}\frac{1}{n+1}\Omega(\Psi_{\text{cov}},M_{n}(\underbrace{\Psi_{\text{cov}},\cdots,\Psi_{\text{cov}}}_{n\text{ times}}))
\end{align}
we obtain the light-cone-gauge effective field theory
\begin{align}
S_{\text{lc}}=&\frac{1}{2}\Omega^{\text{lc}}(\Psi_{\text{lc}},c_{0}L_{0}\Psi_{\text{lc}})
+\frac{1}{3}\Omega^{\text{lc}}(\Psi_{\text{lc}},S^{-1}M_{2}(S\Psi_{\text{lc}},S\Psi_{\text{lc}}))\nonumber\\
&+\frac{1}{4}\Bigg[
\Omega^{\text{lc}}(\Psi_{\text{lc}},S^{-1}M_{3}(S\Psi_{\text{lc}},S\Psi_{\text{lc}},S\Psi_{\text{lc}}))\Bigg.\nonumber\\
&\hspace{7mm}-\Omega^{\text{lc}}\left(\Psi_{\text{lc}},S^{-1}M_{2}\left(S\Psi_{\text{lc}},Q^{+}M_{2}(S\Psi_{\text{lc}},S\Psi_{\text{lc}})\right)\right)
\nonumber\\
&\hspace{7mm}
-\Omega^{\text{lc}}\left(\Psi_{\text{lc}},S^{-1}M_{2}\left(Q^{+}M_{2}(S\Psi_{\text{lc}},S\Psi_{\text{lc}}),S\Psi_{\text{lc}}\right)\right)\Bigg] + \cdots.
\end{align}
Here, we used that the covariant and light-cone-gauge symplectic form are related by the similarity transformation
\begin{align}
\Omega(SA,SB)=\Omega^{\text{lc}}(A,B),\qquad A,B\in\mathcal{H}^{\perp}_{\text{lc}}.
\end{align}
The terms with $Q^{+}$ produce effective vertices generated by integrating out the longitudinal fields.
There is only one effective vertex at quartic order, but we have expressed it symmetrically using the relations,
\begin{align}
&\Omega\left(M_{2}(S\Psi_{\text{lc}},S\Psi_{\text{lc}}),Q^{+}M_{2}(S\Psi_{\text{lc}},S\Psi_{\text{lc}})\right)\nonumber\\
&\
=-\Omega^{\text{lc}}\left(\Psi_{\text{lc}},S^{-1}M_{2}\left(S\Psi_{\text{lc}},Q^{+}M_{2}(S\Psi_{\text{lc}},S\Psi_{\text{lc}})\right)\right)\nonumber\\
&\
=-\Omega^{\text{lc}}\left(\Psi_{\text{lc}},S^{-1}M_{2}\left(Q^{+}M_{2}(S\Psi_{\text{lc}},S\Psi_{\text{lc}}),S\Psi_{\text{lc}}\right)\right).
\end{align}
The light-cone-gauge effective action can be written in the form of Eq.\,(\ref{lc action}):
\begin{align}
S_{\text{lc}}=&
\frac{1}{2}\Omega^{\text{lc}}\left(\Psi_{\text{lc}},c_{0}L_{0}\Psi_{\text{lc}}\right)
+\frac{1}{3}\Omega^{\text{lc}}\left(\Psi_{\text{lc}},M^{\text{eff}}_{2}(\Psi_{\text{lc}},\Psi_{\text{lc}})\right)\nonumber\\
&+\frac{1}{4}\Omega^{\text{lc}}\left(\Psi_{\text{lc}},M^{\text{eff}}_{3}(\Psi_{\text{lc}},\Psi_{\text{lc}},\Psi_{\text{lc}})\right)
+ \cdots .
\end{align}
with
\begin{align}
M^{\text{eff}}_{1}&=c_{0}L_{0},\\
M^{\text{eff}}_{2}&=\mathcal{P}_{\text{lc}}^{\perp}S^{-1}M_{2}\ S\otimes S,\\
M^{\text{eff}}_{3}&=\mathcal{P}_{\text{lc}}^{\perp}S^{-1}M_{3}\ S\otimes S\otimes S
-\mathcal{P}_{\text{lc}}^{\perp}S^{-1}M_{2}\left(
\mathbb{I}\otimes Q^{+}M_{2}+Q^{+}M_{2}\otimes\mathbb{I}
\right)S\otimes S\otimes S\\
&\vdots\ .\nonumber
\end{align}
The light-cone-gauge effective vertices contain two types of contributions: those from original covariant vertices and from integrating out the longitudinal fields.
We illustrate it for the case of four-string vertex in Fig.~\ref{Fig 1}.
Unlike the bosonic string case \cite{Erler:2020beb}, the contribution from the longitudinal degrees of freedom does not vanish but gives new contact vertices in addition to those existing in the original covariant vertices.

We have written the light-cone-gauge effective action explicitly up to quartic order so far.
The all-order structure can be derived using the homotopy transfer technique\cite{M06}.\footnote{
See Refs.~\cite{Masuda:2020tfa,Koyama:2020qfb,Erbin:2020eyc,Arvanitakis:2020rrk,Arvanitakis:2021ecw} for the recent works 
on homotopy transfer techniques deriving effective field theories.
}
We first define maps between covariant vector space $\mathcal{H}_{\textrm{cov}}$ and the genuine light-cone-gauge vector 
space $\mathcal{H}_{\textrm{lc}}^{\perp}$:
\begin{align}
&\iota:\mathcal{H}^{\perp}_{\text{lc}}\rightarrow\mathcal{H}_{\text{cov}}\ ;&\iota&=S,\\
&\tau:\mathcal{H}_{\text{cov}}\rightarrow\mathcal{H}^{\perp}_{\text{lc}}\ ;&\tau&=\mathcal{P}_{\text{lc}}^{\perp}S^{-1}.
\end{align}
These maps are chain maps between the BRST chain complexes $(\mathcal{H}_{\textrm{cov}}, Q)$ and $(\mathcal{H}_{\textrm{lc}}, 
Q^{\textrm{lc}})$,
and also $(\mathcal{H}_{\textrm{cov}}^{\textrm{(res)}}, Q)$ and $(\mathcal{H}_{\textrm{lc}}^{\textrm{(res)}}, Q^{\textrm{lc}})$: 
they satisfy
\begin{align}
Q\iota=\iota Q^{\text{lc}},\hspace{5mm}\tau Q=Q^{\text{lc}}\tau.
\label{chain maps}
\end{align}
The isomorphism of cohomologies follows the existence of a chain homotopy 
$Q^{+}$ defined by Eq.\,(\ref{Q+}), which satisfies
\begin{align}
\{Q, Q^{+}\}&=\mathbb{I}_{\mathcal{H}_{\textrm{cov}}}-\iota\tau.
\label{Q Qplus}
\end{align}
These chain homotopy and chain maps satisfy additional conditions
\begin{align}
(Q^{+})^{2}=0,\hspace{5mm}\tau Q^{+}=Q^{+}\iota=0,
\label{side}
\end{align}
and the chain maps compose the projection operator on both sides:
\begin{align}
(\iota\tau)^2=\iota\tau,\hspace{5mm}(\tau\iota)^2=\tau\iota.
\label{both proj}
\end{align}
The former is the projector onto $\mathcal{H}_{\textrm{DDF}}$,
\begin{align}
\iota\tau\ =&\ S\mathcal{P}_{\textrm{lc}}^{\perp}S^{-1} 
\nonumber\\
=&\ \mathcal{P}_{\text{DDF}}\ =\ \delta(L_{\textrm{long}})\pi^0
        +B_0^-B_0^+\delta(\mathfrak{B}_0)\delta(\Gamma_0)\delta(L_{\textrm{long}})\pi^1,
\end{align}
and the latter is the identity operator on $\mathcal{H}^{\perp}_{\text{lc}}$,
\begin{align}
\tau\iota=\mathbb{I}_{\mathcal{H}_{\textrm{lc}}^{\perp}}.
\label{id perp}
\end{align}

These, $\iota,\ \tau,\ Q$ and $Q^{+}$, are the objects that relate the free gauge-invariant theory to the free light-cone-gauge theory.
Homological perturbation theory (HPT) allows us to deform these objects into those that relate interacting gauge-invariant theory to interacting light-cone-gauge theory.
The setup for it is first to lift up the objects to those acting on the tensor algebras, $\mathcal{TH}_{\textrm{cov}}$ and $\mathcal{TH}_{\textrm{lc}}^{\perp}$ :
\begin{alignat}{2}
\boldsymbol{\iota}&:T\mathcal{H}^{\perp}_{\text{lc}}\rightarrow T\mathcal{H}_{\text{cov}},&\quad 
\boldsymbol{\tau}&:T\mathcal{H}_{\text{cov}}\rightarrow T\mathcal{H}^{\perp}_{\text{lc}},\\
\boldsymbol{Q}^{\text{lc}}&:T\mathcal{H}_{\text{lc}}\rightarrow T\mathcal{H}_{\text{lc}},&\quad
\boldsymbol{Q}&:T\mathcal{H}_{\text{cov}}\rightarrow T\mathcal{H}_{\text{cov}},\quad
\boldsymbol{Q}^{+}:T\mathcal{H}_{\text{cov}}\rightarrow T\mathcal{H}_{\text{cov}}.
\end{alignat}
Here, $\boldsymbol{\iota}$ and $\boldsymbol{\tau}$ are cohomomorphisms defined by
\begin{align}
\boldsymbol{\iota}\pi_{n}&=(\underbrace{\iota\otimes\cdots\otimes\iota}_{n\text{ times}})\pi_{n}
=(\underbrace{S\otimes\cdots\otimes S}_{n\text{ times}})\pi_{n},\\
\boldsymbol{\tau}\pi_{n}&=(\underbrace{\tau\otimes\cdots\otimes\tau}_{n\text{ times}})\pi_{n}
=(\underbrace{\mathcal{P}_{\text{lc}}S^{-1}\otimes\cdots\otimes\mathcal{P}_{\text{lc}}S^{-1}}_{n\text{ times}})\pi_{n},
\end{align}
$\boldsymbol{Q}$ and $\boldsymbol{Q}^{\text{lc}}$ are degree odd coderivations defined by
\begin{align}
\boldsymbol{Q}\pi_{n}&=
\left(\sum^{n-1}_{k=0}\mathbb{I}^{\otimes k}\otimes Q\otimes\mathbb{I}^{\otimes n-k-1}\right)\pi_{n},\\
\boldsymbol{Q}^{\text{lc}}\pi_{n}&=
\left(\sum^{n-1}_{k=0}\mathbb{I}^{\otimes k}\otimes Q^{\text{lc}}\otimes\mathbb{I}^{\otimes n-k-1}\right)\pi_{n},
\end{align}
and 
\begin{align}
\boldsymbol{Q}^{+}\pi_{n}&=
\left(\sum^{n-1}_{k=0}\mathbb{I}^{\otimes k}\otimes Q^{+}\otimes(\iota\tau)^{\otimes n-k-1}\right)\pi_{n}.
\end{align}
Similar relations to Eqs.\,(\ref{chain maps})-(\ref{both proj}), 
and (\ref{id perp}) are also hold on the tensor algebra:
\begin{subequations}\label{und rel}
\begin{gather}
\boldsymbol{Q}\boldsymbol{\iota}=\boldsymbol{\iota}\boldsymbol{Q}^{\text{lc}},\quad
\boldsymbol{\tau}\boldsymbol{Q}=\boldsymbol{Q}^{\text{lc}}\boldsymbol{\tau},\\
\{\boldsymbol{Q},\boldsymbol{Q}^{+}\}=\mathbb{I}_{T\mathcal{H}_{\text{cov}}}-\boldsymbol{\iota\tau},\\
(\boldsymbol{Q}^{+})^{2}=\boldsymbol{\tau Q}^{+}=\boldsymbol{Q}^{+}\boldsymbol{\iota}=0,\\
(\boldsymbol{\iota\tau})^2=\boldsymbol{\iota\tau},\quad
(\boldsymbol{\tau\iota})^2=\boldsymbol{\tau\iota}=\mathbb{I}_{T\mathcal{H}^{\perp}_{\text{lc}}}.
\end{gather}
\end{subequations}

In the interacting theories, $\boldsymbol{Q}$ and $\boldsymbol{Q}^{\textrm{lc}}$ are deformed to the coderivations $\boldsymbol{M}=\boldsymbol{Q}+\boldsymbol{M}_{\textrm{int}}$ and $\boldsymbol{M}^{\textrm{eff}}=\boldsymbol{Q}^{\textrm{lc}}+\boldsymbol{M}^{\textrm{eff}}_{\textrm{int}}$,
and the BRST chain complexes $(\mathcal{H}_{\textrm{cov}}^{\textrm{(res)}}, Q)$ and $(\mathcal{H}_{\textrm{lc}}^{\textrm{(res)}}, Q^{\textrm{lc}})$ are replaced by the $A_\infty$ algebras
$(\mathcal{H}_{\textrm{cov}}^{\textrm{(res)}}, \boldsymbol{M})$ and $(\mathcal{H}_{\textrm{lc}}^{\textrm{(res)}}, \boldsymbol{M}^{\textrm{eff}})$.
The HPT provides the formulae for giving the deformed objects $\boldsymbol{I}=\boldsymbol{\iota}+\delta\boldsymbol{\iota}$, 
$\boldsymbol{T}=\boldsymbol{\tau}+\delta\boldsymbol{\tau}$, and $\boldsymbol{M}^{+}=\boldsymbol{Q}^{+}+\delta\boldsymbol{Q}^+$,
and $\boldsymbol{M}^{\textrm{eff}}$ in terms of $\boldsymbol{M}_{\textrm{int}}$:
\begin{align}
\mathbf{I}\ =&\
\frac{1}{\mathbb{I}_{T\mathcal{H}_{\text{cov}}}+\boldsymbol{Q}^{+}\boldsymbol{M}}_{\textrm{int}}\boldsymbol{\iota},\\
\mathbf{T}\ =&\
\boldsymbol{\tau}\frac{1}{\mathbb{I}_{T\mathcal{H}_{\text{cov}}}+\boldsymbol{M}_{\textrm{int}}\boldsymbol{Q}^{+}},\\
\boldsymbol{M}^{+}\ =&\
\boldsymbol{Q}^{+}\frac{1}{\mathbb{I}_{T\mathcal{H}_{\text{cov}}}+\boldsymbol{M}_{\textrm{int}}\boldsymbol{Q}^{+}},\\
\boldsymbol{M}^{\text{eff}}\ =&\
\boldsymbol{Q}^{\text{lc}}+\boldsymbol{\tau}\boldsymbol{M}_{\textrm{int}}\frac{1}{\mathbb{I}_{T\mathcal{H}_{\text{cov}}}+\boldsymbol{Q}^{+}\boldsymbol{M}_{\textrm{int}}}\boldsymbol{\iota},
\label{M eff}
\end{align}
among which the relations
\begin{align}
\boldsymbol{M}\mathbf{I}=\mathbf{I}\boldsymbol{M}^{\text{eff}},\hspace{5mm}
\mathbf{T}\boldsymbol{M}=\boldsymbol{M}^{\text{eff}}\mathbf{T},\\
\{\boldsymbol{M},\boldsymbol{M}^{+}\}=\mathbb{I}_{T\mathcal{H}_{\text{cov}}}-\mathbf{IT},\\
(\boldsymbol{M}^{+})^{2}=\mathbf{T}\boldsymbol{M}^{+}=\boldsymbol{M}^{+}\mathbf{I}=0,\\
(\mathbf{IT})^2=\mathbf{IT},\hspace{5mm}
(\mathbf{TI})^2=\mathbf{TI}=\mathbb{I}_{T\mathcal{H}^{\perp}_{\text{lc}}},
\end{align}
deformed from those in Eqs.\,(\ref{und rel}) hold.
In particular, the formula in Eq.\,(\ref{M eff}) gives the string products of the light-cone-gauge effective action to all orders in closed form.
The string field $\Psi_{\textrm{cov}}$ in the gauge-invariant theory is written as a function of the
string field $\Psi_{\textrm{lc}}^{\perp}$ in the genuine light-cone-gauge theory using the deformed inclusion map $\boldsymbol{I}$ as
\begin{align}
\Psi_{\text{cov}}=\pi_{1}\mathbf{I}\frac{1}{1-\Psi_{\text{lc}}^{\perp}}.
\end{align}

\section{A consistent light-cone-gauge superstring field theory}\label{cons lgt}

In the previous section, we derive a light-cone-gauge effective theory from generic gauge-invariant theory by fixing the gauge invariance 
and then integrating out nondynamical longitudinal fields. We can derive a genuine light-cone-gauge theory, which is a possible candidate 
for the consistent light-cone-gauge superstring field theory, if we start from a gauge-invariant theory with light-cone-type interaction. 
Such a gauge invariant theory, which we call the Kugo-Zwiebach-type theory, is obtained by inserting PCOs into the light-cone-type interaction 
according to the general prescription given in Appendix~\ref{const M}. 
We denote the interaction of this possible candidate theory as $M_n^{\textrm{lc}}$:
\begin{align}
    M_1^{\textrm{lc}}\ =&\ c_0L_0,\label{M1}\\
    M_2^{\textrm{lc}}\ =&\ \mathcal{P}^{\perp}_{\textrm{lc}}S^{-1}M_2^{\textrm{KZ}}S\otimes S,\label{M2}\\
    M_3^{\textrm{lc}}\ =&\ \mathcal{P}^{\perp}_{\textrm{lc}}S^{-1}M_3^{\textrm{KZ}}S\otimes S \otimes S
    -\mathcal{P}^{\perp}_{\textrm{lc}}S^{-1}M_2^{\textrm{KZ}}\left(\mathbb{I}\otimes Q^{+}M_2^{\textrm{KZ}}+Q^{+}M_2^{\textrm{KZ}}\otimes \mathbb{I}\right)S\otimes S\otimes S,\label{M3}\\
&\vdots\ .\nonumber
\end{align}
Defining the interaction vertices by
\begin{equation}
     V_{n+1}^{\textrm{lc}}(A_1,\cdots,A_{n+1})\ =\ \Omega(A_1,M_n^{\textrm{lc}}(A_2,\cdots,A_{n+1}))\qquad A_1,\cdots,A_{n+1}\in\mathcal{H}_{\textrm{lc}}^{\perp},
\end{equation}
let us concretely examine in what form these interactions are.

\subsection{Evaluating the effective cubic vertices of massless fields}\label{ev cub massless}

Consider first the cubic vertex of massless fields.
From the relation in Eq.\,(\ref{M2}), we have
\begin{equation}
    V_3^{\textrm{lc}}(A, B, C)\ =\ V_3^{\textrm{KZ}}(SA, SB, SC),
\end{equation}
but, unlike the bosonic case, the transfer invariance does not hold in the superstring field theory:
\begin{equation}
    V_3^{\textrm{KZ}}(SA, SB, SC)\ \ne\ V_3^{\textrm{KZ}}(A, B, C).
\end{equation}
To understand the specific form of the interactions, let us first evaluate the cubic vertices of massless fields.

First, consider the vertex for one NS state and two R states:
\begin{align}
V^{\textrm{lc}}_{\textrm{NS-R-R}}(A,B,C)\ =\ V^{\text{KZ}}_3(SA,SB,SC),\qquad
\end{align}
with
\begin{align}
A\ =\ \psi^i_{-1/2}\ket{\ket{k_A}},\qquad
B\ =\ \ket{\ket{(-,a):k_B}},\qquad
C\ =\ \ket{\ket{(-,b):k_C}}.
\end{align}
The similarity transformation of these states gives
\begin{align}
SA\ &=\ \Big(\psi^i-\frac{k^i_A}{k^+_A}\psi^+\Big)(0)\ket{\ket{k_A}},
\nonumber\\
&=\ \Big(\psi^i-\frac{k^i_A}{k^+_A}\psi^+\Big)ce^{-\phi}e^{ik_A\cdot X}(0)\ket{0},\\
SB\ &=\ \ket{\ket{(-,a):k_B}}-\frac{k^j_B}{\sqrt{2}k^+_B}\gamma^j_{ac}\ket{\ket{(+,c):k_B}}
\nonumber\\
&=\ \Big(e^{-\frac{i}{2}H_0}\mathscr{S}_{a}^{-}-\frac{k^j_B}{\sqrt{2}k^+_B}\gamma^j_{ac}\,e^{\frac{i}{2}H_0}\mathscr{S}_c^+\Big)ce^{-\frac{\phi}{2}}e^{ik_B\cdot X}(0)\ket{0},
\label{spinor tf 1}\\
SC\ &=\ \ket{\ket{(-,b):k_C}}-\frac{k^k_C}{\sqrt{2}k^+_C}\gamma^k_{bd}\ket{\ket{(+,d):k_C}}
\nonumber\\
&=\ \Big(e^{-\frac{i}{2}H_0}\mathscr{S}_{b}^{-}-\frac{k^j_C}{\sqrt{2}k^+_C}\gamma^j_{bd}\,e^{\frac{i}{2}H_0}\mathscr{S}_d^+\Big)ce^{-\frac{\phi}{2}}e^{ik_C\cdot X}(0)\ket{0}.
\label{spinor tf 2}
\end{align}
The light-cone-type vertex is defined by the local coordinate maps from
the half-disk, 
\begin{equation}
    0\le|\xi_I|\le 1,\qquad \Im(\xi_I)>0,
\end{equation}
to the light-cone diagram and then the upper half plane: $z=f^{\textrm{lc}}_{(3,I)}(\xi_I)=\rho^{-1}\circ\rho_I(\xi_I)$ with
\begin{align}
\rho(z)\ =\ \sum_{I=A,B,C}\alpha_I\ln{(z-Z_I)},\qquad \rho_I(\xi_I)\ =\ \tau_0+\alpha_I\ln{\xi_I},  
\label{UHP 3}
\end{align}
For simplicity, we denoted $\alpha_0^+$ as $\alpha$ and often use this notation for the string length.
The interaction time $\tau_0$ is determined by the solution $z_0$ of the equation $\frac{d\rho}{dz}(z)=0$ as $\tau_0=\Re\rho(z_0)$. 
We have
\begin{equation}
    \frac{d\rho}{dz}\ =\ \frac{(\sum_{I}\alpha_IZ_I)(z-z_0)}{\prod_{I}(z-Z_I)}.
    \label{drhodz}
\end{equation}
Using these maps, we can evaluate the NS-R-R vertex as a correlation function on the upper half plane as
\begin{align}
&V^{\textrm{lc}}_{\textrm{NS-R-R}}(A,B,C)\nonumber\\
=\ &\bra{V^\textrm{KZ}_3}SA\otimes SB\otimes SC\nonumber\\
=&\ 
\left(f^{\textrm{lc}}_{(3,1)}{}'(0)\right)^{k^2_A}
\left(f^{\textrm{lc}}_{(3,2)}{}'(0)\right)^{k^2_B}
\left(f^{\textrm{lc}}_{(3,3)}{}'(0)\right)^{k^2_C}
\nonumber\\
&\ \times
\Bigg<
\left(\psi^i-\frac{k^i_A}{k^+_A}\psi^+\right)ce^{-\phi}e^{ik_A\cdot X}(Z_1)
\left(
e^{-\frac{i}{2}H_0}\mathscr{S}^-_a
-\frac{k^j_B}{\sqrt{2}k^+_B}\gamma^{j}_{ac}e^{\frac{i}{2}H_0}\mathscr{S}^+_c
\right)ce^{-\frac{\phi}{2}}e^{ik_B\cdot X}(Z_2)
\nonumber\\
&\ \times
\left(
e^{-\frac{i}{2}H_0}\mathscr{S}^-_b
-\frac{k^k_C}{\sqrt{2}k^+_C}\gamma^{k}_{bd}e^{\frac{i}{2}H_0}\mathscr{S}^+_d
\right)ce^{-\frac{\phi}{2}}e^{ik_C\cdot X}(Z_3)
\Bigg>_{\textrm{UHP}}
\nonumber\\
=&\ (2\pi)^{10}\delta^{10}(\Sigma_Ik_I)
\left(f^{\textrm{lc}}_{(3,1)}{}'(0)\right)^{k^2_A}
\left(f^{\textrm{lc}}_{(3,2)}{}'(0)\right)^{k^2_B}
\left(f^{\textrm{lc}}_{(3,3)}{}'(0)\right)^{k^2_C}
|Z_{12}|^{2k_A\cdot k_B}
|Z_{13}|^{2k_A\cdot k_C}
|Z_{23}|^{2k_B\cdot k_C}
\nonumber\\
&\times
\Bigg[
\frac{k^i_A}{k^+_A}\delta_{ab}
-\frac{k^j_B}{2k^+_B}\left(\gamma^j\gamma^i\right)_{ab}
-\frac{k^j_C}{2k^+_C}\left(\gamma^i\gamma^j\right)_{ab}
\Bigg],\label{cubic vertex for NSRR}
\end{align}
where we used:\footnote{
In this section, we will not care about the sign of the calculations as it is not important to the discussion.
}
\begin{align}
\big<c(Z_1)c(Z_2)c(Z_3)\big>_{\textrm{UHP}}
\ =&\ Z_{12}Z_{13}Z_{23},\\
\big<e^{-\phi}(Z_1)e^{-\phi/2}(Z_2)e^{-\phi/2}(Z_3)\big>_{\textrm{UHP}}
\ =&\ Z^{-1/2}_{12}Z^{-1/2}_{13}Z^{-1/4}_{23},\\
\big<
\psi^i(Z_1)\mathscr{S}^{+}_a(Z_2)\mathscr{S}^{-}_b(Z_3)
\big>_{\text{UHP}}
\ =&\ -\frac{1}{\sqrt{2}}\gamma^i_{ab}\,
Z^{-1/2}_{12}Z^{-1/2}_{13}Z^{-1/2}_{23},\\
\big<e^{ik_A\cdot X(Z_1,Z_1)}e^{ik_B\cdot X(Z_2,Z_2)}e^{ik_C\cdot X(Z_3,Z_3)}\big>_{\textrm{UHP}}
\ =&\ (2\pi)^{10}\delta^{10}(\Sigma_Ik_I)
\nonumber\\
&\ \times
|Z_{12}|^{2k_A\cdot k_B}|Z_{13}|^{2k_A\cdot k_C}|Z_{23}|^{2k_B\cdot k_C}.
\end{align}
with $Z_{ij}=Z_i-Z_j$.
The kinematical factor is consistent with the gauge-fermion-fermion interaction of the gauge theory in the light-cone gauge.

Next, we consider the cubic vertex for three NS fields:
\begin{align}
V^{\text{lc}}_{\textrm{NS-NS-NS}}(A,B,C)
\ =\ V^{\text{KZ}}_3(SA,SB,SC),
\end{align}
with
\begin{align}
A\ =\ \psi^i_{-1/2}\ket{\ket{k_A}},\qquad
B\ =\ \psi^j_{-1/2}\ket{\ket{k_B}},\qquad
C\ =\ \psi^k_{-1/2}\ket{\ket{k_C}}.
\end{align}
The similarity transformation of these states becomes
\begin{align}
SA\ =&\ \left(\psi^i-\frac{k^i_A}{k^+_A}\psi^+\right)ce^{-\phi}e^{ik_A\cdot X}(0)\ket{0},\\
SB\ =&\ \left(\psi^i-\frac{k^i_B}{k^+_B}\psi^+\right)ce^{-\phi}e^{ik_B\cdot X}(0)\ket{0},\\
SC\ =&\ \left(\psi^i-\frac{k^i_C}{k^+_C}\psi^+\right)ce^{-\phi}e^{ik_C\cdot X}(0)\ket{0}.
\end{align}
Furthermore, acting the PCO
\begin{align}
X_0=\oint\frac{dz}{2\pi i}\frac{1}{z}\left(G_m e^{\phi}+ \cdots\right),\qquad G_m\ =\ \sqrt{2}\psi^{\mu}i\partial X_{\mu},
\end{align}
on $SA$ gives
\begin{align}
X_0SA\ =\ \left(
\frac{k^i_A}{k^+_A}\sqrt{2}i\partial X^+
-\sqrt{2}i\partial X^i
-(\alpha_0\cdot\psi)\left(
\psi^i-\frac{k^i_A}{k^+_A}\psi^+
\right)
\right)ce^{ik_A\cdot X}(0)\ket{0},
\end{align}
except for the ghost part, which does not contribute to the result.
Evaluating the three-NS vertex using the correlation function on the upper half plane, we find that
\begin{align}
&V^{\text{lc}}_{\textrm{NS-NS-NS}}(A,B,C)\nonumber\\
&=\ \frac{1}{3}\bra{V^\textrm{LPP}_3}\Big(
X_0\otimes\mathbb{I}\otimes\mathbb{I}
+\mathbb{I}\otimes X_0\otimes\mathbb{I}
+\mathbb{I}\otimes\mathbb{I}\otimes X_0
\Big)SA\otimes SB\otimes SC\nonumber\\
&=\ \frac{1}{3}\Bigg<
f^{\textrm{lc}}_{(3,1)}\circ\left(
\left(
\frac{k^i_A}{k^+_A}\sqrt{2}i\partial X^+
-\sqrt{2}i\partial X^i
-(\alpha_0\cdot\psi)\left(\psi^i-\frac{k^i_A}{k^+_A}\psi^+
\right)
\right)ce^{ik_A\cdot X}(0)
\right)\nonumber\\
&\ \times f^{\textrm{lc}}_{(3,2)}\circ\left(
\left(
\psi^j-\frac{k^j_B}{k^+_B}\psi^+
\right)ce^{-\phi}e^{ik_B\cdot X}(0)\right)
f^{\textrm{lc}}_{(3,3)}\circ\left(
\left(
\psi^k-\frac{k^k_C}{k^+_C}\psi^+
\right)ce^{-\phi}e^{ik_C\cdot X}(0)\right)
\Bigg>_{\text{UHP}}
+\cdots\nonumber\\
&=\ \frac{1}{\sqrt{2}}(2\pi)^{10}\delta^{10}(\Sigma_Ik_I)
\left(f^{\textrm{lc}}_{(3,1)}{}'(0)\right)^{k^2_A}\left(f^{\textrm{lc}}_{(3,2)}{}'(0)\right)^{k^2_B}\left(f^{\textrm{lc}}_{(3,3)}{}'(0)\right)^{k^2_C}
|Z_{12}|^{2k_A\cdot k_B}
|Z_{13}|^{2k_A\cdot k_C}
|Z_{23}|^{2k_B\cdot k_C}\nonumber\\
&\ \times\Bigg[
\left(k^k_{AB}-\frac{k^+_{AB}}{k^+_C}k^k_C\right)\delta^{ij}
+\left(k^i_{BC}-\frac{k^+_{BC}}{k^+_A}k^i_A\right)\delta^{jk}
+\left(k^j_{CA}-\frac{k^+_{CA}}{k^+_B}k^j_B\right)\delta^{ik}
\Bigg],
\end{align}
where $\bra{V_3^{\textrm{LPP}}}$ is the vertex without PCO insertion \cite{LeClair:1988sp,LeClair:1988sj} and $k^{\mu}_{IJ}=k^\mu_I-k^\mu_J$.
The kinematical factor is consistent with the cubic gauge interaction of the gauge theory in the light-cone gauge.

\subsection{General cubic vertex}\label{gen cubic}

Now, let us consider general cubic vertices by first evaluating the correlation function of the longitudinal modes.

For the NS-R-R vertex, we consider
\begin{equation}
    V_{\textrm{NS-R-R}}^{\textrm{lc}}(A,B,C)\ =\ V_3^{\textrm{KZ}}(SA,SB,SC),
\end{equation}
with
\begin{align}
    A\ =&\ ce^{-\phi}A^{\perp}e^{ik_A\cdot X}(0)\ket{0}\ \equiv\ ce^{-\phi}\mathscr{A}(0)\ket{0},\\
    B\ =&\ ce^{-\frac{\phi}{2}}e^{-\frac{i}{2}H_0}B^{\perp}e^{ik_B\cdot X}(0)\ket{0}\ \equiv\ ce^{-\frac{\phi}{2}}e^{-\frac{i}{2}H_0}\mathscr{B}(0)\ket{0},\\
    C\ =&\ ce^{-\frac{\phi}{2}}e^{-\frac{i}{2}H_0}C^{\perp}e^{ik_c\cdot X}(0)\ket{0}\ \equiv\ ce^{-\frac{\phi}{2}}e^{-\frac{i}{2}H_0}\mathscr{C}(0)\ket{0},
\end{align}
where $A^{\perp}(0)$, $B^{\perp}(0)$, and $C^{\perp}(0)$ only contain the transverse fields and their derivatives.
When we evaluate the correlation function we can replace $\sqrt{2}i\partial X^+(\xi)$ to the zero mode
$\frac{\alpha_0^+}{\xi}$ in the local coordinate system since $X^-$ does not appear except for the zero-mode $x^-$ \cite{Erler:2020beb}.
Therefore, the similarity transformation in the correlation function becomes $S=e^{-R_{\textrm{F}}}$ with
\begin{equation}
     R_{\textrm{F}}\ =\ \frac{1}{\alpha_0^+}\oint\frac{d\xi}{2\pi i}\xi\psi^{+}G^{\textrm{lc}}(\xi).
\end{equation}
Moreover, considering the conservation of $H_0$ charge, it can be seen that only the first order of $R_{\textrm{F}}$ in $S$ gives 
a nontrivial contribution, and thus,
\begin{align}
V^{\textrm{lc}}_{\textrm{NS-R-R}}\
=&\ \bra{V^\textrm{KZ}_3}SA\otimes SB\otimes SC\nonumber\\
=&\ -\bra{V^\textrm{KZ}_3}\Big(
R_{\textrm{F}}\otimes\mathbb{I}\otimes\mathbb{I}
+\mathbb{I}\otimes R_{\textrm{F}}\otimes\mathbb{I}
+\mathbb{I}\otimes\mathbb{I}\otimes R_{\textrm{F}}
\Big)A\otimes B\otimes C
\nonumber\\
=&\ 
- \left(f^{\textrm{lc}}_{(3,1)}{}'(0)\right)^{\Delta_A}
\left(f^{\textrm{lc}}_{(3,2)}{}'(0)\right)^{\Delta_B}
\left(f^{\textrm{lc}}_{(3,3)}{}'(0)\right)^{\Delta_C}
\nonumber\\
&\ \times\Bigg\{
\Bigg< \oint_{Z_1}\frac{dz}{2\pi i}\left(\frac{d\rho}{dz}\right)^{-1} \psi^+G^{\perp}(z)
ce^{-\phi}\mathscr{A}(Z_1)ce^{-\frac{\phi}{2}}e^{-\frac{i}{2}H_0}\mathscr{B}(Z_2)ce^{-\frac{\phi}{2}}e^{-\frac{i}{2}H_0}\mathscr{C}(Z_3)\Bigg>_{\textrm{UHP}}\nonumber\\
&\hspace{5mm}
+ \Bigg< ce^{-\phi}\mathscr{A}(Z_1)\oint_{Z_2}\frac{dz}{2\pi i}\left(\frac{d\rho}{dz}\right)^{-1} \psi^+G^{\perp}(z)
ce^{-\frac{\phi}{2}}e^{-\frac{i}{2}H_0}\mathscr{B}(Z_2)ce^{-\frac{\phi}{2}}e^{-\frac{i}{2}H_0}\mathscr{C}(Z_3)\Bigg>_{\textrm{UHP}}\nonumber\\
&\hspace{5mm} 
+ \Bigg< ce^{-\phi}\mathscr{A}(Z_1)ce^{-\frac{\phi}{2}}e^{-\frac{i}{2}H_0}\mathscr{B}(Z_2)\oint_{Z_3}\frac{dz}{2\pi i}\left(\frac{d\rho}{dz}\right)^{-1} \psi^+G^{\perp}(z)
ce^{-\frac{\phi}{2}}e^{-\frac{i}{2}H_0}\mathscr{C}(Z_3)\Bigg>_{\textrm{UHP}}
\Bigg\},
\end{align}
where $\Delta_A$, $\Delta_B$, and $\Delta_C$ are the conformal dimensions of $A$, $B$, and $C$, respectively.
Deforming integration contours around the location of the vertex operators, we eventually obtain the contour integral around 
the interaction point, and find that 
\begin{align}
    V^{\textrm{lc}}_{\textrm{NS-R-R}}\
    =&\ \left(f^{\textrm{lc}}_{(3,1)}{}'(0)\right)^{\Delta_A}
\left(f^{\textrm{lc}}_{(3,2)}{}'(0)\right)^{\Delta_B}
\left(f^{\textrm{lc}}_{(3,3)}{}'(0)\right)^{\Delta_C}
\nonumber\\
&\ \times
\Bigg< \oint_{z_0}\frac{dz}{2\pi i}\frac{\prod_{I=1}^3(z-Z_I)}{(\sum_{I=1}^3\alpha_IZ_I)(z-z_0)} e^{iH_0}G^{\perp}(z)
\nonumber\\
&\hspace{10mm} 
ce^{-\phi}\mathscr{A}(Z_1)ce^{-\frac{1}{2}\phi}e^{-\frac{i}{2}H_0}\mathscr{B}(Z_2) ce^{-\frac{1}{2}\phi}e^{-\frac{i}{2}H_0}\mathscr{C}(Z_3)\Bigg>_{\textrm{UHP}}
\nonumber\\
=&\ \left(f^{\textrm{lc}}_{(3,1)}{}'(0)\right)^{\Delta_A}
\left(f^{\textrm{lc}}_{(3,2)}{}'(0)\right)^{\Delta_B}
\left(f^{\textrm{lc}}_{(3,3)}{}'(0)\right)^{\Delta_C}
\left(\frac{Z_{12}}{z_0-Z_2}\right)^{\frac{1}{2}}\left(\frac{Z_{13}}{z_0-Z_3}\right)^{\frac{1}{2}}Z_{23}
\nonumber\\
&\hspace{10mm} \times
\left(\frac{d^2\rho}{dz^2}(z_0)\right)^{-1}\Big< G^{\perp}(z_0)\mathscr{A}(Z_1)\mathscr{B}(Z_2)\mathscr{C}(Z_3)\Big>_{\textrm{UHP}}.
\label{NSRR general}
\end{align}

Similarly, we consider the NS-NS-NS vertex
\begin{equation}
    V^{\text{lc}}_{\textrm{NS-NS-NS}}(A,B,C)\ =\ \frac{1}{3}\bra{V^\textrm{LPP}_3}\Big(
X_0\otimes\mathbb{I}\otimes\mathbb{I}
+\mathbb{I}\otimes X_0\otimes\mathbb{I}
+\mathbb{I}\otimes\mathbb{I}\otimes X_0
\Big)SA\otimes SB\otimes SC,
\end{equation}
with
\begin{align}
    A\ =&\ ce^{-\phi}A^{\perp}e^{ik_A\cdot X}(0)\ket{0}\ \equiv\ ce^{-\phi}\mathscr{A}(0)\ket{0},\\
    B\ =&\ ce^{-\phi}B^{\perp}e^{ik_A\cdot X}(0)\ket{0}\ \equiv\ ce^{-\phi}\mathscr{B}(0)\ket{0},\\
    C\ =&\ ce^{-\phi}C^{\perp}e^{ik_A\cdot X}(0)\ket{0}\ \equiv\ ce^{-\phi}\mathscr{C}(0)\ket{0}.
\end{align}
If we decompose PCO $X_0$ into the four parts,
\begin{align}
    X_0\ =\ \mathcal{X}^{\perp} + \mathcal{X}^{+} + \mathcal{X}^{-} + \mathcal{X}^{\textrm{gh}},
\end{align}
with
\begin{align}
    \mathcal{X}^{\perp}\ =&\ \oint\frac{dz}{2\pi i}\sqrt{2}\psi^ii\partial X^ie^{\phi}(z), \\
    \mathcal{X}^{+}\ =&\ \oint\frac{dz}{2\pi i}\sqrt{2}\psi^+i\partial X^-e^{\phi}(z),\\
    \mathcal{X}^{-}\ =&\ \oint\frac{dz}{2\pi i}\sqrt{2}\psi^-i\partial X^+e^{\phi}(z),
\end{align}
and ghost part $\mathcal{X}^{\textrm{gh}}$, only $\mathcal{X}^{\perp}$ and $\mathcal{X}^{-}$
contribute to the correlation function due to the conservation of $H_0$ charge, and we have
\begin{align}
&\ 
V^{\text{lc}}_{\textrm{NS-NS-NS}}(A,B,C)\nonumber\\
&\ =\ 
\frac{1}{3}\bra{V^\textrm{LPP}_3}\Big(
\mathcal{X}^{\perp}\otimes\mathbb{I}\otimes\mathbb{I}
+\mathbb{I}\otimes \mathcal{X}^{\perp}\otimes\mathbb{I}
+\mathbb{I}\otimes\mathbb{I}\otimes \mathcal{X}^{\perp}
\Big) A\otimes B\otimes C
\nonumber\\
&\hspace{3mm}
- \frac{1}{3}\bra{V^\textrm{LPP}_3}\Big(
\mathcal{X}^{-}R_{\textrm{F}}\otimes\mathbb{I}\otimes\mathbb{I}
+ \mathcal{X}^{-}\otimes R_{\textrm{F}}\otimes\mathbb{I}
+ \mathcal{X}^{-}\otimes\mathbb{I}\otimes R_{\textrm{F}}
\Big) A\otimes B\otimes C
\nonumber\\
&\hspace{3mm}
- \frac{1}{3}\bra{V^\textrm{LPP}_3}\Big(
R_{\textrm{F}}\otimes \mathcal{X}^{-}\otimes\mathbb{I}
+ \mathbb{I}\otimes \mathcal{X}^{-}R_{\textrm{F}}\otimes\mathbb{I}
+ \mathbb{I}\otimes \mathcal{X}^{-}\otimes R_{\textrm{F}}
\Big) A\otimes B\otimes C\nonumber\\
&\hspace{3mm}
- \frac{1}{3}\bra{V^\textrm{LPP}_3}\Big(
R_{\textrm{F}}\otimes\mathbb{I}\otimes \mathcal{X}^{-}
+ \mathbb{I}\otimes R_{\textrm{F}}\otimes \mathcal{X}^{-}
+ \mathbb{I}\otimes\mathbb{I}\otimes \mathcal{X}^{-}R_{\textrm{F}}
\Big) A\otimes B\otimes C
\nonumber\\
&=\ 
- \frac{1}{3}\bra{V^\textrm{LPP}_3}\Big(
R_{\textrm{F}}\mathcal{X}^{-}\otimes\mathbb{I}\otimes\mathbb{I}
+ \mathcal{X}^{-}\otimes R_{\textrm{F}}\otimes\mathbb{I}
+ \mathcal{X}^{-}\otimes\mathbb{I}\otimes R_{\textrm{F}}
\Big) A\otimes B\otimes C
\nonumber\\
&\hspace{6mm}
- \frac{1}{3}\bra{V^\textrm{LPP}_3}\Big(
R_{\textrm{F}}\otimes \mathcal{X}^{-}\otimes\mathbb{I}
+ \mathbb{I}\otimes R_{\textrm{F}}\mathcal{X}^{-}\otimes\mathbb{I}
+ \mathbb{I}\otimes \mathcal{X}^{-}\otimes R_{\textrm{F}}
\Big) A\otimes B\otimes C\nonumber\\
&\hspace{6mm}
- \frac{1}{3}\bra{V^\textrm{LPP}_3}\Big(
R_{\textrm{F}}\otimes\mathbb{I}\otimes \mathcal{X}^{-}
+ \mathbb{I}\otimes R_{\textrm{F}}\otimes \mathcal{X}^{-}
+ \mathbb{I}\otimes\mathbb{I}\otimes R_{\textrm{F}}\mathcal{X}^{-}
\Big) A\otimes B\otimes C,
\end{align}
where we used
\begin{equation}
    [\mathcal{X}^{-}, R_{\textrm{F}}]\ =\ 
    \frac{1}{\alpha}\oint\frac{d\xi}{2\pi i}\sqrt{2}i\partial X^+ G^{\perp}e^{\phi}
    \xrightarrow[\sqrt{2}i\partial X^+\rightarrow\frac{\alpha}{\xi}]{}\ \mathcal{X}^{\perp},
\end{equation}
in the correlation function. Mapping to the upper half plane by Eq.\,(\ref{UHP 3})
and deforming the integration contour of $R_{\textrm{F}}$ to a contour around the interaction point $z_0$,
we find that
\begin{align}
    V^{\textrm{lc}}_{\textrm{NS-NS-NS}}\ =&\ \left(f^{\textrm{lc}}_{(3,1)}{}'(0)\right)^{\Delta_A}
\left(f^{\textrm{lc}}_{(3,2)}{}'(0)\right)^{\Delta_B}
\left(f^{\textrm{lc}}_{(3,3)}{}'(0)\right)^{\Delta_C}
\nonumber\\
&\ \times\frac{1}{3}
    \Bigg< \oint_{z_0}\frac{dz}{2\pi i}\frac{\prod_{I=1}^3(z-Z_I)}{(\sum_{I=1}^3\alpha_IZ_I)(z-z_0)} e^{iH_0}G^{\perp}(z)
\nonumber\\
&\hspace{5mm} \times
\sum_{J=1}^{3}\frac{1}{\alpha_J}\oint_{Z_J}\frac{dz'}{2\pi i}\frac{(\sum_{I=1}^{3}\alpha_IZ_I)^2(z'-z_0)^2}
{\prod_{I=1}^{3}(z'-Z_I)^2}e^{-iH_0}e^{\phi}(z')
\nonumber\\
&\hspace{5mm} \times
ce^{-\phi}\mathscr{A}(Z_1)ce^{-\phi}\mathscr{B}(Z_2)ce^{-\phi}\mathscr{C}(Z_3)
\Bigg>_{\textrm{UHP}}
\nonumber\\
=&\ 
\left(f^{\textrm{lc}}_{(3,1)}{}'(0)\right)^{\Delta_A}
\left(f^{\textrm{lc}}_{(3,2)}{}'(0)\right)^{\Delta_B}
\left(f^{\textrm{lc}}_{(3,3)}{}'(0)\right)^{\Delta_C}
\big(\sum_I\alpha_IZ_I\big)
\nonumber\\
&\ \times 
\left(\frac{d^2\rho}{dz^2}(z_0)\right)^{-1}
\left<G^{\perp}(z_0)\mathscr{A}(Z_1)\mathscr{B}(Z_2)\mathscr{C}(Z_3)
\right>_{\textrm{UHP}}.
\label{NSNSNS general}
\end{align}

These results in Eqs.\,(\ref{NSRR general}) and (\ref{NSNSNS general}) are those expected for a light-cone gauge superstring field theory \cite{Mandelstam:1974hk}. 
As they are, however, they cause a known difficulty due to the divergence coming from the coincidence of $G^{\perp}$s at the interaction points.
Therefore, we consider the theory with interactions with stubs to avoid such collisions of $G^{\perp}$.
In the stubbed theory, the points in the moduli space where the interaction points coincide are within the domain 
of the moduli integral contained in the quartic or higher-order vertices. We next consider whether the quartic vertices in the stubbed theory
contain divergence or not.

\subsection{Quartic vertex in stubbed theory}\label{quartic stubbed}

\begin{figure}[H]
\centering
\begin{minipage}[b]{0.32\columnwidth}
    \centering
    \includegraphics[width=0.9\columnwidth]{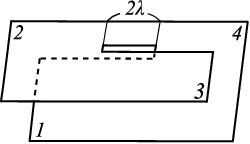}
    \subcaption{extra vertex for stub length $\lambda$}
    \label{extra 1}
\end{minipage}
\begin{minipage}[b]{0.32\columnwidth}
    \centering
    \includegraphics[width=0.9\columnwidth]{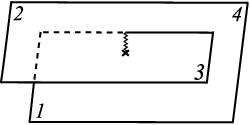}
    \subcaption{genuine quartic vertex}
    \label{genuine}
\end{minipage}
\begin{minipage}[b]{0.32\columnwidth}
    \centering
    \includegraphics[width=0.9\columnwidth]{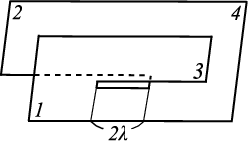}
    \subcaption{extra vertex for stub length $\lambda$}
    \label{extra 2}
\end{minipage}
\caption{The tu-channel light-cone diagram for $\alpha_1,\alpha_2 > 0$ and $\alpha_3,\alpha_4 <0$. }
\label{tu channel}
\end{figure}
%
To investigate whether the quartic vertices in stubbed theory contain divergence or not,
we evaluate 
the four-Ramond vertex of massless states as an example.
In particular, we consider the $tu$-channel scattering, where the collision of interaction points is inevitable.
The $tu$-channel four-string light-cone diagram (Fig.~\ref{tu channel}) transforms to the upper half plane
by the Mandelstam mapping, 
\begin{align}
\rho(z)\ =\ \sum_{I=1}^{4}\alpha_I\ln(z-Z_I),
\end{align}
with $\alpha_1,\alpha_2 > 0,\ \alpha_3,\alpha_4<0$ and $Z_4<Z_2<Z_3<Z_1$.
The four strings sit at the points $\{Z_I\}$. Of the four points where external strings $\{Z_I\}$ sit, three can be fixed using $SL(2)$-invariance, and the remaining one, 
which we choose $Z_3\equiv x$, is a moduli degree of freedom.
Two interaction points, which we denote $z_{\pm}$, are determined by the equation $\frac{d\rho}{dz}=0$ as function of the moduli $x$:
\begin{equation}
\frac{d\rho}{dz}(z)\ =\ \sum_{I=1}^{4}\frac{\alpha_I}{z-Z_I}\ =\ \Big(\sum_{I=1}^{4}\alpha_IZ_I\Big)
\frac{(z-z_+)(z-z_-)}{\prod_{I=1}^4(z-Z_I)}.
\end{equation}

\begin{figure}[H]
\centering
    \includegraphics[scale=0.5]{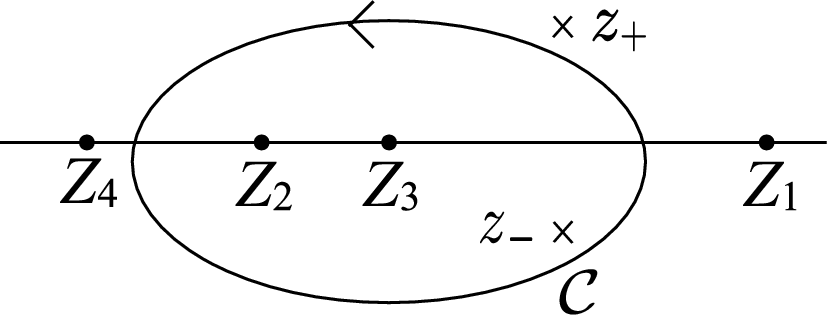}
    \caption{Integration contour of $b$ ghost}
    \label{contour C}
\end{figure}

The light-cone-gauge massless four-Ramond vertex is given by 
\begin{equation}
    V^{\textrm{lc}}_{\textrm{R-R-R-R}}(\hat{V}_1,\hat{V}_2,\hat{V}_3,\hat{V}_4)\ =\ V_{4}^{\textrm{KZ}}(S\hat{V}_1,S\hat{V}_2,S\hat{V}_3,S\hat{V}_4),
\end{equation}
with
\begin{equation}
    \hat{V}_I\ =\ \hat{V}_I(0)\ket{0},\qquad
    \hat{V}_I(\xi)\ =\ ce^{-\frac{\phi}{2}}e^{-\frac{i}{2}H_0}\mathscr{S}_{a}^-e^{ik_I\cdot X}(\xi)\,.
\end{equation}
The transformation of massless Ramond state is given in 
Eq.\,(\ref{spinor tf 1}) or (\ref{spinor tf 2}):
\begin{equation}
    S\hat{V}_I\ =\ \Big(e^{-\frac{i}{2}H_0}\mathscr{S}_{a}^{-}-\frac{k^j_I}{\sqrt{2}k^+_I}\gamma^j_{ac}\,e^{\frac{i}{2}H_0}\mathscr{S}_c^+\Big)ce^{-\frac{\phi}{2}}e^{ik_I\cdot X}(0)\ket{0}\,. 
\end{equation}
The massless four-Ramond vertex is given by
\begin{align}
V^{\textrm{lc}}_{\textrm{R-R-R-R}}    =&\ \prod_I\left(f^{\textrm{lc}}_{(4,I)}{}'(0)\right)^{k_I^2}
\int_{x_a}^{x_b} dx \left(\frac{d\tau}{dx}\right) \Bigg<\oint_{\mathcal{C}}\frac{dz}{2\pi i}\left(\frac{d\rho}{dz}\right)^{-1}b(z)
\nonumber\\
&\times  
\prod_{I=4,2,3,1}
\Big(e^{-\frac{i}{2}H_0}\mathscr{S}_{a}^{-}
-\frac{k^j_I}{\sqrt{2}k^+_I}\gamma^j_{ac}\,e^{\frac{i}{2}H_0}\mathscr{S}_c^+\Big)ce^{-\frac{\phi}{2}}e^{ik_I\cdot X}(Z_I)
\Bigg>_{\textrm{UHP}},
\end{align}
where the $b$-ghost integration contour $\mathcal{C}$ is depicted in Fig.~\ref{contour C}.
\noindent
The moduli $x$ is related to the light-cone time through the difference of two interaction points:
\begin{equation}
 \tau\ =\ \rho(z_+(x)) - \rho(z_-(x)),
\end{equation}
from which we have
\begin{equation}
 \frac{d\tau}{dx}\ =\ \frac{\left(\sum_{I=1}^{4}\alpha_IZ_I\right)(z_+-z_-)}{\prod_{I\ne3}(x-Z_I)}.
\end{equation}
The integration region $[x_a,x_b]$ is determined according to that of the light-cone time $\tau$:
$[x_a,x_b]$ is equal to $[x_-,x_+]$ for the quartic vertex in the theory without stub (Fig.~\ref{genuine}),\footnote{
Note that $\tau$ corresponding to the genuine four-string vertex in Fig.~\ref{genuine} is pure imaginary.
} and 
to $[x_--\delta,x_++\delta]$ for the stubbed theory, which contains extra regions $[x_--\delta, x_-]$ and $[x_+, x_++\delta]$ (Figs.~\ref{extra 1} and \ref{extra 2}).\footnote{
We denote the stub length in the light-cone diagram as $\lambda$ and that in the moduli space as $\delta=\delta(\lambda)$.
}
Evaluating the $bc$ ghost correlation function, the four-string vertex becomes
\begin{align}
V^{\textrm{lc}}_{\textrm{R-R-R-R}}\ =&\ \prod_I\left(f^{\textrm{lc}}_{(4,I)}{}'(0)\right)^{k_I^2}
\nonumber\\
&\times 
\int_{x_a}^{x_b}[dx]
\Bigg<\prod_{I=4,2,3,1}\Big(e^{-\frac{i}{2}H_0}\mathscr{S}_{a}^{-}
-\frac{k^j_I}{\sqrt{2}k^+_I}\gamma^j_{ac}\,e^{\frac{i}{2}H_0}\mathscr{S}_c^+\Big)e^{-\frac{\phi}{2}}e^{ik_I\cdot X}(Z_I)\Bigg>_{\textrm{UHP}}
\nonumber\\
=&\ \prod_I\left(f^{\textrm{lc}}_{(4,I)}{}'(0)\right)^{k_I^2}
\int_{x_a}^{x_b}[dx]\prod_{I>J}|Z_{IJ}|^{2k_I\cdot k_J}
\nonumber\\
&\times  
\frac{1}{2}\Bigg(
\frac{k^{j_3}_3k^{j_1}_1}{k^+_3k^+_1}\frac{\gamma^{j_3}_{a_3c_3}\gamma^{j_1}_{a_1c_1}}{|Z_{34}|^{\frac{1}{2}}|Z_{14}|^{\frac{1}{2}}|Z_{12}|^{\frac{1}{2}}|Z_{23}|^{\frac{1}{2}}}
\Bigg<
\mathscr{S}_{a_4}^{-}(Z_4)\mathscr{S}_{a_2}^{-}(Z_2)\mathscr{S}_{c_3}^+(Z_3)\mathscr{S}_{c_1}^+(Z_1)\Bigg>_{\textrm{UHP}}
\nonumber\\
&\hspace{15mm} 
+ \textrm{5-terms}\Bigg),
\label{massless 4R}
\end{align}
where
\begin{align}
[dx]\ =&\ \frac{dx}{V(sl(2))}\ =\ |Z_{12}||Z_{14}||Z_{24}|dx.
\end{align}
We can further calculate the four-Ramond amplitude (\ref{massless 4R}) by evaluating the correlation functions of the spin operator,
which gives functions of $Z_{IJ}$.
There is no divergence at the moduli points, $x=x_{\pm}$, at which the interaction
points coincide. 

\begin{figure}[H]
\centering
    \includegraphics[scale=1.5]{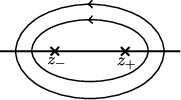}
    \caption{Integration contours of $R_{\textrm{F}}^{\perp}$}
    \label{contour one}
\end{figure}

\begin{figure}[H]
    \centering
    \includegraphics[scale=1.5]{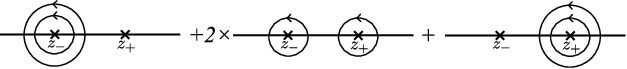}
    \caption{Integration contour on which the correlation function evaluated}
    \label{contour two}
\end{figure}

Let us consider how the divergence that existed 
in the stubless theory vanished. Consider the transformation of states in a slightly different way.
When we evaluate $S\hat{V}_I$ we can replace $\sqrt{2}i\partial X^+(\xi)$ to the zero-mode $\frac{\alpha_0^+}{\xi}$ (and thus $\tilde{X}^+(\xi)$ to zero)
as in the three string vertices. Then, we can replace $S$ with $e^{-R_{\textrm{F}}^{\perp}}e^{-R_{\textrm{F}}^{\textrm{gh}}}$, where
\begin{equation}
    R_{\textrm{F}}^{\perp}\ =\ \frac{1}{\alpha_0^+}\oint\frac{d\xi}{2\pi i}\xi \psi^+G^{\perp}(\xi),\qquad
    R_{\textrm{F}}^{\textrm{gh}}\ =\ \frac{1}{\alpha_0^+}\oint\frac{d\xi}{2\pi i}\xi \psi^+\left(\tilde{\beta}\partial\tilde{c}-\tilde{b}\tilde{\gamma}\right).
\end{equation}
Furthermore, we can show that $R_{\textrm{F}}^{\textrm{gh}}\hat{V}_I=0$ by simple calculation and can put
\begin{equation}
    S\hat{V}_I\ =\ e^{-R_{\textrm{F}}^{\perp}}\hat{V}_I,
\end{equation}
in the correlation function.
The string states $S\hat{V}_I$ are transformed as
\begin{equation}
    f^{\textrm{lc}}_{(4,I)}\circ S\hat{V}_I(0)\ =\ \left(f^{\textrm{lc}}_{(4,I)}{}'(0)\right)^{k_I^2}e^{-R_{\textrm{F}}^{\perp}}\hat{V}_I(Z_I),
\end{equation}
on the upper half plate with
\begin{equation}
R_{\textrm{F}}^{\perp}\ =\ \oint_{Z_I}\frac{dz}{2\pi i} \left(\frac{d\rho}{dz}\right)^{-1}
e^{iH_0}G^{\perp}(z).
\end{equation}
The light-cone-gauge four-Ramond vertex of massless states is now evaluated as
\begin{align}
    V^{\textrm{lc}}_{\textrm{R-R-R-R}}\ =&\ \bra{V^{\textrm{KZ}}_4}S\hat{V}_1\otimes S\hat{V}_2\otimes S\hat{V}_3\otimes S\hat{V}_4
\nonumber\\
=&\ \bra{V^{\textrm{KZ}}_4}\Big(\frac{1}{2}(R_{\textrm{F}}^{\perp})^2\otimes\id\otimes\id\otimes\id+\id\otimes\frac{1}{2}(R_{\textrm{F}}^{\perp})^2\otimes\id\otimes\id+\cdots
\nonumber\\
&\hspace{15mm} + R_{\textrm{F}}^{\perp}\otimes R_{\textrm{F}}^{\perp}\otimes\id\otimes\id+R_{\textrm{F}}^{\perp}\otimes\id\otimes R_{\textrm{F}}^{\perp}\otimes\id+\cdots\Big)\hat{V}_1\otimes \hat{V}_2\otimes \hat{V}_3\otimes \hat{V}_4
\nonumber\\
=&\ \prod_I\left(f^{\textrm{lc}}_{(4,I)}{}'(0)\right)^{k_I^2}
\frac{1}{2}\int_{x_a}^{x_b} dx \left(\frac{d\tau}{dx}\right) \Bigg<R_{\textrm{F}}^{\perp}R_{\textrm{F}}^{\perp}\oint_{\mathcal{C}}\frac{dz}{2\pi i}\left(\frac{d\rho}{dz}\right)^{-1}
b(z)\prod_{I=4,2,3,1}\hat{V}_I(Z_I)\Bigg>_{\textrm{UHP}},
\end{align}
where 
we deform the integration contour of $R_{\textrm{F}}^{\perp}$ from those around $Z_I$ to that depicted in Fig.~\ref{contour one}.
\noindent 
Evaluating the $bc$ correlation function, it becomes
\begin{align}
    V^{\textrm{lc}}_{\textrm{R-R-R-R}}\ =&\ \prod_I\left(f^{\textrm{lc}}_{(4,I)}{}'(0)\right)^{k_I^2}
    \nonumber\\
    &\times \frac{1}{2}\int_{x_a}^{x_b}[dx]\, 
\Bigg<R_{\textrm{F}}^{\perp}R_{\textrm{F}}^{\perp}\prod_{I=4,2,3,1}e^{-\frac{\phi}{2}}e^{-\frac{i}{2}H_0}\mathscr{S}_{a_I}^-e^{ik_I\cdot X}(Z_I)\Bigg>_{\textrm{UHP}}.
\label{4R vertex}
\end{align}
Deforming the integration contours in Fig.~\ref{contour one} to those in Fig.~\ref{contour two},
we evaluate the correlation function in Eq.\,(\ref{4R vertex}) by splitting it into the contributions from the contours I, II, and III in order from the left in Fig.~\ref{contour two}.

First, consider the contribution from the part II. While the longitudinal fields are also propagating, it has a similar structure to the divergent contribution of the stubless theory, 
Calculating the contour integral of $R_{\textrm{F}}^{\perp}$ as
\begin{align}
\oint_{z_{\pm}}\frac{dz}{2\pi i}\left(\frac{d\rho}{dz}\right)^{-1}\psi^+G^{\perp}(z)\ =&\ 
\oint_{z_{\pm}}\frac{dz}{2\pi i}\frac{\prod_I(z-Z_I)}{(\sum_I\alpha_IZ_I)(z-z_+)(z-z_-)}\psi^+G^{\perp}(z)
\nonumber\\
=&\ \pm\frac{\prod_I(z_{\pm}-Z_I)}{(\sum_I\alpha_IZ_I)(z_+-z_-)}\psi^+G^{\perp}(z_{\pm}),
\end{align}
we have
\begin{align}
    V^{\textrm{lc (II)}}_{\textrm{R-R-R-R}}\ =&\ -\frac{\prod_I\left(f^{\textrm{lc}}_{(4,I)}{}'(0)\right)^{k_I^2}}{(\sum_I\alpha_IZ_I)^2}
    \int_{x_a}^{x_b}[dx]\frac{\prod_I(z_+-Z_I)(z_--Z_I)}{(z_+-z_-)^2}
    \nonumber\\
&\times
    \Bigg<e^{iH_0}G^{\perp}(z_+)e^{iH_0}G^{\perp}(z_-)\prod_{I=4,2,3,1}e^{-\frac{\phi}{2}}e^{-\frac{i}{2}H_0}\mathscr{S}_{a_I}^-e^{ik_I\cdot X}(Z_I)\Bigg>_{\textrm{UHP}}.
\end{align}
A supercurrent $G^{\perp}$ sits at each of the two interaction points $z_{\pm}$ and collides at the moduli points $x=x_{\pm}$.
We find that
\begin{align}
    V^{\textrm{lc (II)}}_{\textrm{R-R-R-R}}\ =&\ \frac{\prod_I\left(f^{\textrm{lc}}_{(4,I)}{}'(0)\right)^{k_I^2}}{(\sum_I\alpha_IZ_I)^2}
    \int_{x_a}^{x_b}[dx] \prod_{I>J}|Z_{IJ}|^{2k_I\cdot k_J}\frac{\prod_I(z_+-Z_I)^{\frac{1}{2}}(z_--Z_I)^{\frac{1}{2}}}{z_+-z_-}
        \nonumber\\
&\hspace{0mm} \times\Bigg[
    \frac{1}{(z_+-z_-)^3}
\Bigg<:\psi^i(z_+)\psi^i(z_-):\prod_{I=4,2,3,1}\mathscr{S}^-_{a_I}(Z_I)\Bigg>_{\textrm{UHP}}
\nonumber\\
&\ 
+ \frac{1}{z_+-z_-}\sum_{I}\frac{\sqrt{2}k_{I}^i}{z_+-Z_I}\sum_{J}\frac{\sqrt{2}k_{J}^j}{z_--Z_J}
\Bigg<:\psi^i(z_+)\psi^j(z_-):\prod_{I=4,2,3,1}\mathscr{S}^{-}_{a_I}(Z_I)\Bigg>_{\textrm{UHP}}
\Bigg].
\label{part II}
\end{align}

In the four-string vertex, we have additional contributions from the parts I and III. We find that the contribution from the part I becomes
\begin{align}
    V^{\textrm{lc (I)}}_{\textrm{R-R-R-R}}\ =&\ -\frac{\prod_I\left(f^{\textrm{lc}}_{(4,I)}{}'(0)\right)^{k_I^2}}{2(\sum_I\alpha_IZ_I)^2}
    \int_{x_a}^{x_b}[dx]\frac{\prod_I(z_--Z_I)}{z_+-z_-}    \oint_{z_-}\frac{dz}{2\pi i}\frac{\prod_I(z-Z_I)}{(z-z_-)(z-z_+)}
\nonumber\\
&\times    
    \Bigg<e^{iH_0}G^{\perp}(z)e^{iH_0}G^{\perp}(z_-)\prod_{I=4,2,3,1}e^{-\frac{\phi}{2}}e^{-\frac{i}{2}H_0}\mathscr{S}_{a_I}^-e^{ik_I\cdot X}(Z_I)\Bigg>_{\textrm{UHP}}
\nonumber\\
\end{align}
\begin{align}
\hspace{10mm}
=&\ \frac{\prod_I\left(f^{\textrm{lc}}_{(4,I)}{}'(0)\right)^{k_I^2}}{2(\sum_I\alpha_IZ_I)^2}
    \int_{x_a}^{x_b}[dx]\prod_{I>J}|Z_{IJ}|^{2k_I\cdot k_J}
    \frac{\prod_I(z_--Z_I)^{\frac{1}{2}}}{z_+-z_-}
    \nonumber\\
    &\times 
    \oint_{z_-}\frac{dz}{2\pi i}\frac{\prod_I(z-Z_I)^{\frac{1}{2}}}{z-z_+}\Bigg[\frac{1}{(z-z_-)^3}
    \Bigg<:\psi^i(z)\psi^i(z_-):\prod_{I=4,2,3,1}\mathscr{S}^-_{a_I}(Z_I)\Bigg>_{\textrm{UHP}} 
    \nonumber\\
    &\ 
    + \frac{1}{z-z_-}\sum_{I}\frac{\sqrt{2}k_I^i}{z-Z_I}\sum_{J}\frac{\sqrt{2}k_J^j}{z_--Z_J}
    \Bigg<:\psi^i(z)\psi^j(z_-):\prod_{I=4,2,3,1}\mathscr{S}^{-}_{a_I}(Z_I)\Bigg>_{\textrm{UHP}}\Bigg]
\nonumber\\
=&\ - \frac{\prod_I\left(f^{\textrm{lc}}_{(4,I)}{}'(0)\right)^{k_I^2}}{2(\sum_I\alpha_IZ_I)^2} \int_{x_a}^{x_b}[dx]\prod_{I>J}|Z_{IJ}|^{2k_I\cdot k_J}
\nonumber\\
&\times
\Bigg[\frac{\prod_I(z_--Z_I)}{(z_+-z_-)^3}\Bigg<:\partial\psi^i(z_-)\psi^i(z_-):\prod_{I=4,2,3,1}\mathscr{S}^-_{a_I}(Z_I)\Bigg>_{\textrm{UHP}}
\nonumber\\
&\ 
+\frac{1}{2(z_+-z_-)^2}\Bigg(\sum_J\prod_{I\ne J}(z_--Z_I)\Bigg<:\partial\psi^i(z_-)\psi^i(z_-):\prod_{I=4,2,3,1}\mathscr{S}^-_{a_I}(Z_I)\Bigg>_{\textrm{UHP}}
\nonumber\\
&\hspace{30mm} 
+ \prod_I(z_--Z_I)\Bigg<:\partial^2\psi^i(z_-)\psi^i(z_-):\prod_{I=4,2,3,1}\mathscr{S}^-_{a_I}(Z_I)\Bigg>_{\textrm{UHP}}
\Bigg)\Bigg].
\end{align}
The contribution from the part III can be obtained by exchanging $z_+$ and $z_-$ as
\begin{align}
    V^{\textrm{lc (III)}}_{\textrm{R-R-R-R}}\ =&\ - \frac{\prod_I\left(f^{\textrm{lc}}_{(4,I)}{}'(0)\right)^{k_I^2}}{2(\sum_I\alpha_IZ_I)^2} \int_{x_a}^{x_b}[dx]\prod_{I>J}|Z_{IJ}|^{2k_I\cdot k_J}
\nonumber\\
&\times
\Bigg[- \frac{\prod_I(z_+-Z_I)}{(z_+-z_-)^3}\Bigg<:\partial\psi^i(z_+)\psi^i(z_+):\prod_{I=4,2,3,1}\mathscr{S}^-_{a_I}(Z_I)\Bigg>_{\textrm{UHP}}
\nonumber\\
&\ 
+\frac{1}{2(z_+-z_-)^2}\Bigg(\sum_{J}\prod_{I\ne J}(z_+-Z_I)\Bigg<:\partial\psi^i(z_+)\psi^i(z_+):\prod_{I=4,2,3,1}\mathscr{S}^-_{a_I}(Z_I)\Bigg>_{\textrm{UHP}}
\nonumber\\
&\hspace{30mm} 
+ \prod_I(z_+-Z_I)\Bigg<:\partial^2\psi^i(z_+)\psi^i(z_+):\prod_{I=4,2,3,1}\mathscr{S}^-_{a_I}(Z_I)\Bigg>_{\textrm{UHP}}
\Bigg)\Bigg].
\end{align}
Using the Taylor expansions, 
\begin{align}
(z_+-Z_I)^{\frac{1}{2}}\ =&\ (z_--Z_I)^{\frac{1}{2}} + \frac{1}{2}(z_+-z_-)(z_--Z_I)^{-\frac{1}{2}} -\frac{1}{8}(z_+-z_-)^2(z_--Z_I)^{-\frac{3}{2}} + \cdots ,\\
(z_--Z_I)^{\frac{1}{2}}\ =&\ (z_+-Z_I)^{\frac{1}{2}} - \frac{1}{2}(z_+-z_-)(z_+-Z_I)^{-\frac{1}{2}} -\frac{1}{8}(z_+-z_-)^2(z_+-Z_I)^{-\frac{3}{2}} + \cdots ,\\
    \psi^i(z_+)\ =&\ \psi^i(z_-) + (z_+-z_-)\partial\psi^i(z_-) + \frac{1}{2}(z_+-z_-)^2\partial^2\psi^i(z_-) + \cdots,\\
    \psi^i(z_-)\ =&\ \psi^i(z_+) - (z_+-z_-)\partial\psi^i(z_+) + \frac{1}{2}(z_+-z_-)^2\partial^2\psi^i(z_+) + \cdots,
\end{align}
we can find that
\begin{align}
&\ 
V^{(I)}_{\textrm{R-R-R-R}} + V^{(II)}_{\textrm{R-R-R-R}} + V^{(III)}_{\textrm{R-R-R-R}}\ 
\nonumber\\
&=\ 
    \frac{\prod_I\left(f^{\textrm{lc}}_{(4,I)}{}'(0)\right)^{k_I^2}}{2(\sum_I\alpha_IZ_I)^2} 
    \int_{x_a}^{x_b}[dx]\prod_{I>J}|Z_{IJ}|^{2k_I\cdot k_J}
    \Bigg[ A(x; Z_I) + \sum_{I>J}k_I^ik_J^jB_{IJ}^{ij}(x; Z_I)
 \Bigg]\,,
\end{align}
by adding all the three contributions together, where $A(x; Z_I)$ and $B_{IJ}^{ij}(x; Z_I)$ are functions of $x$ and $Z_I$ which are finite at $x=x_{\pm}$. 
If we explicitly calculate correlation functions of the spin operators, it should reproduce the result in Eq.\,(\ref{massless 4R}).
%
%

Similarly for the general four-Ramond vertices, if we act transformation $S$ upon the light-cone states first, the correlation function is nothing but that of the external DDF states in the gauge-invariant theory, 
so we can find that there is nothing to cause divergence at $x=x_{\pm}$ ($z_{+}=z_{-}$).
This reason can apply not only four-Ramond vertex but also to all the quartic or higher-order vertices in the stubbed theory.
So, it is reasonable to conclude that there is no divergence difficulty in the stubbed theory, and we can take it as a consistent light-cone-gauge superstring field theory.

\section{From light-cone-type interaction to Witten-type interaction}\label{L to W}

We have proposed a consistent light-cone-gauge superstring theory in the previous section
as an effective theory obtained by starting from the stubbed Kugo-Zwiebach-type superstring field theory.
In this section, however, we further relate the (stubbed) Kugo-Zwiebach-type superstring field theory 
and more widely recognized theory based on the (stubbed) Witten-type interaction by a field redefinition 
as in the case of the bosonic string field theory \cite{Erler:2020beb}. 
It (possibly) gives a proof of the unitarity of the Witten-type superstring field theory.

As is given in Appendix~\ref{const M}, the multisuperstring products can be 
obtained by putting PCOs on multistring products without a PCO, which are constructed by applying 
the method used in the bosonic string field theory \cite{LeClair:1988sp,LeClair:1988sj}. 
Following the prescription, 
we can obtain a superstring field theory action where we choose $M^{(0)}_2$ to be a Witten vertex \cite{Witten:1985cc} 
and $M^{(0)}_3,M^{(0)}_4,\dots$ are chosen to vanish. We call it Witten-type superstring 
field theory. If we choose $M^{(0)}_2,M^{(0)}_3$ to be cubic and quartic light-cone vertices 
and $M^{(0)}_4,M^{(0)}_5,\dots$ are chosen to vanish, we can obtain another superstring field 
theory action. Because this is the extension of the Kugo-Zwiebach bosonic string field theory \cite{Kugo:1992md}
($\alpha=p^+$ HIKKO theory) to superstring field theory, we call it Kugo-Zwiebach-type superstring field theory. 
The relation between the Witten-type and the Kugo-Zwiebach-type superstring field theories is obtained
using a theory with a one-parameter family of interactions connecting two types of interactions, which we call
the Kaku-type theory. All of them can easily be extended to the stubbed theory by simply replacing the starting
vertices with those with stubs, where all the $\boldsymbol{M}_{n}^{(0)}$ are non-zero and satisfy the $A_{\infty}$ relations.
The content of this section is independent of whether the theory is the one with or without stubs.

\subsection{Theory with Kaku-type interaction}\label{kaku type}

\begin{figure}
    \centering
    \begin{tikzpicture}
        \draw (-2,0)--(2,0)--(2,2)--(-2,2)--cycle;
        \draw[double] (-2,0)--(2,0);
        \draw[double] (2,2)--(-2,2);
        \draw[dashed] (-2,1/2)--(2,1/2);
        \draw[dashed] (-2,3/2)--(2,3/2);
        \draw[<->,shift={(-0.1,0)}] (-2,3/2+0.1)--(-2,2) node[midway,left] {$l/2$};
        \draw[<->,shift={(-0.1,0)}] (-2,1/2+0.1)--(-2,3/2-0.1) node[midway,left] {$\alpha$};
        \draw[<->,shift={(-0.1,0)}] (-2,0)--(-2,1/2-0.1) node[midway,left] {$l/2$};
    \end{tikzpicture}
    \caption{Diagram of an open string propagator with the length $\alpha+l$}\label{Kakustring}
\end{figure}
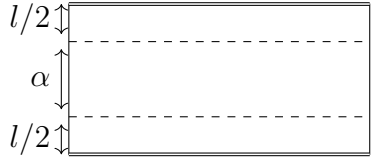

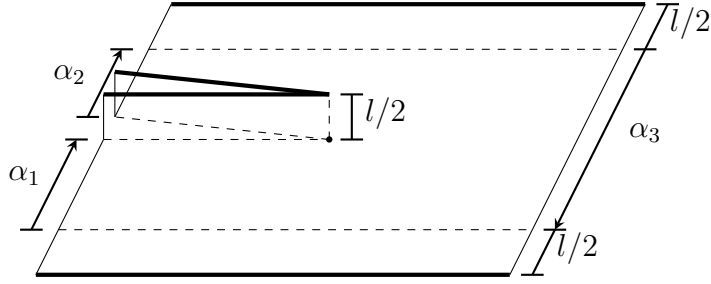
\begin{figure}
    \centering
\begin{tikzpicture}[scale=0.6]
\draw [ultra thick] (1.5,4) -- (12,4);
\draw [thin] (1.5,4) -- (3,7);
\draw [thin] (3.25,7.5) -- (4.5,10);
\draw [ultra thick] (4.5,10) -- (15,10);
\draw [thin] (12,4) -- (15,10);
\draw [dashed] (2,5) -- (12.5,5);
\draw [dashed] (4,9) -- (14.5,9);
\draw [thin] (3,7) -- (3,8);
\draw [thin] (3.25,7.5) -- (3.25,8.5);
\draw [dashed] (3,7) -- (8,7);
\draw [dashed] (8,7) -- (3.25,7.5);
\draw [dashed] (8,7) -- (8,8);
\draw [ultra thick] (3,8) -- (8,8);
\draw [ultra thick] (8,8) -- (3.25,8.5);
\coordinate (O) at (8,7);
\fill (O) circle (2pt);
\draw [thick] (1.15,5) -- (1.65,5);
\draw [thick] (2.15,7) -- (2.65,7);
\draw [thick] (2.4,7.5) -- (2.9,7.5);
\draw [thick] (3.15,9) -- (3.65,9);
\draw [thick] (12.25,4) -- (12.75,4);
\draw [thick] (12.75,5) -- (13.25,5);
\draw [thick] (14.75,9) -- (15.25,9);
\draw [thick] (15.25,10) -- (15.75,10);
\draw [thick] (12.5,4) -- (13,5);
\draw [thick] (15,9) -- (15.5,10);
\draw [thick, ->, >=stealth] (1.4,5) -- (2.4,7);
\draw [thick, ->, >=stealth] (2.65,7.5) -- (3.4,9);
\draw [thick, ->, >=stealth] (15,9) -- (13,5);
\node [above] at (1.25,5.75) {$\alpha_1$};
\node [above] at (2.25,8) {$\alpha_2$};
\node [above] at (13.5,4) {$l/2$};
\node [above] at (16,9) {$l/2$};
\node [above] at (15,6.7) {$\alpha_3$};
\draw [thick] (8.25,8) -- (8.75,8);
\draw [thick] (8.5,7) -- (8.5,8);
\draw [thick] (8.25,7) -- (8.75,7);
\node [above] at (9.25,7) {$l/2$};
\end{tikzpicture}
    \caption{The cubic Kaku vertex when $\alpha_1,\alpha_2>0,\alpha_3<0$.}\label{cubicKaku}
\end{figure}

In \cite{Erler:2020beb}, the field redefinition connecting Witten's bosonic string field 
theory and the Kugo-Zwiebach theory is given through Kaku's bosonic string field theory 
\cite{Kaku:1987jx}. The action of Kaku's bosonic string field theory consists of cubic and 
quartic Kaku vertices in addition to the kinetic term. 
To define the Kaku vertex, we assign the string length $\alpha+l$ to the diagram of the propagator as depicted in Fig.~\ref{Kakustring},
where $\alpha\ (=\sqrt{2}p^+)$ is the light-cone momentum and $l\in[0,\infty]$ is a parameter. Once we fix $l$, a theory is determined.
As is shown in Fig.~\ref{cubicKaku}, the portion of the worldsheet assigned the momentum $\alpha$ splits and joins in the same way as in the light-cone vertex, 
and the portion assigned the parameter $l$ is glued together in the same way as in the Witten vertex.
If we set $l=0$, the Kaku vertex coincides with the light-cone-type vertex.
The portion assigned $l$ shrinks to points and looks like the Chan-Paton factor.
Hence we call the parameter $l$ the Chan-Paton (CP) parameter. 
On the other hand, if we take the limit $l\to\infty$ along with scaling the coordinates by $1/l$, 
the portion assigned the momentum shrinks to a point, and we obtain the Witten vertex. 
In the same way, the quartic Kaku vertex is defined so that it coincides with the quartic light-cone vertex 
in the limit $l\to0$ whereas that vanishes in the limit $l\to\infty$ \cite{Kaku:1987jx,Erler:2020beb}. 
Therefore, in this sense, the Kaku theory is an intermediate 
theory connecting the Kugo-Zwiebach theory and Witten's bosonic string field theory. 

In a similar sense, it is quite natural to expect that the Kaku-type superstring field theory connects 
the Kugo-Zwiebach-type and Witten-type superstring field theories.
In this section, we find a field redefinition connecting the Witten-type superstring field theory and the 
Kugo-Zwiebach-type superstring field theory using the Kaku-type superstring field theory.

\subsection{Field redefinition 
}\label{field redef}

In this section, we construct a field redefinition connecting between the Witten-type and Kugo-Zwiebach-type superstring field theories.
The general discussion is the same for the bosonic case~\cite{Erler:2020beb}.
Let $\Psi_l$ represent the dynamical superstring field of Kaku-type superstring theory with a CP parameter $l$. 
The action of the Kaku-type superstring field theory with a CP parameter $l+\epsilon$, where $\epsilon$ represents 
an infinitesimal parameter, is written as
\begin{align}
    S_{l+\epsilon}=\sum_{n=1}^\infty \frac{1}{n+1}\Omega\left(\,\Psi_{l+\epsilon}\,,
\pi_1\bm{M}^{l+\epsilon}_n\frac{1}{1-\Psi_{l+\epsilon}}\right)\,.
\end{align}
String fields with different lengths $\Psi_{l+\epsilon}$ and $\Psi_l$ are related by 
the infinitesimal field redefinition maps $R^l_2, R^l_3,\dots$ such that
\begin{align}
    \Psi_{l+\epsilon}=\Psi_l-\epsilon\big[\,R^l_2(\Psi_l\,,\Psi_l)+R^l_3(\Psi_l\,,
\Psi_l\,,\Psi_l)+\dots\,\big]\,.
\end{align}
By using an even coderivation $\bd{R}^l=\bd{R}_2^l+\bd{R}^l_3+\dots$, 
this relation can be expressed as
\begin{align}
    \Psi_{l+\epsilon}=\Psi_l-\epsilon\pi_1\bd{R}^l\frac{1}{1-\Psi_l}\,.
\end{align}
Since $\bd{R}_n^l$ must be closed in $\mathcal{H}^{(\textrm{res})}$, its own picture number is equal to the own picture number of the string product $\bd{M}_n^l$, while its own ghost number is one less than the own ghost number of $\bd{M}_n^l$.
For the two actions $S_{l+\epsilon}$ and $S_l$ to give the same physics \cite{Hata:1993gf}\cite{Erler:2020beb}, $\bd{R}^l$ must be cyclic,
\begin{equation}
 \langle\Omega|\pi_2\bd{R}^l\ =\ 0\,,
\end{equation}
and must satisfy the differential equation
\begin{align}
    \pdv{}{l}\bm{M}^l=[\bm{M}^l\,,\bd{R}^l]\,.\label{eq:superfieldredef}
\end{align}
Expanded into component products, the differential equation implies
\begin{align}
    \pdv{}{l}\bm{M}^l_2&=[\bm{Q}\,,\bd{R}^l_2]\,,\\
    \pdv{}{l}\bm{M}^l_3&=[\bm{Q}\,,\bd{R}^l_3]+[\bm{M}^l_2\,,\bd{R}^l_2]\,,\\
    \pdv{}{l}\bm{M}^l_4&=[\bm{Q}\,,\bd{R}^l_4]+[\bm{M}^l_2\,,\bd{R}^l_3]+[\bm{M}^l_3\,,
\bd{R}^l_2]\,,\\
    &\quad\vdots\,.
\end{align}
These conditions are common to both bosonic string and superstring field theories, but note that in the case of bosonic Kaku-type string field theory, $\bd{M}^l_n=0$ for $n>4$. In contrast, in the case of its superstring extension, all the infinitely many $\bd{M}^l_n$ are nonzero. For the superstring field theory, the field redefinition $\bd{R}^l$ must also be closed in $\mathcal{H}^{\textrm{(res)}}$:
\begin{equation}
    [\bd{\eta}, \bd{R}^l]\ =\ 0,\qquad \mathcal{G}\mathcal{G}^{-1}\bd{R}^l\ =\ \bd{R}^l.
    \label{R closed}
\end{equation}
If there exists $\bd{R}^l$ satisfying this differential equation, the field redefinition between 
the string field $\Psi_{\textrm{W}}$ of Witten-type and the string field $\Psi_{\textrm{KZ}}$
of Kugo-Zwiebach-type superstring field theories can be given by 
\begin{align}
    \Psi_{\textrm{W}}=\pi_1\bm{G}\frac{1}{1-\Psi_{\textrm{KZ}}}\,,
\end{align}
where $\bm{G}$ is a cohomomorphism defined by 
\begin{align}
    \bm{G}=\overleftarrow{\mathcal{P}}\exp\left[-\int_0^\infty \dd l\,\bd{R}^l\right]\,,
\end{align}
where $\overleftarrow{\mathcal{P}}$ represents the path-ordered product from right to left in a sequence of 
increasing $l$.
A redefinition $\bd{R}^l$ satisfying these properties can be constructed as follows. 

As is given in Appendix~\ref{const M}, the superstring product $\bd{M}^l$ is constructed
in two steps. We first construct a cyclic $A_\infty$-algebra $(\mathcal{H}_l, \omega_l, \bd{Q}-\bd{\eta}+\bd{A}^l)$
in the large vector space, and then transform it to the product $\bd{M}^l$ with the $A_\infty$ algebra structure 
$(\mathcal{H}^{(\textrm{res})},\Omega,\bd{M}^l)$ using a cohomomorphism.
According to this construction, we first consider an even coderivation 
\begin{equation}
    \bd{\rho}^l(s,t) = \sum_{n,m,r=0}^\infty s^m t^n \bd{\rho}^{l(n)}_{m+n+r+1}\mid^{2r}\ \equiv\ \sum_{n=0}^\infty t^n \bd{\rho}^{(n)}_l(s),
\end{equation}
(with $\bd{\rho}^{l(0)}_1\mid^0\equiv0$) satisfying
\begin{subequations}\label{l der A}
    \begin{align}
    \frac{\partial}{\partial l}\bd{A}^l(s,t)\ =&\ [\bd{Q}, \bd{\rho}^l(s,t)] + [\bd{A}^l(s,t), \bd{\rho}^l(s,t)]^{1}
    + s [\bd{A}^l(s,t), \bd{\rho}^l(s,t)]^2,
    \label{der l A}\\
    [\bd{\eta}, \bd{\rho}^l(s,t)]\ =&\ t [\bd{A}^l(s,t), \bd{\rho}^l(s,t)]^2.
    \label{eta rho}
    \end{align}
\end{subequations}
The picture number of $\bd{\rho}^l_n$ is equal to the picture number of $\bd{A}^l_n$, while 
the ghost number of $\bd{\rho}^l_n$ is one less than the ghost number of $\bd{A}^l$.
These relations (\ref{l der A}) reduce at $t=0$ to those for the products that are closed in $\mathcal{H}^{(res)}$ and
have no own picture number:\footnote{
The relations hold for each power of $s$ corresponding to the different number of R-external lines.}
\begin{subequations}\label{l der bosonic}
    \begin{align}
    \frac{\partial}{\partial l}\bd{A}_l^{(0)}(s),\ =&\ [\bd{Q}, \bd{\rho}^{(0)}_l(s)] + [\bd{A}_l^{(0)}(s), \bd{\rho}_l^{(0)}(s)]^{1} 
    + s [\bd{A}_l^{(0)}(s), \bd{\rho}_l^{(0)}(s)]^2,
    \label{l der bosonic 1}\\
    [\bd{\eta}, \bd{\rho}_l^{(0)}(s)]\ =&\ 0.
\label{l der bosonic 2}
    \end{align}
\end{subequations}
As with $\bd{A}_l^{(0)}(s)$ in Eq.\,(\ref{A(s)}), we can construct such $\bd{\rho}_l^{(0)}(s)$ using the same technique as that for the case of the bosonic string field theory \cite{Erler:2020beb}.\footnote{
It is not difficult to extend the results given in \cite{Erler:2020beb} to the stubbed theory, which can be shown, e.g. by using recent developments
\cite{Schnabl:2023dbv,Schnabl:2024fdx,Erbin:2023hcs}.
} 
Therefore, $\bd{\rho}_l^{(0)}(s)$ that satisfies Eqs.\,(\ref{l der bosonic}) is assumed to be known.
On the other hand, Eqs.\,(\ref{l der A}) reduce at $(s,t)=(0,1)$ to 
\begin{subequations}\label{l der large}
    \begin{align}
        \frac{\partial}{\partial l}\bd{A}^l\ =&\ [\bd{Q}, \bd{\rho}^l] + [\bd{A}^l, \bd{\rho}^l]^1,\\
    [\bd{\eta}, \bd{\rho}^l]\ =&\ [\bd{A}^l, \bd{\rho}^l]^2,
\end{align}
\end{subequations}
with $\bd{A}^l=\bd{A}^l(0,1)$ and $\bd{\rho}^l=\bd{\rho}^l(0,1)$, which are equivalent to a similar relation to Eq.\,(\ref{eq:superfieldredef}) for an $A_\infty$-algebra $(\mathcal{H}_l,\bd{Q}-\bd{\eta}+\bd{A})$ in the large vector space:
\begin{equation}
 \frac{\partial}{\partial l}\bd{A}^l = 
[\bd{Q}-\bd{\eta}+\bd{A}^l, \bd{\rho}^l].
\label{l der Q-eta}
\end{equation}

For an explicit construction, let us show that such a $\bd{\rho}^l(s,t)$ satisfying Eqs.\,(\ref{l der A}) is obtained as a solution of differential equations,
\begin{subequations}\label{diff eqs}
    \begin{align}
    \frac{\partial}{\partial t}\bd{\rho}^l(s,t)\ =&\ [\bd{\rho}^l(s,t), \bd{\mu}^l(s,t)]^{1} + s [\bd{\rho}^l(s,t), \bd{\mu}^l(s,t)]^{2} 
    + \frac{\partial}{\partial l}\bd{\mu}^l(s,t) 
    \nonumber\\
&\hspace{0mm} 
+ [\bd{Q}, \bd{\sigma}^l(s,t)] + [\bd{A}^l(s,t),\bd{\sigma}^l(s,t)]^{1} + s [\bd{A}^l(s,t),\bd{\sigma}^l(s,t)]^{2},\\
[\bd{\eta}, \bd{\sigma}^l(s,t)]\ =&\ \frac{\partial}{\partial s}\bd{\rho}^l(s,t) + t[\bd{\rho}^l(s,t), \bd{\mu}^l(s,t)]^2 + t[\bd{A}^l(s,t),\bd{\sigma}^l(s,t)]^2,
\end{align}
\end{subequations}
by introducing an odd coderivation
\begin{equation}
    \bd{\sigma}^l(s,t)\ =\ \sum_{m,n,r=0}^\infty s^m t^n \bd{\sigma}^{l(n+1)}_{m+n+r+2}\mid^{2r}\ \equiv\ \sum_{n=0}^\infty t^n \bd{\sigma}_l^{(n+1)}(s).
\end{equation}
Here, $\bd{A}^l(s,t)$ and $\bd{\mu}^l(s,t)$ are string and gauge products (see Eqs.\,(\ref{string product}) and (\ref{gauge product})), respectively, for the Kaku-type superstring field theory. The coderivation $\bd{\rho}^l(s,t)$  has the same ghost and picture numbers as those of the gauge product $\bd{\mu}^l(s,t)$. 
If these differential equations hold, we can show that $\bd{I}(s,t)$ and $\bd{J}(s,t)$ defined by
\begin{align}
    \bd{I}(s,t)\ \equiv&\ \frac{\partial}{\partial l}\bd{A}^l(s,t) - [\bd{Q}, \bd{\rho}^l(s,t)] - [\bd{A}^l(s,t), \bd{\rho}^l(s,t)]^{1}
    - s [\bd{A}^l(s,t), \bd{\rho}^l(s,t)]^{2},\\
    \bd{J}(s,t)\ \equiv&\ [\bd{\eta}, \bd{\rho}^l(s,t)] - t [\bd{A}^l(s,t), \bd{\rho}^l(s,t)]^2,
\end{align}
satisfy
\begin{subequations}\label{der I J}
    \begin{align}
    \frac{\partial}{\partial t}\bd{I}(s,t)\ =&\ [\bd{I}(s,t), \bd{\mu}^l(s,t)]^{1} + s [\bd{I}(s,t), \bd{\mu}^l(s,t)]^{2},\\
    \frac{\partial}{\partial t}\bd{J}(s,t)\ =&\ \frac{\partial}{\partial s}\bd{I}(s,t) + t[\bd{I}(s,t), \bd{\mu}^l(s,t)]^2 
    + [\bd{J}(s,t), \bd{\mu}^l(s,t)]^{1} + s [\bd{J}(s,t), \bd{\mu}^l(s,t)]^{2}.
\end{align}
\end{subequations}
Since, as mentioned above, $\bd{I}(s,0)=\bd{J}(s,0)=0$ hold as relations for $\bd{\rho}_l^{(0)}(s)$ that we assumed to be known, Eqs.\,(\ref{der I J}) implies $\bd{I}(s,t)=\bd{J}(s,t)=0$ for ${}^{\forall} t$.

Therefore, we need to solve the differential Eqs.\,(\ref{diff eqs}) that are equivalent to the recurrence relations
\begin{subequations}\label{rec main}
\begin{align}
  (n+1)\bd{\rho}_l^{(n+1)}(s)\ =&\ \sum_{m=0}^n\, \left([\,\bd{\rho}_l^{(n-m)}(s), \bd{\mu}_l^{(m+1)}(s)]^{1} + s [\,\bd{\rho}_l^{(n-m)}(s), \bd{\mu}_l^{(m+1)}(s)]^{2} \right)
  + \frac{\partial}{\partial l}\bd{\mu}_l^{(n+1)}(s) 
    \nonumber\\
    &\    + [\bd{Q}, \bd{\sigma}_l^{(n+1)}(s)] + \sum_{m=0}^{n}\,\left([\bd{A}_l^{(n-m)}(s), \bd{\sigma}_l^{(m+1)}(s)]^{1} 
    + s [\bd{A}_l^{(n-m)}(s), \bd{\sigma}_l^{(m+1)}(s)]^{2}\right),
    \label{rec 1}\\
  [\bd{\eta}, \bd{\sigma}_l^{(n+1)}(s)]\ =&\
\frac{\partial}{\partial s}\bd{\rho}_l^{(n)}(s) 
\nonumber\\
&\
+\sum_{m=0}^{n-1}\left(\,[\bd{\rho}_l^{(n-m-1)}(s), \bd{\mu}_l^{(m+1)}(s)]^2 + [\bd{A}_l^{(n-m-1)}(s), \bd{\sigma}_l^{(m+1)}(s)]^2\right).
\label{rec 2}
\end{align}
\end{subequations}
Let us try to solve them recursively for $n$ like in the case of differential Eqs.\,(\ref{diff eq}). 
For $n=0$, Eq.\,(\ref{rec 2}) becomes
\begin{equation}
    [\bd{\eta}, \bd{\sigma}_l^{(1)}(s)]\ =\ \frac{\partial}{\partial s}\bd{\rho}_l^{(0)}(s).
\end{equation}
Since $\bd{\rho}_l^{(0)}(s)$ is assumed to be known, we can solve it as
\begin{equation}
    \bd{\sigma}_l^{(1)}(s)\ =\ \xi_0\circ\left(\frac{\partial}{\partial s}\bd{\rho}_l^{(0)}(s)\right),
    \label{sigma 1}
\end{equation}
using operation $\xi_0\circ$ defined by Eq.\,(\ref{xi circ}). Next, we find that Eq.\,(\ref{rec 1}) for $n=0$ is
\begin{align}
    \bd{\rho}^{(1)}_l(s)\ =&\ [\bd{\rho}^{(0)}_l(s), \bd{\mu}^{(1)}_l(s)]^{1} + s [\bd{\rho}^{(0)}_l(s), \bd{\mu}^{(1)}_l(s)]^{2} + \frac{\partial}{\partial l}\bd{\mu}^{(1)}_l(s)
\nonumber\\
&\ + [\bd{Q}, \bd{\sigma}_l^{(1)}(s)] + [\bd{A}^{(0)}_l(s), \bd{\sigma}^{(1)}_l(s)]^{1} + s [\bd{A}^{(0)}_l(s), \bd{\sigma}^{(1)}_l(s)]^{2},
\label{rho 1}
\end{align}
in which an unknown quantity of $\frac{\partial}{\partial l}\bd{\mu}_l^{(1)}(s)$ remains on the right-hand side. 
We have to find it independently, which is possible as follows.
By differentiate Eq.\,(\ref{mu 1}) by $l$, we have
\begin{align}
    [\bd{\eta}, \frac{\partial}{\partial l}\bd{\mu}_l^{(1)}(s)]\ =&\ \frac{\partial^2}{\partial l\partial s}\bd{A}_l^{(1)}(s)
    \nonumber\\
    =&\ \frac{\partial}{\partial s}\left(\,[\bd{Q}, \bd{\rho}^{(0)}_l(s)] + [\bd{A}_l^{(0)}(s), \bd{\rho}_l^{(0)}(s)]^{1} + s [\bd{A}_l^{(0)}(s), \bd{\rho}_l^{(0)}(s)]^{2}\right),
\end{align}
where we used Eq.\,(\ref{l der bosonic 1}) in the second equality. Since all the quantities on the right-hand side are known, we have
\begin{equation}
    \frac{\partial}{\partial l}\bd{\mu}_l^{(1)}(s)\ =\ \xi_0\circ\left(\,[\bd{Q}, \bd{\rho}^{(0)}_l(s)] + [\bd{A}_l^{(0)}(s), \bd{\rho}_l^{(0)}(s)]^{1}
    + s [\bd{A}_l^{(0)}(s), \bd{\rho}_l^{(0)}(s)]^{2}\right).
    \label{der l mu 1}
\end{equation}
Substituting this into Eq.\,(\ref{rho 1}), we obtain $\bd{\rho}_l^{(1)}(s)$. To solve the recurrence relations (\ref{rec main}) by repeating this procedure, we need to find $\frac{\partial}{\partial l}\bd{\mu}_l^{(n)}(s)$ systematically through 
the recurrence relation
\begin{align}
    [\bd{\eta}, \frac{\partial}{\partial l}\bd{\mu}^{(n+1)}_l(s)]\ =&\ \partial_{s}\partial_{l}\bd{A}_{l}^{(n)}(s)
    + \sum_{m=0}^{n-1}[\partial_{l}\bd{A}_{l}^{(n-m)}(s), \bd{\mu}_{l}^{(m+1)}(s)]
    \nonumber\\
    &\hspace{0mm}
    + \sum_{m=0}^{n-1}[\bd{A}_l^{(n-m-1)}(s), \frac{\partial}{\partial l}\bd{\mu}_l^{(m+1)}(s)]^2,
    \label{rec 3}
\end{align}
with 
\begin{equation}
        \partial_{l}\bd{A}_{l}^{(n)}(s)\ =\ [\bd{Q}, \bd{\rho}^{(n)}_l(s)] + \sum_{m=0}^n\,\left(
    [\bd{A}^{(n-m)}_l(s), \bd{\rho}_l^{(m)}(s)]^{1} + s [\bd{A}^{(n-m)}_l(s), \bd{\rho}_l^{(m)}(s)]^{2} \right),
\end{equation}
which is obtained from Eqs.\,(\ref{diff A 2}) and (\ref{der l A}) and reduces to Eq.\,(\ref{der l mu 1}) for $n=0$.
However, Eq.\,(\ref{der l A}), which derives Eq.\,(\ref{rec 3}), holds only for $\bd{\rho}^l(s,t)$ that is the solution of Eqs.\,(\ref{rec main}), so we need to carefully solve three recurrence relations in Eqs.\,(\ref{rec 1}), (\ref{rec 2}), and (\ref{rec 3}) simultaneously. We prove it using complete induction as follows. 

For $n=0$, we solved Eqs.\,(\ref{rec 2}) and (\ref{rec 3}) and obtained $\bd{\sigma}_l^{(1)}(s)$ and $\frac{\partial}{\partial l}\bd{\mu}_l^{(1)}(s)$ as in Eqs.\,(\ref{sigma 1}) and (\ref{der l mu 1}), respectively. Then, substituting the latter into Eq.\,(\ref{rec 1}), we obtained $\bd{\rho}_l^{(1)}(s)$.
Next, suppose that we obtained $\bd{\rho}_l^{(n)}(s)$, $\bd{\sigma}_l^{(n)}(s)$, and $\frac{\partial}{\partial l}\bd{\mu}_l^{(n)}(s)$ for ${}^{\forall}n\le k$. Consider Eqs.\,(\ref{rec 1}), (\ref{rec 2}), and (\ref{rec 3}) for $n=k$. 
We can obtain $\bd{\sigma}_l^{(k+1)}(s)$ by acting $\xi_0\circ$ on the right-hand side of the second recursive relation. 
The third recursive relation holds since it consists only known quantities, namely, the solutions of Eqs.\,(\ref{rec 1}) and (\ref{rec 2}), and thus, we can obtain $\frac{\partial}{\partial l}\bd{\mu}_l^{(k+1)}(s)$ by solving it.
Then, we obtain $\bd{\rho}_l^{(k+1)}(s)$ from the first recursive relation. Hence, we can obtain $\bd{\rho}_l^{(n)}(s)$ and $\bd{\sigma}_l^{(n)}(s)$ (and also $\frac{\partial}{\partial l}\bd{\mu}_l^{(n)}(s)$) for ${}^\forall n$, which determine $\bd{\rho}^l(s,t)$ and $\bd{\sigma}^l(s,t)$ (and also $\frac{\partial}{\partial l}\bd{\mu}^l(s,t)$). 

Now, we have $\bd{\rho}^l$ satisfying Eq.\,(\ref{l der Q-eta}). The next step is to transform it by a cohomomorphism for Kaku-type theory:
\begin{equation}
    \pi_1\hat{\bd{F}}_l^{-1}\ =\ \pi_1\id - \Xi \pi^1_1\bd{A}^l.
\end{equation}
It transforms the relation (\ref{l der Q-eta}) to
\begin{equation}
    \frac{\partial}{\partial l}\bd{M}^l\ =\ [\,\bd{M}^l - \bd{\eta}, \tilde{\bd{\rho}}^l],
    \label{M-eta}
\end{equation}
with 
\begin{equation}
    \tilde{\bd{\rho}}^l\ =\ \left(\hat{\bd{F}}_l^{-1}\bd{\rho}^l\hat{\bd{F}}_l\right)-\left(\frac{\partial}{\partial l}\hat{\bd{F}}_l^{-1}\right)\!\hat{\bd{F}}_l.
\end{equation}
Here, $\tilde{\bd{\rho}}^l$ is an even coderivation but is not closed in $\mathcal{H}^{(\textrm{res})}$. Instead, simple calculation gives
\begin{align}
    \pi_1^0\tilde{\bd{\rho}}^l\ =&\ \pi_1^0\bd{\rho}^l\hat{\bd{F}}_l
    \nonumber\\
    =&\ \pi_1^0\left(\bd{\rho}^l\hat{\bd{F}}_l + \bd{A}^l \hat{\bd{F}}_l\bd{\mathfrak{D}}(\Xi\pi_1^1\bd{\rho}^l\hat{\bd{F}}_l)\right) - \pi_1^0\bd{A}^l \hat{\bd{F}}_l\bd{\mathfrak{D}}(\Xi\pi_1^1\bd{\rho}^l\hat{\bd{F}}_l),\\
\pi_1^1\tilde{\bd{\rho}}^l\ =&\ \pi_1^1\bd{\rho}^l\hat{\bd{F}}_l + \Xi\pi_1^1\left(\partial_l\bd{A}^l-\bd{A}^l\bd{\rho}^l\right)\hat{\bd{F}}_l
\nonumber\\
=&\ X\pi_1^1\bd{\rho}^l\hat{\bd{F}}_l - (Q-\eta)\Xi\pi_1^1\bd{\rho}^l\hat{\bd{F}}_l-\Xi\pi_1^1\bd{\rho}^l\hat{\bd{F}}_l(\bd{M}^l-\bd{\eta})
\nonumber\\
=&\ X\pi_1^1\left(\bd{\rho}^l\hat{\bd{F}}_l + \bd{A}^l\hat{\bd{F}}_l\bd{\mathfrak{D}}(\Xi\pi_1^1\bd{\rho}^l\hat{\bd{F}}_l)\right)
\nonumber\\
&\ 
- (Q-\eta)\Xi\pi_1^1\bd{\rho}^l\hat{\bd{F}}_l-\Xi\pi_1^1\bd{\rho}^l\hat{\bd{F}}_l(\bd{M}^l-\bd{\eta})
-X\pi_1^1\bd{A}^l\hat{\bd{F}}_l\bd{\mathfrak{D}}(\Xi\pi_1^1\bd{\rho}^l\hat{\bd{F}}_l),
\end{align}
from which we have
\begin{equation}
    \tilde{\bd{\rho}}^l\ =\ \bd{R}^l - \left[\bd{M}^l-\bd{\eta},\ \bd{\mathfrak{D}}(\Xi\pi_1^1\bd{\rho}^l\hat{\bd{F}}_l)\right],
    \label{rho tilde}
\end{equation}
with
\begin{align}
    \pi_1\bd{R}^l\ =&\ \mathcal{G}\pi_1\bd{r}^l\,\\
    \bd{r}^l\ =&\ \bd{\mathfrak{D}}\left(\pi_1\left(
    \bd{\rho}^l\hat{\bd{F}}_l+\bd{A}^l\hat{\bd{F}}_l\bd{\mathfrak{D}}(\Xi\pi_1^1\bd{\rho}^l\hat{\bd{F}}_l)\right)\right).
    \label{true R}
\end{align}
Here, $\bd{\mathfrak{D}}(f)$ denotes the coderivation derived by a mapping $f:\mathcal{TH}\rightarrow\mathcal{H}$.
Therefore, the relation (\ref{M-eta}) can be rewritten as
\begin{equation}
    \frac{\partial}{\partial l}\bd{M}^l\ =\ [\bd{M}^l-\bd{\eta}, \bd{R}^l],
    \label{M-eta 2}
\end{equation}
or equivalently
\begin{equation}
    \frac{\partial}{\partial l}\bd{M}^l\ =\ [\bd{M}^l, \bd{R}^l],\qquad [\bd{\eta}, \bd{R}^l]\ =\ 0,
\end{equation}
because of $[\bd{M}^l-\bd{\eta}, \bd{M}^l-\bd{\eta}]=0$ and $\bd{R}^l$ defined by Eq.\,(\ref{true R}) has no picture number deficit.
Since it is obvious by definition (\ref{true R}) that $\mathcal{G}\mathcal{G}^{-1}\bd{R}^l=\bd{R}^l$, the remaining task is to show that $\bd{R}^l$ is cyclic with respect to $\Omega$, which is equivalent to $\bd{r}^l$ being cyclic with respect to $\omega$ because of
\begin{align}
    \langle\Omega|\pi_2\bd{R}^l\ =&\ \langle\omega|(\id\otimes\mathcal{G}^{-1})(\pi_1\otimes\mathcal{G}\pi_1\bd{r}^l + \mathcal{G}\pi_1\bd{r}^l\otimes\pi_1)
    \nonumber\\
    =&\ \langle\omega|(\pi_1\otimes\pi_1\bd{r}^l+\pi_1\bd{r}^l\otimes\pi_1)\ =\ 
    \langle\omega|\pi_2\bd{r}^l,
\end{align}
where we used the BPZ even property of $\mathcal{G}$. 
Since $\bd{r}^l$ is closed in $\mathcal{H}$, $[\bd{\eta}, \bd{r}^l]=0$, however, it is sufficient to show that $\bd{r}^l$ is cyclic with respect 
to the symplectic form in the large vector space $\omega_l$ \cite{Kunitomo:2019glq}. 
We show it by using the method given in \cite{Erler:2017onq}. First, we have
\begin{align}
    \langle\omega_l|\pi_2\bd{\mathfrak{D}}(\pi_1(\bd{\rho}^l\hat{\bd{F}}_l))\ =&\
    \langle\omega_l|\nabla\left(\pi_1\bd{\rho}^l\hat{\bd{F}}_l\otimes'\pi_1 + \pi_1\otimes'\pi_1\bd{\rho}^l\hat{\bd{F}}_l\right)\Delta
    \nonumber\\
    =&\ \langle\omega_l|\nabla\left(\pi_1\bd{\rho}^l\otimes'\pi_1\hat{\bd{F}}_l^{-1{}} + \pi_1\hat{\bd{F}}_l^{-1}\otimes'\pi_1\bd{\rho}^l\right)\Delta\hat{\bd{F}}_l
\nonumber\\
=&\ \langle\omega_l|\nabla\left(\pi_1\bd{\rho}^l\otimes'\pi_1(\id-\Xi\pi^1_1\bd{A}^l) + \pi_1(\id-\Xi\pi^1_1\bd{A}^l)\otimes'\pi_1\bd{\rho}^l\right)\Delta\hat{\bd{F}}_l
\nonumber\\
=&\ \langle\omega_l|\pi_2\bd{\rho}^l\hat{\bd{F}}_l
- \langle\omega_l|\nabla\left(\pi_1\bd{\rho}^l\otimes'\Xi\pi^1_1\bd{A}^l + \Xi\pi^1_1\bd{A}^l\otimes'\pi_1\bd{\rho}^l\right)\Delta\hat{\bd{F}}_l,
\label{cyc 1}
\end{align}
where $\nabla$ is the product, $\Delta$ is the coproduct, and $\otimes'$ denotes the tensor product of tensor algebras.
Similarly, using the fact that $\bd{\mathfrak{D}}(\pi_1\bd{A}^l\hat{\bd{F}}_l)$ is cyclic with respect to $\omega_l$, 
\begin{equation}
    \langle\omega_l|\left(\pi_1\bd{A}^l\hat{\bd{F}}_l\otimes \pi_1+\pi_1\otimes \pi_1\bd{A}^l\hat{\bd{F}}_l\right)\ =\ 0,
\end{equation}
we have
\begin{align}
&\     \langle\omega_l|\pi_2\bd{\mathfrak{D}}(\pi_1(\bd{A}^l\hat{\bd{F}}_l\bd{\mathfrak{D}}(\Xi\pi_1^1\bd{\rho}^l\hat{\bd{F}}_l)))
\nonumber\\ 
&=\ \langle\omega_l|\nabla\left(\pi_1(\bd{A}^l\hat{\bd{F}}_l\bd{\mathfrak{D}}(\Xi\pi_1^1\bd{\rho}^l\hat{\bd{F}}_l))\otimes'\pi_1+ \pi_1\otimes'\pi_1(\bd{A}^l\hat{\bd{F}}_l\bd{\mathfrak{D}}(\Xi\pi_1^1\bd{\rho}^l\hat{\bd{F}}_l))
     \right)\Delta
     \nonumber\\
&=\
\langle\omega_l|\nabla\left((\pi_1\bd{A}^l\hat{\bd{F}}_l\otimes'\pi_1)(\bd{\mathfrak{D}}(\Xi\pi_1^1\bd{\rho}^l\hat{\bd{F}}_l)\otimes'\id) + (\pi_1\otimes'\pi_1\bd{A}^l\hat{\bd{F}}_l)(\id\otimes'\bd{\mathfrak{D}}(\Xi\pi_1^1\bd{\rho}^l\hat{\bd{F}}_l))\right) \Delta    
     \nonumber\\
&=\
\langle\omega_l|\left((\pi_1\bd{A}^l\hat{\bd{F}}_l\otimes\pi_1)\nabla(\bd{\mathfrak{D}}(\Xi\pi_1^1\bd{\rho}^l\hat{\bd{F}}_l)\otimes'\id)\Delta + (\pi_1\otimes\pi_1\bd{A}^l\hat{\bd{F}}_l)\nabla(\id\otimes'\bd{\mathfrak{D}}(\Xi\pi_1^1\bd{\rho}^l\hat{\bd{F}}_l))\Delta\right)
\nonumber\\
&=\
- \langle\omega_l|\left((\pi_1\otimes\pi_1\bd{A}^l\hat{\bd{F}}_l)\nabla(\bd{\mathfrak{D}}(\Xi\pi_1^1\bd{\rho}^l\hat{\bd{F}}_l)\otimes'\id)\Delta + (\pi_1\bd{A}^l\hat{\bd{F}}_l\otimes\pi_1)\nabla(\id\otimes'\bd{\mathfrak{D}}(\Xi\pi_1^1\bd{\rho}^l\hat{\bd{F}}_l))\Delta\right)
\nonumber\\
&=\
\langle\omega_l|\nabla\left((\Xi\pi_1^1\bd{\rho}^l\hat{\bd{F}}_l\otimes'\pi_1\bd{A}^l\hat{\bd{F}}_l) - (\pi_1\bd{A}^l\hat{\bd{F}}_l\ \otimes'\Xi\pi_1^1\bd{\rho}^l\hat{\bd{F}}_l)\right)\Delta 
\nonumber\\
&=\
\langle\omega_l|\nabla\Big((\Xi\pi_1^1\bd{\rho}^l\otimes'\pi_1\bd{A}^l) - (\pi_1\bd{A}^l\ \otimes'\Xi\pi_1^1
\bd{\rho}^l)\Big)\Delta\hat{\bd{F}}_l.
\label{cyc 2}
\end{align}
Then, summing up Eqs.\,(\ref{cyc 1}) and (\ref{cyc 2}), we can show that
\begin{alignat}{2}
\langle\omega_l|\pi_2\bd{r}^l\ 
=&\ \langle\omega_l|\pi_2\bd{\rho}^l\hat{\bd{F}}_l
&&- \langle\omega_l|\nabla\left(\pi_1\bd{\rho}^l\otimes'\Xi\pi^1_1\bd{A}^l - \Xi\pi_1^1\bd{\rho}^l\otimes'\pi_1\bd{A}^l
\right)\Delta\hat{\bd{F}}_l
\nonumber\\
&  
&&- \langle\omega_l|\nabla\left(\Xi\pi^1_1\bd{A}^l\otimes'\pi_1\bd{\rho}^l + \pi_1\bd{A}^l\ \otimes'\Xi\pi_1^1\bd{\rho}^l\right)\Delta\hat{\bd{F}}_l
\nonumber\\
=&\ \langle\omega_l|\pi_2\bd{\rho}^l\hat{\bd{F}}_l
&&+ \langle\omega_l|\left( (\Xi\otimes\id-\id\otimes\Xi)\nabla\left(\pi_1\bd{\rho}^l\otimes'\pi^1_1\bd{A}^l
\right)\right)\Delta\hat{\bd{F}}_l
\nonumber\\
&
&&- \langle\omega_l|(\Xi\otimes\id-\id\otimes\Xi)\nabla\left(\pi^1_1\bd{A}^l\otimes'\pi_1\bd{\rho}^l\right)\Delta\hat{\bd{F}}_l\ =\ 0\,.
\end{alignat}
The first term vanishes because of the cyclicity of $\bd{\rho}^l$, which follows from the construction, and
the second and third terms also vanish since $\Xi$ is BPZ-even.
Thus, $\bd{R}^l$ can be shown to be cyclic with respect to $\Omega$, resulting the field redefinition that we construct.
The construction in this section is applicable even for the stubbed theory as long as it has an $A_{\infty}$ algebra structure.

\section{Conclusion and discussion}\label{concl}

In this paper, extending a recent development in the bosonic string field theory,
we have constructed a map from the Witten-type gauge-invariant superstring field theory based on an $A_{\infty}$ structure 
to a light-cone-gauge superstring field theory via the Kaku-type and Kugo-Zwiebach-type theories
as intermediate steps. 
Our construction of a chain map $S$ between BRST complexes $(\mathcal{H}_{\textrm{cov}}, Q)$ and $(\mathcal{H}_{\textrm{lc}}, Q^{\textrm{lc}})$ 
provides an alternative proof of the No-Ghost Theorem that is simpler and more concrete, in particular for the Rammond sector, than any other known proofs.\footnote{
See, for example, Refs.~\cite{Ito:1985qa}\cite{Lian:1989cy}\cite{Henneaux:1987ux,Figueroa-OFarrill:1989vhl}\cite{Dedushenko:2012ui}
}. Using an $A_{\infty}$ structure as a guiding principle, we have 
integrated out the longitudinal fields and have constructed a non-linear extension of $S$ 
called homotopy transfer.
In this construction, we have found that a naive extension only gives us an inconsistent
light-cone-gauge theory that sufferers from the well-known difficulty caused by divergence. 
However, we have also found that this divergence problem may be resolved by considering the stubbed theory and 
proposed it as a consistent light-cone-gauge superstring field theory.
In addition, we have connected the Kugo-Zwiebach-type superstring field theory to the Witten-type superstring field theory by a field redefinition 
via the Kaku-type theory. It possibly gives a proof of the unitarity of the Witten-type superstring field theory.

We should note that the absence of divergence in the stubbed theory shows another interesting fact.
In the first place, where does divergence come from although it does not exist in the gauge-invariant theory we start with?
If the stubbed theory is divergence-free, it should reproduce the correct amplitudes, just as its equivalent gauge-invariant theory would.
It means that contributions producing divergence cancel between those coming from the propagation of the transverse mode and the integration 
of the longitudinal degrees of freedom, and their sum connects smoothly to the contribution from the genuine (stubbed) quartic interaction.
From another point of view, the effective quartic interaction obtained by integrating out the longitudinal degrees of freedom also diverges in 
the stubless theory. If we consider the stub as regularization, the contribution from the
effective interaction provides a counter-term to make the amplitude finite.
In this viewpoint, we can consider that the effective vertices $\boldsymbol{M}^{\textrm{eff}}$ (\ref{M eff}) provide an infinite number of counter-terms 
with the help of $A_{\infty}$ algebra structure.
It is an interesting problem to directly verify the cancellation of divergence in  the four-string amplitudes
although it is difficult due to the use of different propagators for transverse and longitudinal degrees of freedom. 
It is necessary to develop a method to project the intermediate states onto each of the transverse and longitudinal 
fields separately and calculate them independently. This remains an issue that should be resolved in the future.

\section*{Acknowledgments}
The authors would like to thank the Yukawa Institute for Theoretical Physics (YITP) at Kyoto University 
and participants of the YITP workshop YITP-W-23-07 on\lq\lq Strings and Fields 2023" for discussion
useful to complete this work. 
Y.A. would like to thank the Atom research fellow program at YITP, Kyoto University for kind hospitality during part of this research. 
The work of Y.A. was partially supported by JST, the establishment of university fellowships towards the creation of science technology innovation, Grant Number JPMJFS2106.
The work of R.F was partially supported by JST, the establishment of university fellowships towards the creation of science technology innovation, Grant Number JPMJFS2105.
The work of H.K. is supported in part by JSPS Grant-in-Aid for Scientific Research (C) Grant Number JP18K03645.
The work of J.~T-Y. is supported in part by JSPS KAKENHI Grant Number JP23KJ1311.
\newpage
\appendix

\section{Conventions}\label{conventions}

\subsection{Coordinates}\label{coord}
Light-cone coordinates are denoted $x^{+},x^{-},x^{1},\cdots.x^{8}$.
There are related to Minkowski coordinates $x^{0},x^{1},\cdots,x^{9}$ through
\begin{align}
x^{\pm}=\frac{1}{\sqrt{2}}(\pm x^{0}+x^{9}).\label{lc coord}
\end{align}
Choosing the mostly plus metric in Minkowski space, $\eta^{\mu\nu}=\text{diag}(-,+,\cdots.+)$, the metric in light-cone coordinates is
\begin{align}
ds^{2}=2dx^{+}dx^{-}+(dx^{1})^{2}+\cdots+(dx^{8})^{2}.
\end{align}
Following Eq.\,(\ref{lc coord}), we define
\begin{equation}
    X^{\pm}\ =\ \frac{1}{\sqrt{2}}\left(\pm X^{0}+X^{9}\right),\qquad
    \psi^{\pm}\ =\ \frac{1}{\sqrt{2}}\left(\pm \psi^{0}+\psi^{9}\right).
\end{equation}

\subsection{Oscillator conventions}\label{oscillators}
In this paper, we use oscillator conventions
\begin{align}
X^{\mu}(z,\bar{z})&=X^{\mu}(z)+X^{\mu}(\bar{z}),\nonumber\\
X^{\mu}(z)&=\frac{1}{2}x^{\mu}-ip^{\mu}\ln{z}+\frac{i}{\sqrt{2}}\sum_{n\neq0}\frac{\alpha^{\mu}_{n}}{n}z^{-n},\nonumber\\
\sqrt{2}i\partial X^{\mu}(z)&= \sum_{n}\alpha^{\mu}_{n}z^{-n-1},\\
\psi^{\mu}(z)&=\sum_{r}\psi^{\mu}_{r}z^{-r-\frac{1}{2}},\\
b(z)&=\sum_{n}b_{n}z^{-n-2}, \hspace{10mm}c(z)=\sum_{n}c_{n}z^{-n+1},\\
\beta(z)&=\sum_{r}\beta_{r}z^{-r-\frac{3}{2}}, \hspace{10mm}\gamma(z)=\sum_{r}\gamma_{r}z^{-r+\frac{1}{2}},
\end{align}
where $n\in\mathbb{Z}$ and $r\in\mathbb{Z}+\kappa$ with $\kappa=1/2$ for the NS sector and $\kappa=0$ for the R sector.
We set $\alpha'$ and the string field coupling constant to one.
We also define a zero-mode $\alpha^{\mu}_{0}=\sqrt{2}p^{\mu}$.
The (anti-)commutation relations of the oscillators are
\begin{align}
[x^{\mu},p^{\nu}]&=i\eta^{\mu\nu},\\
[\alpha^{\mu}_{n},\alpha^{\nu}_{m}]&=n\eta^{\mu\nu}\delta_{n+m,0}, &\{\psi^{\mu}_{r},\psi^{\nu}_{s}\}&=\eta^{\mu\nu}\delta_{r+s,0},\\
\{b_{n},c_{m}\}&=\delta_{n+m,0}, &[\gamma_{r},\beta_{s}]&=\delta_{r+s,0}.
\end{align}
The hermiticity relations of the oscillators are 
\begin{align}
(x^{\mu})^{\dagger}&=x^{\mu}, &(p^{\mu})^{\dagger}&=p^{\mu},\nonumber\\
(\alpha^{\mu}_{n})^{\dagger}&=\alpha^{\mu}_{-n}, &(b_{n})^{\dagger}&=b_{-n}, &(c_{n})^{\dagger}&=c_{-n},\nonumber\\
(\psi^{\mu}_{r})^{\dagger}&=\psi^{\mu}_{-r}, &(\beta_{r})^{\dagger}&=-\beta_{-r}, &(\gamma_{r})^{\dagger}&=-\gamma_{-r}.
\end{align}
These conventions provide the operator product expansions (OPEs) as
\begin{align}
X^{\mu}(z,\bar{z})X^{\nu}(w,\bar{w})&\sim-\frac{1}{2}\eta^{\mu\nu}\Big(\ln|z-w|^2+\ln|z-\bar{w}|^2\Big),\nonumber\\
\partial X^{\mu}(z)\partial X^{\nu}(w)&\sim -\frac{1}{2}\frac{\eta^{\mu\nu}}{(z-w)^2},\qquad\quad
\psi^{\mu}(z)\psi^{\nu}(w)\sim\frac{\eta^{\mu\nu}}{z-w},\nonumber\\
b(z)c(w)&\sim\frac{1}{z-w}\sim c(z)b(w),\qquad
\gamma(z)\beta(w)\sim\frac{1}{z-w}\sim-\beta(z)\gamma(w).
\end{align}

\subsection{Spin operator}\label{spin op}

We define a bosonization of the world-sheet fermion as
\begin{subequations}
\begin{align}\label{spinor combinations}
\psi^{\pm}=\frac{1}{\sqrt{2}}\left(\pm\psi^{0}+\psi^{9}\right)=e^{\pm iH_{0}}\ ,\\
\frac{1}{\sqrt{2}}\left(\psi^{1}\pm i\psi^{5}\right)=e^{\pm iH_{1}}\ ,\\
\frac{1}{\sqrt{2}}\left(\psi^{2}\pm i\psi^{6}\right)=e^{\pm iH_{2}}\ ,\\
\frac{1}{\sqrt{2}}\left(\psi^{3}\pm i\psi^{7}\right)=e^{\pm iH_{3}}\ ,\\
\frac{1}{\sqrt{2}}\left(\psi^{4}\pm i\psi^{8}\right)=e^{\pm iH_{4}}\ ,
\end{align}
\end{subequations}
where $H_{I}\ (I=0,1,2,3,4)$ are chiral bosons and satisfy
\begin{align}
H_{I}(z)H_{J}(w)\sim-\delta_{IJ}\ln(z-w)\ .
\end{align}
The spin operator is defined as:\footnote{
Precisely speaking, we need appropriate Klein factors. See, for example, \cite{Kostelecky:1986xg,Asada:2017ykq}}
\begin{align}
\mathscr{S}_{\alpha}(z)=\exp\left\{\frac{i}{2}\sum^{4}_{I=0}\epsilon_{I}H_{I}(z)\right\},\hspace{10mm}(\alpha=1,\cdots,32)
\end{align}
where $\alpha=\{\epsilon_I\}$ with $\epsilon_{I}=\pm1$.
The OPE between two spin operators is normalized as
\begin{equation}
\mathscr{S}_{\alpha}(z)\mathscr{S}_{\beta}(w)\sim (z-w)^{-5/4}\mathcal{C}_{\alpha\beta},
\end{equation}
where $\mathcal{C}$ is the charge conjugate matrix satisfying
\begin{align}
\mathcal{C}^{T}=\mathcal{C}^{-1}=\mathcal{C}^{\dagger}=-\mathcal{C},\hspace{10mm}
\mathcal{C}^{\dagger}\mathcal{C}=1,\hspace{10mm}
\mathcal{C}\Gamma^{\mu}\mathcal{C}^{-1}=-(\Gamma^{\mu})^{T}.
\end{align}
The OPE between $\psi^{\mu}$ and $\mathscr{S}_{\alpha}$ is 
\begin{align}
\psi^{\mu}(z)\mathscr{S}_{\alpha}(w)\sim\frac{1}{\sqrt{2}}\frac{1}{(z-w)^{1/2}}\Gamma^{\mu}_{\alpha\beta}\mathscr{S}_{\beta}(w),
\end{align}
with the 10-dimensional gamma matrices  $\Gamma^{\mu}$ satisfying $\{\Gamma^{\mu},\Gamma^{\nu}\}=2\eta^{\mu\nu}$. We take 
a representation:
\begin{align}
    \Gamma^{0}=-i\sigma_{2}\otimes\mathbb{I}_{16} ,\hspace{10mm}
\Gamma^{i}=\sigma_{3}\otimes\gamma^{i},\hspace{10mm}
\Gamma^{9}=\sigma_{1}\otimes\mathbb{I}_{16},
\end{align}
where $\gamma^{i}$ is the 8-dimensional gamma matrices: $\{\gamma^{i},\gamma^{j}\}=2\delta^{ij}$.
We define $\Gamma_{11}=\Gamma^{0}\cdots\Gamma^{9}$ and take $\mathcal{C}=-\Gamma^0$. 

\subsection{Fock space representation}\label{Fock sp}

We take the Fock space representation of the vector space 
$\mathcal{H}=\mathcal{H}_{\textrm{mat}}\otimes\mathcal{H}_{bc}\otimes\mathcal{H}_{\beta\gamma}$ as follows.

\subsubsection{Matter sector $\mathcal{H}_{\textrm{mat}}$}

On the oscillator ground state
\begin{equation}
    \alpha_{n}^{\mu}\ket{0}\ =\ 0,\qquad n\ge0,\qquad \psi^{\mu}_{r}\ket{0}\ =\ 0,\qquad r>0,
\end{equation}
we can define the momentum eigenstate as
\begin{equation}
  \alpha_0^{\mu}\ket{k}\ =\ \sqrt{2}k^{\mu}\ket{k},\qquad \ket{k}\ =\ e^{ik\cdot X(0,0)}\ket{0}.
\end{equation}
In the NS sector, we identify the state $\ket{k}$ the matter ground state.
The vector space $\mathcal{H}_{\textrm{mat}}^{\textrm{NS}}$ is spanned by Fock states
\begin{equation}
    \psi_{-r_1}^{\nu_1}\cdots\psi_{-r_l}^{\nu_{j}}\alpha_{-n_1}^{\mu_1}\cdots\alpha_{-n_m}^{\mu_m}\ket{k}.
\end{equation}
The ground state of the R sector is given by
\begin{align}
\ket{\alpha;\, k}=\mathscr{S}_{\alpha}(0)\ket{k},
\end{align}
which is degenerated and the (space-time) spinor representation of the zero-mode $\psi_0^\mu$:
\begin{equation}
    \psi_0^{\mu}\ket{\alpha;k}\ =\ \frac{1}{\sqrt{2}}\Gamma^{\mu}_{\alpha\beta}\ket{\beta,k}.
\end{equation}
We can show that $\Gamma_{11}\ket{\alpha}=\tilde{\Gamma}_{11}\ket{\alpha}=\prod_{I}\epsilon_{I}\ket{\alpha}$ 
with $\tilde{\Gamma}_{11}\ =\ 2^5 \psi^0_0\psi^1_0\psi^1_0\cdots\psi^9_0$.
We restrict to $\prod_{I}\epsilon_{I}=1$ or $-1$ for a Weyl representation.
The vector space $\mathcal{H}_{\textrm{mat}}^{\textrm{R}}$ is spanned by Fock states
\begin{equation}
    \psi_{-r_1}^{\nu_1}\cdots\psi_{-r_l}^{\nu_{l}}\alpha_{-n_1}^{\mu_1}\cdots\alpha_{-n_m}^{\mu_m}\ket{\alpha;k}.
\end{equation}

The spinor ground states are decomposed to 
\begin{equation}
    |(+, a);\,k\rangle\ =\ e^{\frac{i}{2}H_0}\mathscr{S}_a^+(0)\ket{k},\qquad
    |(-, a);\,k\rangle\ =\ e^{-\frac{i}{2}H_0}\mathscr{S}_a^-(0)\ket{k},
\end{equation}
where
\begin{equation}
    \mathscr{S}_a^{\pm}\ =\ 
\exp\left\{\frac{i}{2}\sum^{4}_{i=1}\epsilon_{i}H_{i}(z)\right\},\hspace{10mm}(a=1,\cdots,16),
\end{equation}
with $a=\{\epsilon_i\}=\{\epsilon_1,\cdots,\epsilon_4\}$.
If $\Gamma_{11}=1\ (-1)$ on the ground state, $\prod_i\epsilon_i=\pm 1\ (\mp 1)$ for $\mathscr{S}^{\pm}_a$.
They satisfy
\begin{alignat}{2}
\tilde{\Gamma}^{\pm}|(\mp, a)\rangle\ =&\ |(\pm, a)\rangle,&\qquad     
\tilde{\Gamma}^{\pm}|(\pm, a)\rangle\ =&\ 0,
\end{alignat}
with $\tilde{\Gamma}^{\pm}=\frac{1}{\sqrt{2}}(\pm\psi_0^{0}+\psi_0^{9})$.
We have
\begin{equation}
    \psi_0^i\ket{(\pm,a)}\ =\ \mp\frac{1}{\sqrt{2}}\gamma^i_{ab}\ket{(\pm,b)},
\end{equation}
and
\begin{alignat}{2}
    \braket{(+,a)}{(+,b)}\ =&\ 0,\qquad& \braket{(+,a)}{(-,b)}\ =&\ \delta_{a,b},\\
    \braket{(-,a)}{(+,b)}\ =&\ -\delta_{a,b},\qquad& \braket{(-,a)}{(-,b)}\ =&\ 0.
\end{alignat}

\subsubsection{Ghost sectors $\mathcal{H}_{bc}\otimes\mathcal{H}_{\beta\gamma}$}

The ground state of conformal ghost $(b,c)$ is doubly degenerate,
\begin{equation}
    b_{n}\ket{\downarrow}\ =\ c_{n}\ket{\downarrow}\ =\ 0,\qquad    
    b_{n}\ket{\uparrow}\ =\ c_{n}\ket{\uparrow}\ =\ 0,\qquad n>0,\qquad    
\end{equation}
which are a representation of the zero-modes:
\begin{equation}
    |\!\downarrow\,\rangle\ =\ b_0|\!\uparrow\rangle\,,\qquad
    |\!\uparrow\,\rangle\ =\ c_0|\!\downarrow\rangle\,.
\end{equation}
Using the conformal vacuum $\ket{0}$, they can be written as
\begin{equation}
    |\!\downarrow\,\rangle\ =\ c_1|0\rangle,\qquad |\!\uparrow\,\rangle\ =\ c_0c_1|0\rangle.
\end{equation}
In this paper, we consider $|\!\downarrow\rangle$ as a \lq\lq ground state" unless otherwise specified.

There are an infinite number of choices for a \lq\lq ground state" of superconformal ghost $(\beta, \gamma)$, which are labeled by the picture number $l$:
\begin{equation}
\begin{split}
    \beta_{r}\ket{l}&=0,\qquad \text{for }\ r\geq-l-\frac{1}{2},\\
    \gamma_{r}\ket{l}&=0,\qquad \text{for }\ r\geq l+\frac{3}{2},
\end{split}
\end{equation}
where $r\in\mathbb{Z}+\kappa$ and $l$ is an integer for the NS sector and a half-integer for the R sector.
For the superstring field theory, we take $(-1)$-picture $(l=-1)$ for the NS sector and $(-1/2)$-picture $(l=-1/2)$ for the R sector, 
which are natural in that they have the lowest 2D energy. 

\subsubsection{$\mathcal{H}=\mathcal{H}_{\textrm{mat}}\otimes\mathcal{H}_{bc}\otimes\mathcal{H}_{\beta\gamma}$}

The vector space $\mathcal{H}$ is obtained by the direct product
of $\mathcal{H}_{\textrm{mat}}$, $\mathcal{H}_{bc}$, and $\mathcal{H}_{\beta\gamma}$.
For example, the massless vector state in the NS sector is given by
\begin{equation}
    \psi_{-1/2}^{\mu}\ket{\ket{k}}\ =\ \psi_{-1/2}^{\mu}\ket{k}\otimes\ket{\downarrow}\otimes\ket{-1}.
\end{equation}
The massless spinor state in the R sector is given by
\begin{equation}
    \ket{\ket{\alpha;\,k}}\ =\ \ket{\alpha;\,k}\otimes\ket{\downarrow}\otimes\ket{-1/2}.
\end{equation}
The string field is spanned by the states restricted by the GSO projection.

\subsection{Super conformal field theory}\label{SCFT}
In our conventions, the matter EM tensor and matter supercurrent are given by
\begin{align}
T_{\textrm{m}}(z)&=-\partial X^{\mu}\partial X_{\mu}-\frac{1}{2}\psi^{\mu}\partial\psi_{\mu},\\
G_{\textrm{m}}(z)&=\sqrt{2}\psi^{\mu}i\partial X_{\mu},
\end{align}
which satisfy superconformal algebra with central charge $c_{m}=15$.
The ghost EM tensor and ghost supercurrent are given by
\begin{align}
T_{\textrm{gh}}(z)&=-\frac{3}{2}\beta\partial\gamma-\frac{1}{2}\partial\beta\gamma-2b\partial c-\partial bc,\\
G_{\textrm{gh}}(z)&=\frac{3}{2}\beta\partial c+\partial\beta c-2b\gamma,
\end{align}
which also satisfy the superconformal algebra with $c=-15$.
The total central charge, total EM  tensor, and total supercurrent are given by the sum of those for the matter and ghost,
\begin{subequations}\label{T and G}
    \begin{align}
c_{\textrm{tot}}&=c_{\textrm{m}}+c_{\textrm{gh}}=0,\\
T(z)&=T_{\textrm{m}}(z)+T_{\textrm{gh}}(z),\\
G(z)&=G_{\textrm{m}}(z)+G_{\textrm{gh}}(z).
\end{align}
\end{subequations}

The light-cone-gauge EM tensor and supercurrent are given by Eqs.\,(\ref{lc EM perp}), (\ref{lc EM para}) and (\ref{lc SC para}).
We summarize here the form of their mode expansion:
\begin{align}
L^{\perp}_{n}=&
\frac{1}{2}\sum_{m}:\alpha^{i}_{n-m}\alpha^{i}_{m}:
+\frac{1}{2}\sum_{r\in\mathbb{Z}+\kappa}(r-\frac{n}{2}):\psi^{i}_{n-r}\psi^{i}_{r}:,\\
L^{\parallel}_{n}=&
\sum_{m\neq0,n}:\alpha^{+}_{n-m}\alpha^{-}_{m}:
+\sum_{r\in\mathbb{Z}+\kappa}(r-\frac{n}{2}):\psi^{+}_{n-r}\psi^{-}_{r}:\nonumber\\
&+\sum_{m\neq0}m:b_{n-m}c_{m}:
+\sum_{r\in\mathbb{Z}+\kappa}(r-\frac{n}{2}):\beta_{n-r}\gamma_{r}:,\\
G^{\perp}_{r}=&
\sum_{m}\psi^{i}_{r-m}\alpha^{i}_{m},\\
G^{\parallel}_{r}=&
\sum_{m\neq0,n}\psi^{\pm}_{r-m}\alpha^{\mp}_{m}
-\sum_{m}\gamma_{r-m}b_{m}
-\sum_{m\neq0}m\beta_{r-m}c_{m}.
\end{align}

\subsection{GSO projection}\label{GSO}

The GSO projection operator for the NS sector is defined by
\begin{align}
        \mathcal{P}^{\textrm{GSO}}_{\textrm{NS}}\ =&\ \frac{1}{2}\left(1+(-1)^{F_{\textrm{NS}}}\right),
        \label{GSO NS}\\
        F_{\textrm{NS}}\ =&\ \sum_{r>0}\left(\psi^\mu_{-r}\psi_{\mu\,r}+\beta_{-r}\gamma_r-\gamma_{-r}\beta_r\right)+1.
\end{align}
The light-cone-gauge vector space $\mathcal{H}_{\textrm{lc}}^{\textrm{NS}}$ (with the ghost number one) is spanned by the states
\begin{equation}
     \psi^{i_1}_{-r_1}\cdots\psi^{i_l}_{-r_l}\alpha^{j_1}_{-n_1}\cdots\alpha^{j_m}_{-n_m}\ket{\ket{k}}, 
\end{equation}
with $l=$ odd.

The GSO projection operator for the R sector is defined by
\begin{align}
        \mathcal{P}^{\textrm{GSO}}_{\textrm{R}}\ =&\ \frac{1}{2}\left(1+(-1)^{F_{\textrm{R}}}\tilde{\Gamma}_{11}\right),\\
        F_{\textrm{R}}\ =&\ \sum_{n=1}^\infty\left(\psi^\mu_{-n}\psi_{\mu\,n}+\beta_{-n}\gamma_n-\gamma_{-n}\beta_n\right)+\gamma_0\beta_0.
\end{align}
The light-cone-gauge Hilbert space $\mathcal{H}_{\textrm{lc}}^{\textrm{R}}$ is spanned by the states
\begin{align}
         &\ \psi^{i_1}_{-r_1}\cdots\psi^{i_l}_{-r_l}\alpha^{j_1}_{-n_1}\cdots\alpha^{j_m}_{-n_m}\ket{\ket{(-, a);\,k}}, \quad \textrm{for}\ l=\ \textrm{even},   \\
         &\ \psi^{i_1}_{-r_1}\cdots\psi^{i_l}_{-r_l}\alpha^{j_1}_{-n_1}\cdots\alpha^{j_m}_{-n_m}\ket{\ket{(-, \bar{a});\,k}}, \quad \textrm{for}\ l=\ \textrm{odd},   
\end{align}
where $a=\{\epsilon_i\}$ with $\prod_i\epsilon_i=-1$ and $\bar{a}=\{\epsilon_i\}$ with $\prod_i\epsilon_i=1$.


\section{Evidence for the Conjecture}\label{proof}
For any operator $\mathcal{O}$, the similarity transformation by $S=e^{-R}$ can be written as
\begin{align}\label{Similarity tr. of transverse op.}
&S\mathcal{O}S^{-1}
=\mathcal{O}
-[R,\mathcal{O}]
+\frac{1}{2}[R,[R,\mathcal{O}]]
-\frac{1}{6}[R,[R,[R,\mathcal{O}]]]
+\cdots.
\end{align}
We can calculate the action of $R$ on fields $(\sqrt{2}i\tilde{X}^+,\psi^+)$ and $(\sqrt{2}i\partial X^i,\psi^i)$ as
\begin{subequations}\label{act on field}
\begin{align}
&[R,\sqrt{2}i\tilde{X}^{+}(z)]=\frac{1}{\alpha_0^{+}}
\left\{
z\sqrt{2}i\tilde{X}^{+}\sqrt{2}i\partial\tilde{X}^{+}(z)
\right\},\\
&[R,\psi^{+}(z)]=\frac{1}{\alpha_0^{+}}
\left\{
\partial\left(z\sqrt{2}i\tilde{X}^{+}\psi^{+}(z)\right)
-\frac{1}{2}\partial\left(z\sqrt{2}i\tilde{X}^{+}\right)\psi^{+}(z)
+z\psi^{+}\sqrt{2}i\partial\tilde{X}^{+}(z)
\right\},
\end{align}
\end{subequations}
and
\begin{subequations}\label{R X psi}
\begin{align}
&[R,\sqrt{2}i\partial X^{i}(z)]=\frac{1}{\alpha_0^{+}}
\left\{
\partial\left(z\sqrt{2}i\tilde{X}^{+}\sqrt{2}i\partial X^{i}(z)\right)
+\partial\left(z\psi^{+}\psi^{i}(z)\right)
\right\},\label{R on pX}\\
&[R,\psi^{i}(z)]=\frac{1}{\alpha_0^{+}}
\left\{
\partial\left(z\sqrt{2}i\tilde{X}^{+}\psi^{i}(z)\right)
-\frac{1}{2}\partial\left(z\sqrt{2}i\tilde{X}^{+}\right)\psi^{i} (z)
+z\psi^{+}\sqrt{2}i\partial X^{i}(z)
\right\},\label{R on psi}
\end{align}
respectively.
From Eqs.\,(\ref{R X psi}), we have
\begin{align}
-[R,\alpha^{i}_{n}]
=
\frac{n}{\alpha_0^{+}}
\oint\frac{dz}{2\pi i}
z^{n}
\left\{
\sqrt{2}i\tilde{X}^{+}\sqrt{2}i\partial X^{i}(z)+\psi^{+}\psi^{i}(z)
\right\}.
\end{align}
\end{subequations}
Using Eqs.\,(\ref{act on field}), double and triple commutators become
\begin{align}
[R,[R,\alpha^{i}_{n}]]
&=\left(\frac{n}{\alpha_0^{+}}\right)^{2}
\oint\frac{dz}{2\pi i}
z^{n}
\left\{
(\sqrt{2}i\tilde{X}^{+})^{2}\sqrt{2}i\partial X^{i}(z)+2\sqrt{2}i\tilde{X}^{+}\psi^{+}\psi^{i}(z)
\right\},
\end{align}
and
\begin{align}
-[R,[R,[R,\alpha^{i}_{n}]]]
&=
\left(\frac{n}{\alpha_0^{+}}\right)^{3}
\oint\frac{dz}{2\pi i}
z^{n}
\left\{
(\sqrt{2}i\tilde{X}^{+})^{3}\sqrt{2}i\partial X^{i}(z)+3(\sqrt{2}i\tilde{X}^{+})^{2}\psi^{+}\psi^{i}(z)
\right\},
\end{align}
respectively.
It must be clear that this process exponentiates the factor $i\frac{n}{p^{+}}\tilde{X}^{+}$ and so Eq.\,(\ref{Similarity tr. of transverse op.}) gives
the DDF operator $A_n^i$:
\begin{align}
S\alpha^{i}_{n}S^{-1}
&=
\oint\frac{dz}{2\pi i} z^{n}
\left\{
\sqrt{2}i\partial X^{i}+\frac{n}{\alpha^{+}_{0}}\psi^{+}\psi^{i}
\right\}e^{i\frac{n}{p^{+}}\tilde{X}^{+}(z)}\nonumber\\
&=
e^{-i\frac{n}{2p^{+}}x^{+}}
\oint\frac{dz}{2\pi i}
\left\{
\sqrt{2}i\partial X^{i}+\frac{n}{\alpha^{+}_{0}}\psi^{+}\psi^{i}
\right\}e^{i\frac{n}{p^{+}}X^{+}(z)}.
\end{align}
The single and double commutators with a fermionic transverse oscillator $\psi_r^i$ are a little complicated, but
can be calculated as
\begin{align}
   -[R,\psi_r^i]\ =&\   \frac{1}{\sqrt{\alpha_0^+}}\oint\frac{dz}{2\pi i}z^r\Bigg\{
   \left(\frac{\alpha_0^+}{z}\right)^{\frac{1}{2}}\left(i\frac{r}{p^+}\tilde{X}^+ + \frac{1}{2}\left(\frac{z}{\alpha_0^+}\right)\sqrt{2}i\partial\tilde{X}^+\right)\psi^i
   \nonumber\\
&\hspace{30mm}
   -\left(\frac{z}{\alpha_0^+}\right)^{\frac{1}{2}}\sqrt{2}\psi^+i\partial X^i
   \Bigg\},
\end{align}
and
\begin{align}
   [R,[R,\psi_r^i]]\ =&\ 
   \frac{1}{\sqrt{\alpha_0^+}}\oint\frac{dz}{2\pi i}z^r\Bigg\{
\left(\frac{\alpha_0^+}{z}\right)^{\frac{1}{2}}\Bigg(-\frac{1}{4}\left(\frac{z}{\alpha_o^+}\sqrt{2}i\partial\tilde{X}^+\right)^2
+\left(i\frac{r}{p^+}\tilde{X}^+\right)^2\nonumber\\
&\hspace{30mm}
+\left(i\frac{r}{p^+}\tilde{X}^+\right)\left(\frac{z}{\alpha_0^+}\sqrt{2}i\partial\tilde{X}^+\right)\Bigg)\psi^i
\nonumber\\
&\hspace{30mm}
- \left(\frac{z}{\alpha_0^+}\right)^{\frac{1}{2}}\left(-\left(\frac{z}{\alpha_0^+}\sqrt{2}i\partial\tilde{X}^+\right)+2\left(i\frac{r}{p^+}\tilde{X}^+\right)\right)
\sqrt{2}\psi^+i\partial X^i 
\nonumber\\
&\hspace{30mm}
+ \left(\frac{z}{\alpha_0^+}\right)^{\frac{3}{2}}\psi^+\partial\psi^+\psi^i
   \Bigg\},
\end{align}
respectively, which are consistent with the DDF operator $B_r^i$.
For the other fields except for $(\sqrt{2}i\partial\tilde{X}^-,\psi^-)$, 
the same relations as in Eqs.\,(\ref{R X psi}) hold by replacing $(\sqrt{2}i\partial X^i,\psi^i)$ with $(\tilde{b},-\tilde{\beta})$
or $(\partial\tilde{c},-\tilde{\gamma})$.\footnote{
We can see that the same is true for $(\sqrt{2}i\partial\tilde{X}^+,\psi^+)$ although the relation for $\sqrt{2}i\partial\tilde{X}^+$ becomes simpler
due to $(\psi^+)^2=0$.} 

To show that it is a superconformal transformation,
let us reconsider it using superfields.
The similarity transformation $S$ is represented by superfields as
\begin{align}
S=e^{-R},\hspace{5mm}
R=\frac{i}{p^+}\oint dZ
zV(z,\theta)\boldsymbol{T}^{\text{lc}}(z,\theta),
\end{align}
where 
\begin{align}
V(z,\theta)=(1+\theta \partial_\theta)\bm{\tilde{X}}^+(z,\theta).
\end{align}
The light-cone EM tensor superfield $\boldsymbol{T}^{\text{lc}}(z,\theta)$ is given by
\begin{align}
\boldsymbol{T}^{\text{lc}}
&=\frac{1}{2}G^{\text{lc}}+\theta T^{\text{lc}}\nonumber\\
&=-D_{\theta}\boldsymbol{X}^i
\partial_z\boldsymbol{X}^i
-D_{\theta}\tilde{\boldsymbol{X}}^+
\partial_z\tilde{\boldsymbol{X}}^-
-D_{\theta}\tilde{\boldsymbol{X}}^-\partial_z\tilde{\boldsymbol{X}}^+
-\frac{1}{2}\tilde{\boldsymbol{\beta}}
\partial_z\tilde{\boldsymbol{c}}
+\frac{1}{2}D_{\theta}\tilde{\boldsymbol{\beta}}
D_{\theta}\tilde{\boldsymbol{c}}.
\end{align}
Suppose that $\Phi(z,\theta)$ is a weight-$h$ primary superfield with respect to the superconformal transformation
generated by $\boldsymbol{T}^{\textrm{lc}}$.
The OPE between $\boldsymbol{T}^{\textrm{lc}}$ and $\Phi$ is 
\begin{align}
\tilde{\boldsymbol{T}}^{\text{lc}}(z_1,\theta_1)\Phi(z_2,\theta_2)
\sim
h\frac{\theta_{12}}{z^2_{12}}\Phi(z_2,\theta_2)
+\frac{1}{2}\frac{1}{z_{12}}D_{\theta_2}\Phi(z_2,\theta_2)
+\frac{\theta_{12}}{z_{12}}\partial_2\Phi(z_2,\theta_2),
\end{align}
where 
\begin{align}
\theta_{12}&=\theta_1-\theta_2,\qquad
z_{12}=z_1-z_2-\theta_1\theta_2.
\end{align}
We can calculate a single commutator between $\Phi$ and $R$ as
\begin{align}
[\Phi(z,\theta),R]
&=\lambda
\left[
(zV)\partial\Phi+\frac{1}{2}D_\theta(zV)D_\theta\Phi+h\partial(zV)\Phi\right]
\nonumber\\
&=\lambda
\left[
\ell^{(h)}_0\left(V\Phi\right)
-\frac{1}{2}D_{\theta}
\left(
zD_{\theta}V\Phi
\right)
+\left(h-\frac{1}{2}\right)
z\partial V\Phi\right],
\end{align}
with
\begin{align}
\lambda=-\frac{i}{p^{+}},\hspace{5mm}
\ell^{(h)}_0=(h+z\partial+\frac{1}{2}\theta\partial_\theta)\,,
\end{align}
which coincides with the infinitesimal superconformal transformation with the parameter $zV(z,\theta)$.
We can also calculate the commutator between $V$ and $R$ as
\begin{equation}
    [V(z,\theta),R]
=\lambda
\left[
V\ell^{(0)}_0V
+\frac{1}{2}z\theta\partial VD_{\theta}V
\right],
\end{equation}
which yields the double commutator
\begin{align}
[[\Phi(z,\theta),R],R]
=
\lambda^2
&\left[
(l^{(h)}_0)^2(V^2\Phi)
+2(h-1)zl^{(h+1)}_0(V\partial V\Phi)
+zl^{(h+1)}_0(VD_{\theta}VD_{\theta}\Phi)
\right.\nonumber\\
&-\theta l^{(h+1/2)}_0(VD_{\theta}V\Phi)
+h(h-1)z^2(\partial V)^2\Phi\nonumber\\
&+\frac{h}{2}z^2\theta\partial(\partial VD_{\theta}V)\Phi
-\frac{h}{2}z^2\partial D_{\theta}VD_{\theta}V\Phi
+\frac{1}{2}z^2\theta \partial VD_{\theta} V\partial\Phi\nonumber\\
&\left.+hz^2\partial VD_{\theta} VD_{\theta}\Phi
-\frac{1}{4}z^2\theta D_{\theta}(\partial V
D_{\theta} V)D_{\theta} \Phi
\right].
\end{align}

On the other hand, a weight-$h$ primary superfield $\Phi(z,\theta)$ is transformed under 
the superconformal transformation $(z',\theta')=(F_z(z,\theta),F_\theta(z,\theta))$ as
\begin{align}
    \Phi(z',\theta')=\big(D_\theta F_\theta^{-1}(z',\theta')\big)^{2h}\Phi\big(F_z^{-1}(z',\theta'),F_\theta^{-1}(z',\theta')\big)\,.
\end{align}
So, we take
\begin{align}
    F_z(z,\theta;\lambda)&=\exp[-\lambda\left(\bm{X}^+(z,\theta)-\frac{x^+}{2}\right)]=z\exp[-\lambda\bm{\tilde{X}}^+(z,\theta)]\,,\\
    F_\theta(z,\theta;\lambda)&=\frac{D_\theta F_z(z,\theta;\lambda)}{\big(\partial_z F_z(z,\theta;\lambda)\big)^{1/2}}\,,
\end{align}
as Eq.\,(\ref{sctr.}) in the Conjecture and set
\begin{align}
    S^{(h)}[\Phi(z,\theta)](\lambda)\equiv\big(D_\theta F_\theta^{-1}(z,\theta;\lambda)\big)^{2h}\Phi\big(F_z^{-1}(z,\theta;\lambda),F_\theta^{-1}(z,\theta;\lambda)\big)\,.
    \label{Sh lambda}
\end{align}
Expanding $S^{(h)}[\Phi(z,\theta)](\lambda)$ in powers of $\lambda$,
\begin{equation}
    S^{(h)}[\Phi(z,\theta)](\lambda)=\sum_{n=0}^\infty\frac{\lambda^n}{n!}S^{(h)}_{n}[\Phi(z,\theta)],
\end{equation}
the Conjecture leads to
\begin{equation}
    \lambda^nS^{(h)}_{n}[\Phi(z,\theta)]=[\Phi(z,\theta),\underbrace{R],\cdots,R],R}_{n}].
    \label{Sh n}
\end{equation}
By differentiate Eq.\,(\ref{Sh lambda}), we can show that
\begin{align}
    \dv{S^{(h)}[\Phi(z,\theta)]}{\lambda}
    &=l^{(h)}_0\Big((1+\theta\partial_\theta)\bm{\tilde{X}}^+(F_z^{-1}(z,\theta),F_\theta^{-1}(z,\theta))S^{(h)}[\Phi(z,\theta)]\Big)\nonumber\\
    &-\frac{1}{2}D_\theta\Big(zD_\theta(1+\theta \partial_\theta)\bm{\tilde{X}}^+(F_z^{-1}(z,\theta),F_\theta^{-1}(z,\theta))S^{(h)}[\Phi(z,\theta)]\Big)\nonumber\\
    &+(h-\frac{1}{2})z\partial_z(1+\theta\partial_\theta)\bm{\tilde{X}}^+(F_z^{-1}(z,\theta),F_\theta^{-1}(z,\theta))S^{(h)}[\Phi(z,\theta)]\,,
\end{align}
which allows us to check Eq.\,(\ref{Sh n}) efficiently at some lower orders in $\lambda$.
At $\mathcal{O}(\lambda^0)$, we find
\begin{align}
    \lambda S^{(h)}_1[\Phi(z,\theta)]
    =\ &\lambda\left[l^{(h)}_0(V\Phi)-\frac{1}{2}D_\theta(zD_\theta V\Phi)+(h-\frac{1}{2})z\partial_zV\Phi\right]\ =\ [\Phi(z,\theta),R],\label{Sh one}
\end{align}
by noting that $S^{(h)}_0[\Phi(z,\theta)]=\Phi(z,\theta)$.
For the next order, we have
\begin{align}
    F_z^{-1}(z,\theta)&= z+\lambda z \bm{\tilde{X}}^++\mathcal{O}(\lambda^2)\,,\\
    F_\theta^{-1}(z,\theta)&= \theta+\lambda z\partial_\theta \bm{\tilde{X}}^+ +\frac{\lambda}{2}\theta\partial_z(z\bm{\tilde{X}}^+) + \mathcal{O}(\lambda^2)\,,
\end{align}
which leads to
\begin{align}
    (1+\theta \partial_\theta)\bm{\tilde{X}}^+(F_z^{-1}(z,\theta),F_\theta^{-1}(z,\theta))= V+
    \lambda Vl^{(0)}_0V+\frac{\lambda}{2} z\theta\partial_z V D_\theta V
    +\mathcal{O}(\lambda^2)\,.
\end{align}
Therefore, we get
\begin{align}
    S^{(h)}_2[\Phi(z,\theta)]=&l^{(h)}_0\Big[VS^{(h)}_1+\big(Vl^{(0)}_0V+\frac{1}{2}z\theta \partial_z VD_\theta V\big)\Phi\Big]\nonumber\\
    &+(h-\frac{1}{2})z\Big[\partial_z VS^{(h)}_1+\partial_z\big(Vl^{(0)}_0V+\frac{1}{2}z\theta \partial_z VD_\theta V\big)\Phi\Big]\nonumber\\
    &-\frac{1}{2}D_\theta\Big[ zD_\theta V S^{(h)}_1+zD_\theta \big(Vl^{(0)}_0V+\frac{1}{2}z\theta \partial_z VD_\theta V\big)\Phi\Big]\,,
\end{align}
at $\mathcal{O}(\lambda)$.

Substituting the previous result (\ref{Sh one}), we get
\begin{align}
    \lambda^2S^{(h)}_2[\Phi(z,\theta)]
    =&\lambda^2\Big[(l^{(h)}_0)^2(V^2\Phi)+2(h-1)zl^{(h+1)}_0(V\partial V\Phi)+zl^{(h+1)}_0(VD_\theta V D_\theta \Phi)\nonumber\\
    &-\theta l^{(h+1/2)}_0(VD_\theta V\Phi)+h(h-1)z^2(\partial V)^2\Phi\nonumber\\
    &+\frac{h}{2}z^2\theta \partial(\partial VD_\theta V)\Phi-\frac{h}{2}z^2\partial D_\theta VD_\theta V\Phi+\frac{1}{2}z^2\theta \partial VD_\theta V\partial\Phi\nonumber\\
    &+hz^2\partial VD_\theta VD_\theta \Phi -\frac{1}{4}z^2\theta D_\theta (\partial V D_\theta V)D_\theta \Phi\Big]
    =\big[\big[\Phi(z,\theta),R\big],R\big]\,,
\end{align}
after some calculations.

\sectiono{Superstring products with picture number}\label{const M}

We summarize the method to construct superstring products 
with their own picture numbers while respecting the $A_\infty$ algebra structure. 
The method is applicable regardless of the type of interactions provided, which gives a correct triangulation 
of the moduli space.
If we apply the method to the theory with the Witten-type interaction, or in general 
the one with stubs, we obtain the superstring field theory with $A_\infty$ 
algebra structure \cite{Erler:2016ybs,Kunitomo:2020xrl}.
A superstring extension of the Kugo-Zwiebach-type theory is obtained by applying 
it to the theory with light-cone interactions.
We will use an extension of Kaku theory 
achieved by applying it to the theory with Kaku-type interactions.
Their stubbed theories can also constructed.


We first consider a degree odd coderivation,
\begin{equation}
 \bd{A}(s,t)\ =\ \sum_{m,n,r=0}^\infty s^mt^n\bd{A}^{(n)}_{m+n+r+1}\mid^{2r}\ \equiv\ \sum_n t^n \bd{A}^{(n)}(s),\
 \label{string product}
\end{equation}
with $\bd{A}^{(0)}_1\mid^0\equiv0$, acting on a tensor algebra $\mathcal{TH}_l$, 
where the number $n$ in superscript parentheses denotes the picture number that the product has,
and the superscript after the vertical line deonotes the cyclic Ramond number defined by
the total number of related Ramond string fields: the sum of the number of Ramond inputs 
and Ramond output. If $\bd{A}(s,t)$ satisfies the differential equations
\begin{subequations} \label{diff eq}
 \begin{align}
 \frac{\partial}{\partial t}\bd{A}(s,t)\ =&\ [\bd{Q}, \bd{\mu}(s,t)] + [\bd{A}(s,t),\bd{\mu}(s,t)]^1+s[\bd{A}(s,t),\bd{\mu}(s,t)]^2,
 \label{dif A 1}\\
 [\bd{\eta},\bd{\mu}(s,t)]\ =&\ \frac{\partial}{\partial s}\bd{A}(s,t)\ + t[\bd{A}(s,t),\bd{\mu}(s,t)]^2,
\label{diff A 2}
\end{align}
\end{subequations}
with a degree even coderivation (gauge products)
\begin{equation}
 \bd{\mu}(s,t)\ =\ \sum_{m,n,r=0}^\infty s^mt^n\bd{\mu}^{(n+1)}_{m+n+r+2}\mid^{2r}\ \equiv\ \sum_{n=0}^\infty t^n \bd{\mu}^{(n+1)}(s),
 \label{gauge product}
\end{equation}
with $\bd{\mu}^{(0)}_2\mid^0\equiv0$, it satisfies
\begin{align}
  &\ [\bd{Q},\bd{A}(s,t)] + \frac{1}{2}[\bd{A}(s,t),\bd{A}(s,t)]^{1} + \frac{s}{2}[\bd{A}(s,t),\bd{A}(s,t)]^{2}\ =\ 0,
 \label{Q A}\\
&\ [\bd{\eta},\bd{A}(s,t)] - \frac{t}{2}[\bd{A}(s,t),\bd{A}(s,t)]^2\ =\ 0.
\label{eta A}
\end{align}
Here, the (graded) commutators $[\bd{A}, \bd{B}]^1$ and $[\bd{A}, \bd{B}]^2$ are defined by
\begin{align}
    [\bd{A}, \bd{B}]^1\ =&\ \bd{\mathfrak{D}}\left(\pi_1\left(\bd{A}\bd{\mathfrak{D}}\left(\pi_1^0\bd{B}\right)-(-1)^{AB}\bd{B}\bd{\mathfrak{D}}\left(\pi_1^0\bd{A}\right)\right)\right),\\
        [\bd{A}, \bd{B}]^2\ =&\ \bd{\mathfrak{D}}\left(\pi_1\left(\bd{A}\bd{\mathfrak{D}}\left(\pi_1^1\bd{B}\right)-(-1)^{AB}\bd{B}\bd{\mathfrak{D}}\left(\pi_1^1\bd{A}\right)\right)\right).
\end{align}
At $(s,t)=(0,1)$, these reduce to the $A_\infty$ relation of coderivation $\bd{Q}-\bd{\eta}+\bd{A}(0,1)$, and thus,
we obtain a cyclic $A_\infty$ algebra $(\mathcal{H}_l,\omega_l,\bd{Q}-\bd{\eta}+\bd{A})$ in the large vector space $\mathcal{H}_l$. 
Here and hereafter, we denote $\bd{A}(0,1)$ as simply $\bd{A}$. 
On the other hand, at $t=0$, the products reduce to those without their own picture number, 
\begin{equation}
 \bd{A}(s,0)\ =\ \sum_{m,r=0}^\infty s^m\bd{A}^{(0)}_{m+r+1}\mid^{2r}\ =\ \bd{A}^{(0)}(s),
 \label{A(s)}
\end{equation}
satisfying
\begin{align}
 &\ [\bd{Q},\bd{A}^{(0)}(s)] + \frac{1}{2}[\bd{A}^{(0)}(s),\bd{A}^{(0)}(s)]^1+\frac{s}{2}[\bd{A}^{(0)}(s),\bd{A}^{(0)}(s)]^2\ =\ 0,\\
&\ [\bd{\eta},\bd{A}^{(0)}(s)]\ =\ 0.
\end{align}
Except for the point that the parameter $s$ distinguishes the number of R-external lines, 
this is essentially the same as those of bosonic string-products. 
Thus we can construct them according to the general method given in Refs.\cite{LeClair:1988sp,LeClair:1988sj}.
Therefore, we solve the differential Eqs.\,(\ref{diff eq}) under the initial condition $\bd{A}^{(0)}(s)$.
It is achieved by rewriting Eqs.\,(\ref{diff eq}) as recursive relations
    \begin{align}
  (n+1)\bd{A}^{(n+1)}(s)\ =&\ 
 [\bd{Q}, \bd{\mu}^{(n+1)}(s)] \nonumber\\
&\  + \sum_{m=0}^n\,\left([\bd{A}^{(n-m)}(s), \bd{\mu}^{(m+1)}(s)]^{1} + s [\bd{A}^{(n-m)}(s), \bd{\mu}^{(m+1)}(s)]^{2} \right),
\label{rec A 1}\\
[\bd{\eta}, \bd{\mu}^{(n+1)}(s)]\ =&\ 
\frac{\partial}{\partial s}\bd{A}^{(n)}(s) + \sum_{m=0}^{n-1}\,[\bd{A}^{(n-m-1)}(s), \bd{\mu}^{(m+1)}(s)]^2.
\label{rec A 2}
\end{align}
For $n=0$, Eq.\,(\ref{rec A 2}) becomes
\begin{equation}
    [\bd{\eta}, \bd{\mu}^{(1)}(s)]\ =\ \frac{\partial}{\partial s}\bd{A}^{(0)}(s),
\end{equation}
which can be solved as
\begin{equation}
    \bd{\mu}^{(1)}(s)\ =\ \xi_0\circ\left(\frac{\partial}{\partial s}\bd{A}^{(0)}(s)\right).
    \label{mu 1}
\end{equation}
Here, $\xi_o\circ \bd{A}$ is defined as an operation on each linear mapping $A_n=\pi_1\bd{A}\pi_n$ as
\begin{equation}
    \xi_0\circ A_n\ =\ \frac{1}{(n+1)}\left(\xi_0A_n - \sum_{k=0}^{n-1}A_n(\id^{\otimes n-k-1}\otimes\xi_0\otimes\id^{\otimes k} )\right).
    \label{xi circ}
\end{equation}
Then, substituting it into Eq.\,(\ref{rec A 1}) for $n=0$,
\begin{equation}
    \bd{A}^{(1)}(s)\ =\ [\bd{Q}, \bd{\mu}^{(1)}(s)] + [\bd{A}^{(0)}(s), \bd{\mu}^{(1)}(s)]^{1} + s [\bd{A}^{(0)}(s), \bd{\mu}^{(1)}(s)]^{2},
\end{equation}
we obtain $\bd{A}^{(1)}(s)$. Repeating this procedure recursively gives us $\bd{A}^{(n)}(s)$ for arbitrary $n$, or in other words $\bd{A}(s,t)$ 
\cite{Kunitomo:2019glq,Kunitomo:2020xrl}. Since we define the operation $\xi_0\circ$ to respect cyclicity, we can show that $\bd{A}(s,t)$ is cyclic with respect to 
the symplectic form $\omega_l$. 

The superstring products $\bd{M}=\bd{Q}+\bd{M}_{\textrm{int}}$ is obtained by transforming $\bd{A}$ using a cohomomorphism
\begin{equation}
 \pi_1\hat{\bd{F}}^{-1}\ =\ \pi_1\id - \Xi\pi_1^1\bd{A},
 \label{cohomo}
\end{equation}
as 
\begin{align}
\bd{M}\ =\ \hat{\bd{F}}^{-1}(\bd{Q}+\bd{A})\hat{\bd{F}},\qquad
\pi_1\bd{M}_{\textrm{int}}\ =\ \mathcal{G}\pi_1\bd{A}\hat{\bd{F}}. 
\end{align}


Since the factor $\mathcal{G}$ in front of $\bd{M}_{\textrm{int}}$ cancels the factor $\mathcal{G}^{-1}$ in the symplectic form
$\Omega$, we can show that $\bd{M}$ is cyclic with respect $\Omega$.
Taking the Witten-type, light-cone-type, or Kaku-type string
products as an initial condition $\bd{A}(s,0)$ of the differential Eqs.\,(\ref{diff eq}), 
we can obtain the superstring extension of the Witten-type, Kugo-Zwiebach-type, and Kaku-type
superstring field theory, respectively. In the case of the Kaku-type product, we specify the parameter $l$,
such as $\bd{A}^l$, $\bd{M}^l$, and so on. We can extend them to the stubbed theories
by choosing the corresponding initial conditions \cite{Kunitomo:2020xrl}.

\end{document}